\documentclass[a4paper,11pt]{article}
\pdfoutput=1 

\usepackage{amssymb}
\usepackage{dsfont}
\usepackage[T1]{fontenc} 
\usepackage{physics}
\usepackage{comment}
\usepackage{amsmath}
\usepackage{graphicx}
\usepackage{mathtools}
\usepackage{array}
\usepackage{makecell}
\usepackage{float}
\setcellgapes{3pt}
\usepackage{xcolor}
\usepackage{booktabs}

\usepackage[normalem]{ulem}
\usepackage[numbers,sort&compress]{natbib}
\usepackage{geometry}
\usepackage{pdflscape}
\usepackage{stackrel}

\definecolor{azure}{rgb}{0.0, 0.5, 1.0}
\definecolor{darkblue}{rgb}{0.15,0.35,0.7}
\definecolor{reddish}{rgb}{0.65, 0.2, 0.2}
\definecolor{brandeisblue}{rgb}{0.0, 0.44, 1.0}
\definecolor{ceruleanblue}{rgb}{0.16, 0.32, 0.75}
\definecolor{indigo(dye)}{rgb}{0.0, 0.25, 0.42}
\definecolor{grey}{rgb}{0.9,0.9,0.9}
\definecolor{dgrey}{rgb}{0.3,0.3,0.3}
\definecolor{dgreen}{rgb}{0.345098, 0.596078, 0.14902}

\usepackage[linktocpage=true]{hyperref}
\hypersetup{
colorlinks=true,
citecolor=ceruleanblue,
linkcolor=ceruleanblue,
urlcolor=ceruleanblue,
pdfauthor={},
pdftitle={},
pdfsubject={}
}

\usepackage{colortbl}
\definecolor{dgreen}{rgb}{0, 0.55, 0}
\definecolor{llightyellow}{rgb}{1.0, 0.95, 0.7}
\definecolor{llightblue}{rgb}{0.7, 0.9, 1.0}
\definecolor{llightpink}{rgb}{1.0, 0.85, 0.95}
\definecolor{llightgreen}{rgb}{0.7, 1.0, 0.4}
\colorlet{lightyellow}{llightyellow!50!white}
\colorlet{lightblue}{llightblue!50!white}
\colorlet{lightgreen}{llightgreen!50!white}
\colorlet{lightpink}{llightpink!50!white}

\usepackage{cleveref}
\usepackage{bm}
\crefname{lem}{lemma}{lemmas}
\crefname{thm}{theorem}{theorems}
\crefname{cor}{corollary}{corollaries}
\crefname{rem}{remark}{remarks}
\crefname{prop}{proposition}{propositions}

\usepackage{colortbl}
\definecolor{dgreen}{rgb}{0, 0.55, 0}
\definecolor{llightyellow}{rgb}{1.0, 0.95, 0.7}
\definecolor{llightblue}{rgb}{0.7, 0.9, 1.0}
\definecolor{llightpink}{rgb}{1.0, 0.85, 0.95}
\definecolor{llightgreen}{rgb}{0.7, 1.0, 0.4}
\colorlet{lightyellow}{llightyellow!50!white}
\colorlet{lightblue}{llightblue!50!white}
\colorlet{lightgreen}{llightgreen!50!white}
\colorlet{lightpink}{llightpink!50!white}

\usepackage{tikzit}

\usepackage[framemethod=TikZ]{mdframed} 
\usepackage{tikz-cd} 
\usetikzlibrary{external}
\usetikzlibrary{arrows,snakes,shapes.arrows,decorations.markings,shapes.geometric}

     \tikzset{>=triangle 90}
     \tikzstyle{bbc}=[draw,circle,fill=black,scale=.75]
     \tikzstyle{rc}=[circle,fill=red,scale=.6]
     \tikzstyle{wc}=[draw,circle,scale=.75]

\tikzset{snake it/.style={decorate, decoration=snake}}

\tikzset{
	on each segment/.style={
		decorate,
		decoration={
			show path construction,
			moveto code={},
			lineto code={
				\path [#1]
				(\tikzinputsegmentfirst) -- (\tikzinputsegmentlast);
			},
			curveto code={
				\path [#1] (\tikzinputsegmentfirst)
				.. controls
				(\tikzinputsegmentsupporta) and (\tikzinputsegmentsupportb)
				..
				(\tikzinputsegmentlast);
			},
			closepath code={
				\path [#1]
				(\tikzinputsegmentfirst) -- (\tikzinputsegmentlast);
			},
		},
	},
	mid arrow/.style={postaction={decorate,decoration={
				markings,
				mark=at position .5 with {\arrow[#1]{stealth}}
	}}},
}

\usetikzlibrary{decorations.markings}

\tikzset{line/.style={line width=0.25mm},
curve/.style={line,smooth,tension=1},
->-/.style={decoration={
  markings,
  mark=at position #1 with {\arrow[>=stealth]{>}}},postaction={decorate}},
-<-/.style={decoration={
  markings,
  mark=at position #1 with {\arrow[>=stealth]{<}}},postaction={decorate}},
}

\tikzset{bg/.style={opacity=.5}}

\tikzstyle{red dot}=[fill={rgb,255: red,240; green,165; blue,165}, draw=black, thick, shape=circle, minimum size=3mm, inner sep=0.2mm,font=\small]
\tikzstyle{green dot}=[fill={rgb,255: red,216; green,248; blue,216}, draw=black, thick, shape=circle,minimum size=3mm, inner sep=0.2mm,font=\small]
\tikzstyle{had}=[fill=yellow, draw=black, shape=rectangle]

\tikzstyle{tiny red dot}=[fill={rgb,255: red,240; green,165; blue,165}, draw=black, thick, shape=circle, minimum size=2.2mm, inner sep=0.12mm,font=\small]
\tikzstyle{tiny green dot}=[fill={rgb,255: red,216; green,248; blue,216}, draw=black, thick, shape=circle,minimum size=2.2mm, inner sep=0.12mm,font=\small]

\tikzstyle{thick line}=[fill=none, thick, draw=black, <-]
\tikzstyle{narrow}=[fill=none, thick, draw=black, ->-=0.5]

\newcommand{\fixme}[1]{\textcolor{red}{[#1]}}

\tikzset{
    partial ellipse/.style args={#1:#2:#3}{
        insert path={+ (#1:#3) arc (#1:#2:#3)}
    }
}

\makeatletter
\renewcommand\section{\@startsection {section}{1}{\z@}%
                               {-3.5ex \@plus -1ex \@minus -.2ex}
                               {2.3ex \@plus.2ex}%
                               {\normalfont\large\bfseries}}
\renewcommand\subsection{\@startsection{subsection}{2}{\z@}%
                                 {-3.25ex\@plus -1ex \@minus -.2ex}%
                                 {1.5ex \@plus .2ex}%
                                 {\normalfont\bfseries}}
\makeatother

\let\non\nonumber

\newfont{\goth}{ygoth.tfm scaled 1200}                   
\newcommand{\Goth}[1]{\ensuremath{\bf\goth #1}}      

\newcommand{\hph}[1]{{\hphantom{#1}}}

\renewcommand{\thesection}{\arabic{section}}
\renewcommand{\theequation}{\thesection.\arabic{equation}}
\numberwithin{equation}{section}

\renewcommand\baselinestretch{1.25}

\newcommand{\be}{\begin{equation}}
\newcommand{\ee}{\end{equation}}
\newcommand{\bee}{\begin{equation} \begin{aligned}}
\newcommand{\eee}{\end{aligned} \end{equation}}

\newcommand\IA{\mathbb{A}}
\newcommand\IB{\mathbb{B}}
\newcommand\IC{\mathbb{C}}
\newcommand\ID{\mathbb{D}}
\newcommand\IE{\mathbb{E}}
\newcommand\IF{\mathbb{F}}
\newcommand\IG{\mathbb{G}}
\newcommand\IH{\mathbb{H}}
\newcommand\II{\mathbb{I}}
\newcommand\IK{\mathbb{K}}
\newcommand\IL{\mathbb{L}}
\newcommand\IM{\mathbb{M}}
\newcommand\IN{\mathbb{N}}
\newcommand\IO{\mathbb{O}}
\newcommand\IP{\mathbb{P}}
\newcommand\IQ{\mathbb{Q}}
\newcommand\IR{\mathbb{R}}
\newcommand\IS{\mathbb{S}}
\newcommand\IT{\mathbb{T}}
\newcommand\IU{\mathbb{U}}
\newcommand\IV{\mathbb{V}}
\newcommand\IW{\mathbb{W}}
\newcommand\IX{\mathbb{X}}
\newcommand\IY{\mathbb{Y}}
\newcommand\IZ{\mathbb{Z}}

\newcommand\CA{\mathcal{A}}
\newcommand\CB{\mathcal{B}}
\newcommand\CC{\mathcal{C}}
\newcommand\CD{\mathcal{D}}
\newcommand\CE{\mathcal{E}}
\newcommand\CF{\mathcal{F}}
\newcommand\CG{\mathcal{G}}
\newcommand\CH{\mathcal{H}}
\newcommand\CI{\mathcal{I}}
\newcommand\CJ{\mathcal{J}}
\newcommand\CK{\mathcal{K}}
\newcommand\CL{\mathcal{L}}
\newcommand\CM{\mathcal{M}}
\newcommand\CN{\mathcal{N}}
\newcommand\CO{\mathcal{O}}
\newcommand\CP{\mathcal{P}}
\newcommand\CQ{\mathcal{Q}}
\newcommand\CR{\mathcal{R}}
\newcommand\CS{\mathcal{S}}
\newcommand\CT{\mathcal{T}}
\newcommand\CU{\mathcal{U}}
\newcommand\CV{\mathcal{V}}
\newcommand\CW{\mathcal{W}}
\newcommand\CX{\mathcal{X}}
\newcommand\CY{\mathcal{Y}}
\newcommand\CZ{\mathcal{Z}}

\newcommand{\sfT}{\mathsf{T}}
\newcommand{\sfS}{\mathsf{S}}
\newcommand{\sfH}{\mathsf{H}}
\newcommand{\cz}{\mathsf{CZ}}
\newcommand{\cx}{\mathsf{CX}}
\newcommand{\sfP}{\mathsf{P}}
\newcommand{\cnot}{\mathsf{CNOT}}
\newcommand{\swap}{\mathsf{SWAP}}
\newcommand{\kw}{\mathsf{KW}}
\newcommand{\spt}{\mathsf{SPT}}

\newcommand{\VEC}{\operatorname{Vec}}
\newcommand\Ker{\operatorname{Ker}}
\newcommand\DW{\operatorname{DW}}
\newcommand\Spin{\operatorname{Spin}}
\newcommand\wo{\widehat{\omega}}
\newcommand{\GL}{\mathrm{GL}}

\newcommand{\bg}{\mathbf{g}}
\newcommand{\bh}{\mathbf{h}}
\newcommand{\dsi}{\mathds{1}}

\newcommand{\otau}{\overline{\tau}}
\newcommand{\ochi}{\overline{\chi}}
\newcommand{\oomega}{\overline{\omega}}
\newcommand{\oQ}{\overline{Q}}
\newcommand{\oq}{\overline{q}}
\newcommand{\oL}{\overline{L}}
\newcommand{\oj}{\overline{j}}
\newcommand{\oV}{\overline{V}}
\newcommand{\oT}{\overline{T}}
\newcommand{\oeta}{\overline{\eta}}
\newcommand{\red}{\textcolor{red}}

\newcommand{\ii}{\mathsf{i}}
\newcommand{\opartial}{\overline{\partial}}
\newcommand{\oh}{\overline{h}}
\newcommand\rbar{\overline{r}}
\newcommand\Rep{\operatorname{Rep}}
\newcommand\TY{\operatorname{TY}}
\newcommand\Hom{\operatorname{Hom}}
\newcommand\He{\operatorname{He}}
\newcommand\p{\operatorname{p}}

\newcommand{\figref}[1]{Fig.\,\ref{#1}}
\newcommand{\tabref}[1]{Tab.\,\ref{#1}}
\newcommand{\secref}[1]{Sec.\,\ref{#1}}
\newcommand{\appref}[1]{Appendix\,\ref{#1}}

\newcommand{\dcl}[1]{{\color{orange}{[(DCL) #1]}}}
\newcommand{\xf}[1]{{\color{blue}{[(XF) #1]}}}
\newcommand{\ccc}[1]{{\color{orange}{#1}}}
\newcommand{\zhengdi}[1]{{\color{red}{ZS: #1}}}
\newcommand{\yzy}[1]{{\textcolor[rgb]{0.567426, 0.32317, 0.729831}{[\textsf{YZY:} #1]}}}

\newcommand{\asym}{\sigma_\text{sym}}
\newcommand{\asymi}{\sigma^{-1}_\text{sym}}
\newcommand{\splv}[3]{V_{\CL_{#3}}^{\CL_{#1},\CL_{#2}}}
\newcommand{\tw}{{}^\text{tw}}
\newcommand{\twi}[1]{{}^{#1}}

\newcommand{\cond}{\text{cond}}
\newcommand{\gauge}{\text{gauge}}
\newcommand{\initial}{\text{initial}}
\newcommand{\final}{\text{final}}
\newcommand{\se}{\text{e}}
\newcommand{\so}{\text{o}}

\newcommand{\veca}{\mathbf{a}}
\newcommand{\vecb}{\mathbf{b}}
\newcommand{\Aut}{\operatorname{Aut}}

\newcommand\thetatilde{\widetilde{\theta}}
\newcommand\phitilde{\widetilde{\phi}}
\newcommand\Ctilde{\widetilde{C}}
\newcommand{\cst}[1]{{\color{orange}\st{#1}}}
\newcommand{\tocite}[1]{[{\color[rgb]{0.13,0.55,0.13}{#1}}]}
\newcommand{\ha}{\widehat{a}}
\newcommand{\hA}{\widehat{A}}
\newcommand{\hZ}{\widehat{Z}}

\newcommand{\tX}{\widetilde{X}}
\newcommand{\tZ}{\widetilde{Z}}
\newcommand{\tA}{\widetilde{A}}
\newcommand{\tB}{\widetilde{B}}
\newcommand{\tH}{\widetilde{H}}
\newcommand{\teta}{\widetilde{\eta}}
\newcommand{\TG}{\mathsf{TG}}
\newcommand{\tri}{\mathsf{Tri}}
\newcommand{\oCQ}{\overline{\mathcal{Q}}}
\newcommand{\pal}{\mathsf{P}}

\newcommand{\sdg}{\text{diag}}

\newcommand{\bkS}{\mathrm{S}}
\newcommand{\bkT}{\mathrm{T}}
\newcommand{\bkR}{\mathrm{R}}

\newcommand{\prb}{Phys. Rev. B}
\newcommand{\prd}{Phys. Rev. D}
\newcommand{\prl}{Phys. Rev. Lett.}

\newcommand{\Hm}{\widetilde{\mathsf{H}}}
\newcommand{\Cm}{\widetilde{\mathsf{C}}}
\newcommand{\auto}{\mathrm{auto}}
\newcommand{\bphi}{\bar{\phi}}
\newcommand{\hCO}{\widehat{\CO}}
\newcommand{\simp}{\text{Irr}}
\newcommand{\bCS}{\overline{\CS}}

\newcommand{\ketx}[1]{|#1\rangle}
\newcommand{\brax}[1]{\langle #1|}
\newcommand{\braketx}[2]{\langle {#1}|{#2} \rangle}

\newcommand{\blfootnote}[1]{%
	\begingroup
	\renewcommand\thefootnote{}%
	\footnote{#1}%
	\addtocounter{footnote}{-1}
	\endgroup
}

\newcommand{\dagfootnote}[1]{%
	\begingroup
	\renewcommand\thefootnote{\dag}%
	\footnote{#1}%
	\addtocounter{footnote}{-1}%
	\endgroup
}
\newcommand{\starfootnote}[1]{%
	\begingroup
	\renewcommand\thefootnote{*}%
	\footnote{#1}%
	\addtocounter{footnote}{-1}%
	\endgroup
}

\definecolor{pink}{rgb}{1.0, 0.2, 0.6}
\definecolor{cyan}{rgb}{0.0, 0.6, 1.0}

\makeatletter
\providecommand{\leftsquigarrow}{%
  \mathrel{\mathpalette\reflect@squig\relax}%
}
\newcommand{\reflect@squig}[2]{%
  \reflectbox{$\m@th#1\rightsquigarrow$}%
}
\makeatother

\begin{document}

\begin{titlepage}
\begin{center}

\hfill         \phantom{xxx}

\vskip 2 cm {\Large \bf Strange correlator and string order parameter for non-invertible symmetry protected topological phases in 1+1d}

\vskip 1.25 cm {\bf Da-Chuan Lu${}^{1,2}$\dagfootnote{dclu137@gmail.com}, Fu Xu${}^{3}$
	 \blfootnote{The first two authors contribute equally.}, Yi-Zhuang You${}^{4}$}\non\\

\vskip 0.2 cm
 {\it ${}^{1}$ Department of Physics and Center for Theory of Quantum Matter, University of Colorado, Boulder, Colorado 80309, USA}
\vskip 0.2 cm
 {\it ${}^{2}$ Department of Physics, Harvard University, Cambridge, MA 02138, USA}

\vskip 0.2 cm
 {\it ${}^{3}$ Department of Physics, Nanjing University, Nanjing, Jiangsu 210093, China}

\vskip 0.2 cm
 {\it ${}^{4}$ Department of Physics, University of California, San Diego, CA 92093, USA}

\vskip 0.2 cm

\end{center}
\vskip 0.8 cm

\begin{abstract}
\noindent
\baselineskip=15pt
We construct strange correlators and string order parameters for non-invertible symmetry protected topological phases (NISPTs) in 1+1d spin chains. These diagnostics and the entanglement spectrum are connected to the interface algebra between two NISPTs. To compare phases, we first use finite-depth circuits of locally symmetric gates to minimize the bond dimension of each matrix product state. The resulting action tensors determine local strange charged operators and the charge decorations of truncated symmetry operators. The strange correlator probes the representation of the interface algebra realized by the two states. Higher-dimensional irreducible components yield nontrivial strange charged operators. For the distinct phases studied here, we identify such operators with long-range strange correlations. The full interface algebra admits a one-dimensional representation precisely when the phases coincide. For an on-site symmetry with a symmetric product reference state, the dimensions of the irreducible representations of the interface algebra also constrain degeneracies in the entanglement spectrum. We use unitary Drinfeld twists of a finite-dimensional Hopf $C^*$-algebra $H$ to construct general fixed point $\mathrm{Rep}(H)$ SPTs, action tensors and string order parameters. Our examples include the three $\mathrm{Rep}(D_8)$ NISPTs and intrinsically mixed $G\times\mathrm{Rep}(G)$ SPTs with explicit Hamiltonians for $G=S_3$. For the self-dual Hopf algebra $A_{12}$, we construct Hamiltonians and string order parameters for the two $\mathrm{Rep}(A_{12})$ SPT phases and evaluate a strange correlator between them. We also outline a geometric extension to higher dimensions.
\end{abstract}
\end{titlepage}

\tableofcontents

\flushbottom

\section{Introduction}
\begingroup
\color{black}
\hypersetup{citecolor=ceruleanblue,linkcolor=ceruleanblue}
Symmetry constrains quantum phases of matter and their transitions. Generalized symmetries extend this role beyond group actions to include higher-form and non-invertible symmetries \cite{gaiotto:generalizedsym,McGreevy:2022review,Schafer-Nameki:2023review,Shao:2023review}. In 1+1d, a finite non-invertible symmetry is described by a fusion category $\CC$: its simple objects label symmetry lines, and their fusion replaces group multiplication. Familiar examples are Kramers-Wannier duality at the Ising critical point and $\Rep(G)$ symmetry in group-valued spin chains, whose lines are labeled by irreducible representations of a finite non-abelian group $G$ \cite{kramers1941kw,Frohlich:2004ef,seiberg2023lsm,nat2023gcluster,sakura2024reps3,arkya2024reps3}. Lattice realizations make these symmetries explicit and allow their consequences for quantum states to be studied microscopically \cite{Aasen:2016dop,verstraete2023mpo,Inamura2022nigap,lan2024mpo}.

For ordinary group symmetry, symmetric gapped states with a unique ground state can belong to distinct symmetry protected topological (SPT) phases: they cannot be adiabatically connected without closing the gap while preserving the symmetry \cite{xgw2009spt,frank2012mpsspt,xie2011mpsspt,xie2011mpsspt2,chen2011z2spt,pollmann2010entanglement,cirac2011spts,xgw2013spt}. An anomaly-free fusion category likewise admits symmetric gapped phases with a unique ground state, which we call non-invertible symmetry protected topological phases (NISPTs) \cite{Thorngren:2019iar,Choi:2023xjw,kansei2024nispt}. The Ising fusion category is anomalous \cite{Thorngren:2019iar,Choi:2023xjw,sahand2024kw}, whereas $\Rep(G)$ is anomaly free for any finite $G$. We focus on lattice realizations of NISPTs in 1+1d and ask how different phases can be distinguished using local operator insertions on a closed chain.

Ordinary SPTs can be characterized by symmetry-protected boundary modes, string order parameters and entanglement-spectrum degeneracies \cite{ian1987edge,pollmann2010entanglement,den1989stringo,tasaki1992stringo,frank2008stringo,pollmann2012stringo,cirac2012sptorder}. Extending these diagnostics to NISPTs requires care. A symmetry-preserving open chain requires compatible boundary terms and may need additional boundary degrees of freedom \cite{kansei2024nispt,lan2024mpo}. Some NISPTs can also be mapped to symmetry-breaking phases by gauging or Kennedy-Tasaki transformations \cite{Sahand2024cluster,lu2025symset}, but a general mapping need not be available for non-group-theoretical symmetries, and spatial symmetry can complicate its implementation. 
String order parameters have been constructed for several $\Rep(G)$ examples \cite{nat2023gcluster,alison2024repd8s}. Here we give a systematic construction for general NISPTs. We study diagnostics on a closed chain with periodic boundary conditions, without choosing boundary terms or introducing additional boundary degrees of freedom. Their construction uses the symmetry action at an interface between two NISPTs, encoded by representations of the interface algebra.

The distinction between NISPT phases is encoded by fiber functors. A fiber functor assigns finite-dimensional Hilbert spaces to symmetry lines, together with unitary fusion maps compatible with the category's associativity \cite{Thorngren:2019iar,kansei2024nispt}. These maps retain the fusion information. To compare two NISPT states, we study the symmetry action at their interface, described by the interface algebra \cite{kitaev2012gapbdy,choi2024bdytube,kansei2024nispt}. Its representations provide the symmetry data used in the diagnostics developed below.

The strange correlator was introduced as a diagnostic of ordinary SPTs \cite{yizhuang2014strange}. It is a normalized matrix element of local operators between two symmetric states on the same Hilbert space with the same symmetry action,
\begingroup
\color{black}
\begin{equation}\label{eq:scdef}
    SC(I,J) \equiv \frac{\bra{B}\bar{\CS}_I\CS_J\ket{A}}{\bra{B}\ket{A}}
\end{equation}
\par\endgroup
where $\ket{B}$ is the state of interest, $\ket{A}$ is a reference state, and $\CS_I$ is supported near $I$. We consider nonzero overlap $\bra{B}\ket{A}$, even when it becomes small with increasing system size. The local insertions need not themselves be symmetry invariant: their transformation properties provide information that is not contained in the overlap alone.

The physical interpretation follows from Wick rotation: the overlap becomes an interface between the two states, and the strange correlator probes correlations along that interface \cite{yizhuang2014strange}. Suitable local operators can therefore reveal the interface modes through long-range or power-law correlations. However, being charged under the symmetry is not sufficient to make an operator a useful diagnostic. Even for ordinary SPTs, some charged operators fail to distinguish a chosen pair of phases \cite{paganelli2023stranget,qi2023strange1}. This motivates constructing the local insertions from the interface symmetry data, rather than testing unrelated operators individually.

Our main contribution is an algebraic construction connecting the interface algebra to strange correlators and string order parameters. {\color{black}\hypersetup{citecolor=ceruleanblue,linkcolor=ceruleanblue}Matrix product operators (MPOs) describe the symmetry, and matrix product states (MPSs) describe the states \cite{cirac2021mpsreview,jose2023mpoclas,kansei2024nispt}. To diagnose the phases, we fix the symmetry action. For each chosen MPS, we minimize the bond dimension over states reachable from it by finite-depth circuits of locally symmetric gates, allowing symmetric product ancillas. Each gate commutes with every component of each symmetry MPO restricted to its support. From the action tensors of the resulting representatives we construct the local strange charged operators and the operators dressing a truncated symmetry MPO. These observables can be used in numerical studies of NISPTs.} After a common finite blocking makes both MPS tensors injective, the action-tensor contraction can be represented by a local physical operator. The construction is illustrated below, with the green squares denoting the local insertions; its detailed form is given in \eqref{eq:strangecharge}.
\begingroup
\color{black}
\begin{align}
    &\bra{B}\ket{A} = \begin{tikzpicture}
    		\tikzset{baseline=(current bounding box.center)}
    		\draw[thick, cyan] (-0.5,0) -- (0.5,0);
    		\draw[thick, orange] (-0.5,0.8) -- (0.5,0.8);
    	\end{tikzpicture}\   =\  \begin{tikzpicture}
    		\tikzset{baseline=(current bounding box.center)}
    		\draw[thick, cyan] (-0.5,0) -- (0.5,0);
    		\draw[thick, orange] (-0.5,0.8) -- (0.5,0.8);
    		\draw[thick, black, dashed] (-0.5,0.4) -- (0.5,0.4);
    	\end{tikzpicture}\  \propto\  
    	\begin{tikzpicture}
    		\tikzset{baseline=(current bounding box.center)}
    		\draw[thick, cyan] (-0.5,0) -- (0.5,0);
    		\draw[thick, orange] (-0.5,0.8) -- (0.5,0.8);
    		\draw[thick, black, dashed] (-0.5,0.4) -- (-0.2,0.4);
    		\draw[thick, black, dashed] (0.5,0.4) -- (0.2,0.4);
    		\filldraw[fill=white, draw=black, thick] (0,0.4) circle (0.2);
    		\node at (0.3,0.6) {$x$};
    	\end{tikzpicture}\  =\  \begin{tikzpicture}
    		\tikzset{baseline=(current bounding box.center)}
    		\draw[thick, cyan] (-0.5,0) -- (0.5,0);
    		\draw[thick, orange] (-0.5,0.8) -- (0.5,0.8);
    		\draw[thick, black, dashed] (-0.5,0.4) -- (-0.2,0.4);
    		\draw[thick, black, dashed] (0.5,0.4) -- (0.2,0.4);
    		\draw[thick,black] plot[smooth] coordinates {(0.15,0) (0.2,0.4) (0.15,0.8)};
    		\draw[thick,black] plot[smooth] coordinates {(-0.15,0) (-0.2,0.4) (-0.15,0.8)};
    		\node at (0.3,0.6) {$x$};
    	\end{tikzpicture}\   \cong\   \begin{tikzpicture}
    		\tikzset{baseline=(current bounding box.center)}
    		\draw[thick, cyan] (-0.5,0) -- (0.5,0);
    		\draw[thick, orange] (-0.5,0.8) -- (0.5,0.8);
    		\draw[thick,gray] (0.2,0) -- (0.2,0.8);
    		\draw[thick,gray] (-0.2,0) -- (-0.2,0.8);
    		\node[draw, rectangle,minimum width=2mm, minimum height=2mm, inner sep=0mm,thick, fill=dgreen] at (0.2,0.4) {};
    		\node[draw, rectangle,minimum width=2mm, minimum height=2mm, inner sep=0mm,thick, fill=dgreen] at (-0.2,0.4) {};
    	\end{tikzpicture}=\bra{B}\CS_I \CS_J \ket{A} 
\end{align}
\par\endgroup
Starting from the overlap of the orange and blue states, we insert an identity line and expand it into a symmetry loop. Absorbing the loop into the two states produces local action tensors, whose contraction defines the strange charged operator. This realizes the interface-algebra action through local insertions in a closed-chain overlap. For the representatives studied below, suitable insertions give a nonzero strange correlator at long distances. We also obtain long-range order for a symmetry-preserving deformation of the Even state by reconstructing the insertions from its blocked MPS. Its behavior for other pairs must be checked separately; the construction does not imply decay in every unmatched phase. However, representatives within the same SPT phase can realize a higher-dimensional irreducible representation of the interface algebra and exhibit long-range strange correlations. Such correlations alone do not imply distinct SPT phases, as illustrated by the $\Rep(H_8)$ example in \secref{sec:interalg}.

We also construct string order parameters by dressing a truncated symmetry MPO with local charge decoration operators. The action tensors determine these operators and the contraction of their symmetry indices. For a matched fixed point MPS, this gives a string with a nonzero long-distance expectation value. A string designed for one phase can nevertheless remain nonzero in another phase, so distinguishing phases requires a suitable set of observables. The $S_3\times\Rep(S_3)$ example makes this explicit: two distinct fixed point states share the same sign-string expectation value, while the selected collection of string and strange correlators distinguishes all three states considered.

Finally, for the entanglement spectrum of the NISPTs, one can examine the non-invertible symmetry action on the left half Schmidt eigenstates. In an on-site realization with a symmetric product reference, these states transform under the interface algebra between the NISPT and the reference. The dimensions of its irreducible representations therefore constrain the entanglement-spectrum degeneracies. This argument requires a consistent restriction of the symmetry to one side of the cut and does not directly apply to a general non-on-site realization \cite{lan2024mpo,david2024nononsite,lu2024nononsite,sahand2025nononsite}. A chosen representative may exhibit entanglement-spectrum degeneracy from a higher-dimensional representation even when the interface algebra admits one-dimensional representations. Such degeneracy need not distinguish SPT phases.

{\color{black}We also consider a spin chain whose local Hilbert space is a finite-dimensional Hopf $C^*$-algebra $H$. The irreducible representations of $H$ label the symmetry MPOs, and a unitary Drinfeld twist of $H$ specifies the SPT phase. From the twist we construct the action tensors that describe how each symmetry MPO acts on the MPS bonds. We assign a physical state to each pair of left and right bond labels so that applying the MPO at a site reproduces its action on the bonds. Contracting the bonds between neighboring sites gives a fixed point MPS. The same action tensors determine which local operators to place at the two ends of a truncated symmetry MPO to obtain a string order parameter.}

We illustrate the method with the ordinary $\IZ_2\times\IZ_2$ SPTs and the three $\Rep(D_8)$ NISPTs\footnote{\color{black}$D_{2n}$ denotes the dihedral group of order $2n$, with presentation $D_{2n}=\langle r,s\mid r^n=s^2=1,\ srs=r^{-1}\rangle$.} \cite{Thorngren:2019iar,Sahand2024cluster,kansei2024nispt}. We also study intrinsically mixed $G\times\Rep(G)$ SPTs, whose protection involves both symmetry factors, and give explicit spin-chain Hamiltonians and interface algebras for $G=S_3$. {\color{black}\hypersetup{citecolor=ceruleanblue,linkcolor=ceruleanblue}The Hopf-algebra construction is illustrated by $\Rep(D_8)$ and $\Rep(A_{12})$, where $A_{12}$ is a twelve-dimensional self-dual Hopf algebra \cite{david2011kac}. Gauging its non-invertible $\Rep(A_{12})$ symmetry with the canonical fiber functor therefore returns the same symmetry category, $\Rep(A_{12}^*)\simeq\Rep(A_{12})$ \cite{lu2026hopfising}. On a chain of $A_{12}$ qudits, we construct Hamiltonians for both phases, evaluate string order in each ground state and compute a strange correlator between the two ground states.} These examples show how the same interface data enter the different diagnostics. We further discuss the geometric extension of the strange-operator construction to higher dimensions, where a symmetry surface can give rise to a loop-like insertion. A general higher-dimensional construction is left for future work.

\secref{sec:review} reviews the tensor-network framework and interface algebra. \secref{sec:strangecor} and \secref{sec:stringopn} develop strange correlators and string order parameters, respectively. \secref{sec:grepgspt} studies $G\times\Rep(G)$ SPTs. \secref{sec:hopftwist} constructs $\Rep(H)$ SPT states, their action tensors and their string order parameters from Drinfeld twists. \secref{sec:concl} discusses the conclusions and entanglement spectrum. The appendices provide the categorical data, ZX-calculus calculations and the $\Rep(H_8)$ test.
\par\endgroup

\section{Review of MPO and MPS for non-invertible symmetry in 1+1d}\label{sec:review}
\begingroup
\color{black}
\hypersetup{citecolor=ceruleanblue,linkcolor=ceruleanblue}
We review the MPO and MPS descriptions of fusion category symmetry and the interface algebra \cite{cirac2021mpsreview,frank2012mpsspt,xie2011mpsspt,xie2011mpsspt2,chen2011z2spt,nick2015mpoanyon}. Symmetry lines are objects of the fusion category, their fusion is the tensor product, and the $F$-symbols relate the two ways of fusing three lines. These lattice descriptions connect symmetry operators and gapped states to their categorical data \cite{jose2023mpoclas,verstraete2023mpo,jose2022mpoweakhopf,verstraete2024mpo,delcamp2024nonsimp,rubio2025stringano,blanik2025mpsgauging}.

NISPTs correspond, up to equivalence, to indecomposable module categories with one simple object up to isomorphism, or equivalently to fiber functors of the fusion category \cite{etingof2015tensor,Thorngren:2019iar,kansei2024nispt}. We consider unitary fusion categories and unitary fiber functors. A fusion category admitting a fiber functor is anomaly free and tensor equivalent to $\Rep(H)$ for a finite-dimensional semisimple Hopf algebra $H$ \cite{etingof2015tensor}. All Hopf algebras considered below are finite-dimensional and semisimple. A fiber functor assigns finite-dimensional Hilbert spaces to symmetry lines, together with unitary fusion maps compatible with the category's associativity. These maps retain the fusion information. For a fixed $H$, the canonical forgetful functor $\Rep(H)\rightarrow\VEC$ is one example \footnote{\color{black}Throughout, $\VEC$ denotes the category of finite-dimensional Hilbert spaces and linear maps, also written $\operatorname{Hilb}$; $\VEC_G$ denotes its $G$-graded version, also written $\operatorname{Hilb}_G$.} : it forgets the $H$-action on representations while preserving tensor products. On a closed chain, each symmetry MPO acts on the symmetric ground state by the quantum dimension of its line. This scalar action does not determine the SPT phase; the fiber functor's fusion maps, expressed by the $L$-symbols up to gauge transformations, encode its boundary symmetry data \cite{kansei2024nispt}.

For a general fusion category, stacking does not give a canonical operation on NISPTs with the same symmetry: the diagonal objects of $\CC\boxtimes\CC$ need not form a fusion subcategory equivalent to $\CC$ \cite{Thorngren:2019iar,Sahand2024cluster,robbins2024catdiag}. A choice of on-site symmetry realization specifies a reference phase, but the fusion category alone does not select a canonical trivial phase \cite{lan2024mpo}.
\par\endgroup

\begingroup
\color{black}
\hypersetup{citecolor=ceruleanblue,linkcolor=ceruleanblue}
Broader aspects of non-invertible symmetry, including topological defects, gauging and anomalies, are discussed in \cite{Thorngren:2019iar,Thorngren:2021yso,Choi:2023xjw,clay2023tyanomaly,Zhang:2023wlu,Benini2023anomalyTY,Bhardwaj2017finitesym,Tachikawa2017gaugefiniteg,Chang:2018iay,Fuchs:2002cm,zuber2001twistedp,Frohlich:2004ef,Frohlich:2006ch,Frohlich:2009gb,ingo2012completion,plencner2015discretetorsion,Chang:2022hud,Lanzetta:2022lze,Arias-Tamargo:2022nlf,Hayashi:2022fkw,Choi:2022jqy,Cordova:2022ieu,Damia:2022rxw,2023triality,yichul2024selfdual,lu2025symset,delzotto2023symcat,tong2025monopole,nagoya2023nisym,sakura2023classify,sakura2023classify2,choi2024bdytube,choi2024entangle,quella2024entangle,das2024entangle,pal2024entangle,xueda2025stringop,diatlyk2023gauging,sharpe2024noninvertg,Perez-Lona:2024sds,Popov:2025cha}. Lattice constructions provide complementary microscopic descriptions \cite{Aasen:2016dop,Aasen:2020jwb,Ji:2019jhk,arkya2022gensym,verstraete2023mpo,jose2023mpoclas,jose2022mpoweakhopf,verstraete2024mpo,kansei2024nispt,delcamp2024nonsimp,rubio2025stringano,blanik2025mpsgauging,robin2022strange,Inamura2022nigap,kansei2024fusion2,roy2023latt3potts,zheng2024ni21d,lan2024mpo,seifnashri2023lieb,seiberg2023lsm,shuheng2024maj,sahand2024kw,sahand2025gaugempo,lu2024latticep,sal2024modulatedni,sal2024tdual,sal2025projni,arkya2024reps3,sakura2024reps3,apoorv2024reps3,weiguang2023subsym,weiguang2024nisym,weiguang2024multip,weiguang2025cossym,linhao2023kt,linhao2023kt2,yabo2024kt,yabo2024nispt,ruochen2024subalg,nat2023prestate,ning2024ggraded,lingyan2024lattsymtft,lingyan2024lattsymtft2,corey2024fusionspin,corey2024qspin,alison2024repd8s,shuheng2025haagdual,chen2024sequential,nat2024zxtensor,nat2023gcluster,eck2024fusionsurface,delcamp2025twg,delcamp2024latt,ingo2024strange}.
\par\endgroup

\subsection{Fusion category symmetry from the matrix product operators}
Given a non-invertible symmetry described by the fusion category $\CC$, one can construct the matrix product operator (MPO) representations to realize such non-invertible symmetry on the lattice. Following the notation in \cite{kansei2024nispt} (also see \cite{jose2023mpoclas,lan2024mpo,sahand2025gaugempo}), we consider the set of injective MPOs $\hCO_x,x\in \simp(\CC)$ that generate the fusion category symmetry $\CC$, where $\simp(\CC)$ are the collection of simple objects in the fusion category $\CC$. The MPO $\hCO_x$ is obtained by contracting the 4-leg tensors $\CO_x$ as,
\begin{align}
    &\hCO_x = \cdots\begin{tikzpicture}
    \tikzset{baseline=(current bounding box.center)}
    \draw[thick, black, ->-=.3,->-=0.9] (-1.5,0) -- (-0.5,0);
    \draw[thick,gray, ->-=0.25,->-=0.95] (-1,-0.5) -- (-1,0.5);
    \filldraw[fill=gray!30, draw=black, thick] (-1,0) circle (0.12);
    \draw[thick, black, ->-=.3,->-=0.9] (-0.5,0) -- (0.5,0);
    \draw[thick,gray, ->-=0.25,->-=0.95] (0,-0.5) -- (0,0.5);
    \filldraw[fill=gray!30, draw=black, thick] (0,0) circle (0.12);
    \draw[thick, black, ->-=.3,->-=0.9] (0.5,0) -- (1.5,0);
    \draw[thick,gray, ->-=0.25,->-=0.95] (1,-0.5) -- (1,0.5);
    \filldraw[fill=gray!30, draw=black, thick] (1,0) circle (0.12);
    \node[above] at (-1.5,0) {$\CO_x$};
\end{tikzpicture}\cdots \\
&= \sum_{i_1,\cdots,i_L} \sum_{j_1,\cdots,j_L} \tr(\CO_x^{i_1,j_1}\CO_x^{i_2,j_2}\cdots \CO_x^{i_L,j_L}) \ket{i_1,i_2,\cdots,i_L} \bra{j_1,j_2,\cdots,j_L}.
\end{align}
where the gray legs act on the physical Hilbert space and the black legs act on the virtual bond Hilbert space. We assume a closed chain with $L$ sites and translation symmetry such that $\CO_x$ are the same on different sites. Following from the arrow, each MPO tensor $\CO_x$ is a map from $\CH^x_{\text{bond}}\otimes \CH^x_{\text{phys}}$ to $\CH^x_{\text{phys}}\otimes \CH^x_{\text{bond}}$. See \cite{cirac2021mpsreview} review for more details.

To represent the fusion category symmetry, the MPO tensor $\CO_x$ should further satisfy,
\begin{equation}
    \begin{tikzpicture}
    \tikzset{baseline=(current bounding box.center)}
    \draw[thick, black, ->-=.3,->-=0.8] (-1,0.5) -- (1,0.5);
    \draw[thick, black, ->-=.3,->-=0.8] (-1,-0.5) -- (1,-0.5);
    \draw[thick,gray, ->-=0.55,->-=0.12,->-=0.95] (0,-1) -- (0,1);
    \filldraw[fill=gray!30, draw=black, thick] (0,0.5) circle (0.15);
    \filldraw[fill=gray!30, draw=black, thick] (0,-0.5) circle (0.15);
    \node at (0.6,0.8) {$\mathcal{O}_x$};
    \node at (0.6,-0.2) {$\mathcal{O}_y$};
\end{tikzpicture}  = \sum_{z\in \simp(\CC)} \sum_{1\le\mu \le N_{xy}^z} \ \begin{tikzpicture}
\tikzset{baseline=(current bounding box.center)}
    \draw[thick, black, ->-=.3,->-=0.8] (-1,0) -- (1,0);
    \draw[thick,gray, ->-=0.25,->-=0.95] (0,-0.5) -- (0,0.5);
    \filldraw[fill=gray!30, draw=black, thick] (0,0) circle (0.15);
    \draw[thick, black] (1,0) -- (1,0.5);
    \draw[thick, black, ->-=.6] (1,0.5) -- (1.5,0.5);
    \draw[thick, black] (1,0) -- (1,-0.5);
    \draw[thick, black, ->-=.6] (1,-0.5) -- (1.5,-0.5);
    \draw[thick, black] (-1,0) -- (-1,0.5);
    \draw[thick, black, -<-=.6] (-1,0.5) -- (-1.5,0.5);
    \draw[thick, black] (-1,0) -- (-1,-0.5);
    \draw[thick, black, -<-=.6] (-1,-0.5) -- (-1.5,-0.5);
    \filldraw[fill=white, draw=black, thick] (1,0) circle (0.08);
    \filldraw[fill=black, draw=black, thick] (-1,0) circle (0.08);
    \node at (0.6,0.3) {$\mathcal{O}_z$};
    \node[left] at (-1.0,0.0) {$(\phi_{xy}^z)_\mu$};   \node[right] at (1.0,0.0) {$(\bphi_{xy}^z)_\mu$};
    \node at (1,-0.7) {};
\end{tikzpicture}
\end{equation}
where $N_{xy}^z$ is the fusion coefficient. The fusion and splitting tensor $(\phi_{xy}^z)_\mu$, $(\bphi_{xy}^z)_\mu$ should satisfy the orthogonality condition,
\begin{equation}
\begin{tikzpicture}
\tikzset{baseline=(current bounding box.center)}
    \draw[thick, black, ->-=.7] (-1.5,0) -- (-0.5,0);
    \draw[thick, black] (-0.5,0) -- (-0.5,0.5);
    \draw[thick, black] (-0.5,0) -- (-0.5,-0.5);
    \draw[thick, black,->-=.6] (-0.5,0.5) -- (0.5,0.5);
    \draw[thick, black,->-=.6] (-0.5,-0.5) -- (0.5,-0.5);
    \draw[thick, black] (0.5,0) -- (0.5,0.5);
    \draw[thick, black] (0.5,0) -- (0.5,-0.5);
    \draw[thick, black, -<-=.7] (1.5,0) -- (0.5,0);
    \filldraw[fill=white, draw=black, thick] (-0.5,0) circle (0.08);
    \filldraw[fill=black, draw=black, thick] (0.5,0) circle (0.08);
    \node[above] at (-1,0) {$z$};
    \node[above] at (1,0) {$z'$};
    \node[right] at (-0.5,0) {$\mu$};
    \node[left] at (0.5,0) {$\nu$};
    \node[above] at (0,0.5) {$x$};
    \node[below] at (0,-0.5) {$y$};
\end{tikzpicture}  \ = \ \delta_{\mu,\nu}\delta_{z,z'}   \begin{tikzpicture}
\tikzset{baseline=(current bounding box.center)}
    \draw[thick, black, ->-=.6] (-0.5,0) -- (0.5,0);
    \node[right] at (0.5,0) {$z$};
\end{tikzpicture}
\end{equation}
where the $\mu,\nu$ denote the splitting and fusion tensor $(\bphi_{xy}^z)_\mu$, $(\phi_{xy}^z)_\nu$.

Following from the above conditions, the multiplication of MPOs $\hCO_x,x\in \simp(\CC)$ follows from the fusion rule,
\begin{equation}
    \hCO_x \times \hCO_y = \sum_z N_{xy}^z \hCO_z.
\end{equation}
One can easily extract the $F$-symbols of the fusion category $\CC$ from the fusion tensors $(\phi_{xy}^z)_\mu$ according to the diagram,
\begin{equation}\label{eq:Fsymbol}
\begin{tikzpicture}
\tikzset{baseline=(current bounding box.center)}
    \draw[thick, black, ->-=.7] (-1.5,0.5) -- (-0.5,0.5);
    \draw[thick, black, ->-=.7] (-1.5,-0.5) -- (-0.5,-0.5);
    \draw[thick, black] (-0.5,-0.5) -- (-0.5,0.5);
    \draw[thick, black, ->-=.7] (-0.5,0) -- (0.5,0);
    \draw[thick, black, ->-=.7] (-1.5,-1) -- (0.5,-1);
    \draw[thick, black] (0.5,0) -- (0.5,-1);
    \draw[thick, black, ->-=.7] (0.5,-0.5) -- (1.5,-0.5);
    \filldraw[fill=black, draw=black, thick] (-0.5,0) circle (0.08);
    \filldraw[fill=black, draw=black, thick] (0.5,-0.5) circle (0.08);
    \node[left] at (-1.5,0.5) {$x$};
    \node[left] at (-1.5,-0.5) {$y$};
    \node[left] at (-1.5,-1) {$z$};
    \node[left] at (-0.5,0) {$\mu$};
    \node[left] at (0.5,-0.5) {$\nu$};
    \node[above] at (0,0) {$u$};
    \node[right] at (1.5,-0.5) {$w$};
\end{tikzpicture}   \ = \ \sum_{v,\rho,\sigma} (F_{w}^{xyz})_{(u,\mu,\nu),(v,\rho,\sigma)} \begin{tikzpicture}
\tikzset{baseline=(current bounding box.center)}
    \draw[thick, black, ->-=.7] (-1.5,0.5) -- (0.5,0.5);
    \draw[thick, black, ->-=.7] (-1.5,0) -- (-0.5,0);
    \draw[thick, black] (-0.5,0) -- (-0.5,-1);
    \draw[thick, black, ->-=.7] (-0.5,-0.5) -- (0.5,-0.5);
    \draw[thick, black, ->-=.7] (-1.5,-1) -- (-0.5,-1);
    \draw[thick, black] (0.5,0.5) -- (0.5,-0.5);
    \draw[thick, black, ->-=.7] (0.5,0) -- (1.5,0);
    \filldraw[fill=black, draw=black, thick] (-0.5,-0.5) circle (0.08);
    \filldraw[fill=black, draw=black, thick] (0.5,0) circle (0.08);
    \node[left] at (-1.5,0.5) {$x$};
    \node[left] at (-1.5,0) {$y$};
    \node[left] at (-1.5,-1) {$z$};
    \node[left] at (-0.5,-0.5) {$\rho$};
    \node[left] at (0.5,0) {$\sigma$};
    \node[below] at (0,-0.5) {$v$};
    \node[right] at (1.5,0) {$w$};
\end{tikzpicture}
\end{equation}
The $F$-symbol for the group $G$ case is the 3-cocycle $\omega\in H^3(G,U(1))$. Note that the above calculation is similar to the local defect fusion in \cite{sahand2025gaugempo}. In \appref{app:FLRepD8}, we present the calculation of $F$-symbols for $\Rep(D_8)$ using the ZX calculus.

\begingroup
\color{black}
\hypersetup{citecolor=ceruleanblue,linkcolor=ceruleanblue}
For anomaly free fusion category $\CC_\text{free}$, the quantum dimension of every line operator is an integer $d_x\in \IZ,x\in \CC_\text{free}$. The ``on-site'' condition for the MPO of $x\in\CC_\text{free}$ requires that the bond dimension equals the quantum dimension of $x$, i.e. $\dim(\CH^x_\text{bond})=d_x$ \cite{jose2023mpoclas,kansei2024nispt,lan2024mpo}.\footnote{\color{black}The on-site condition is sufficient but not necessary for a symmetric product state to exist; some non-on-site anomaly-free group symmetries also admit such states.} For group-like elements, the on-site MPOs have bond dimension $1$. In other words, one can always find a MPO representation of the anomaly-free fusion category $\CC_\text{free}$, such that the MPO satisfies the on-site condition. All the $\CC_\text{free}$ are representation category of Hopf algebra $H$, and the on-site MPO can be constructed from the Hopf algebra $H$ and its irreducible representation. Hence, we require the MPO $\hCO_x$ satisfies the on-site condition. Therefore, the MPO further satisfies the completeness condition,
\par\endgroup
\begin{equation}
    {\color{black}\sum_{z\in\simp(\CC)}\sum_{\mu=1}^{N_{xy}^{z}}}\;\;\begin{tikzpicture}
\tikzset{baseline=(current bounding box.center)}
    \draw[thick, black,->-=.6] (-0.5,0.5) -- (0.5,0.5);
    \draw[thick, black,->-=.6] (-0.5,-0.5) -- (0.5,-0.5);
    \draw[thick, black] (0.5,0) -- (0.5,0.5);
    \draw[thick, black] (0.5,0) -- (0.5,-0.5);
    \draw[thick, black, -<-=.6] (1.5,0) -- (0.5,0);
    \draw[thick, black,->-=.6] (1.5,0.5) -- (2.5,0.5);
    \draw[thick, black,->-=.6] (1.5,-0.5) -- (2.5,-0.5);
    \draw[thick, black] (1.5,0) -- (1.5,0.5);
    \draw[thick, black] (1.5,0) -- (1.5,-0.5);

    \filldraw[fill=white, draw=black, thick] (1.5,0) circle (0.08);
    \filldraw[fill=black, draw=black, thick] (0.5,0) circle (0.08);
    \node[above] at (1,0) {$z$};
    \node[right] at (1.5,0) {$\mu$};
    \node[left] at (0.5,0) {$\mu$};
    \node[above] at (0,0.5) {$x$};
    \node[below] at (0,-0.5) {$y$};
    \node[above] at (2,0.5) {$x$};
    \node[below] at (2,-0.5) {$y$};
\end{tikzpicture} \ = \ \begin{tikzpicture}
\tikzset{baseline=(current bounding box.center)}
    \draw[thick, black,->-=.6] (-0.5,0.5) -- (0.5,0.5);
    \draw[thick, black,->-=.6] (-0.5,-0.5) -- (0.5,-0.5);
    \node[above] at (0,0.5) {$x$};
    \node[below] at (0,-0.5) {$y$};
\end{tikzpicture}
\end{equation}

\begingroup
\color{black}
\hypersetup{citecolor=ceruleanblue,linkcolor=ceruleanblue}
In a Hopf-algebra realization, a symmetric product reference can be constructed from the Haar integral of $H^*$ \cite{jia_generalized_2024}. Its existence depends on the chosen realization, so a product state should not be assumed to be symmetric for an arbitrary MPO representation.
\par\endgroup

\subsection{Matrix product states as fusion category symmetric gapped phases}
The matrix product state (MPS) is given by contracting the 3-leg tensor $A$,
\begin{equation}
    \ket{A} = \cdots\begin{tikzpicture}
    \tikzset{baseline=(current bounding box.center)}
    \draw[thick, cyan, -<-=.3,-<-=0.9] (-1.5,0) -- (-0.5,0);
    \draw[thick,gray, ->-=0.7] (-1,0) -- (-1,0.5);
    \filldraw[fill=blue!30, draw=black, thick] (-1,0) circle (0.12);
    \draw[thick, cyan, -<-=.3,-<-=0.9] (-0.5,0) -- (0.5,0);
    \draw[thick,gray, ->-=0.7] (0,0) -- (0,0.5);
    \filldraw[fill=blue!30, draw=black, thick] (0,0) circle (0.12);
    \draw[thick, cyan, -<-=.3,-<-=0.9] (0.5,0) -- (1.5,0);
    \draw[thick,gray, ->-=0.7] (1,0) -- (1,0.5);
    \filldraw[fill=blue!30, draw=black, thick] (1,0) circle (0.12);
    \node[above] at (-1.5,0) {$A$};
\end{tikzpicture}\cdots = \sum_{i_1,\cdots,i_L} \tr(A^{i_1}A^{i_2}\cdots A^{i_L}) \ket{i_1,i_2,\cdots,i_L}.
\end{equation}
where each MPS tensor $A$ is a map from the virtual bond Hilbert space $V$ to $V\otimes \CH_{\text{phys}}$. We use a tilde for operators on the MPS virtual bond, for example $\widetilde{X}$ and $\widetilde{Z}$, and write $X$ and $Z$ for the corresponding physical operators. We assume translation invariance throughout the paper and consider MPSs that become injective after grouping finitely many sites. In the local tensor diagrams of this subsection, $A$ denotes the injective blocked tensor when blocking is required, with the MPO tensors blocked in the same way. When the block size is explicit, $A^{\otimes n}$ denotes $n$ consecutive tensors of the original MPS with their internal virtual bonds contracted; each tensor may already represent a unit cell. For a $\CC$-symmetric MPS satisfying this assumption, there exist 3-leg action tensors $(\phi_x)_i,(\bphi_x)_j$ such that,
\begin{equation}\label{eq:oactonA}
    \CO_x A = \begin{tikzpicture}
    \tikzset{baseline=(current bounding box.center)}
    \draw[thick, black, ->-=.3,->-=0.9] (-0.5,0) -- (0.5,0);
    \draw[thick,gray, ->-=0.35,->-=0.95] (0,-0.5) -- (0,0.5);
    \filldraw[fill=gray!30, draw=black, thick] (0,0) circle (0.12);
    \draw[thick, cyan, -<-=.3,-<-=0.9] (-0.5,-0.5) -- (0.5,-0.5);
    \filldraw[fill=blue!30, draw=black, thick] (0,-0.5) circle (0.12);
    \node[above] at (-0.5,0) {$\CO_x$};
    \node[below] at (0,-0.6) {$A$};
\end{tikzpicture} \ =\ \sum_{1\le i \le d_x}  \begin{tikzpicture}
    \tikzset{baseline=(current bounding box.center)}
    \draw[thick,gray, ->-=0.6] (0,-0.5) -- (0,0.5);
    \draw[thick, cyan, -<-=.3,-<-=0.9] (-0.5,-0.5) -- (0.5,-0.5);
    \draw[thick, cyan, -<-=0.6] (-1,-0.5) -- (-0.6,-.5);
    \draw[thick, cyan, -<-=0.6] (0.6,-0.5) -- (1,-.5);
    \draw[thick, black] (0.6,-0.5) -- (0.6,0);
    \draw[thick, black] (-0.6,-0.5) -- (-0.6,0);
    \draw[thick, black,->-=0.6] (0.6,0) -- (1,0);
    \draw[thick, black,->-=0.6] (-1,0) -- (-0.6,0);
    \filldraw[fill=blue!30, draw=black, thick] (0,-0.5) circle (0.12);
    \filldraw[fill=white, draw=black, thick] (0.6,-0.5) circle (0.08);
    \filldraw[fill=black, draw=black, thick] (-0.6,-0.5) circle (0.08);
    \node[above] at (-0.3,-0.5) {$A$};
    \node[below] at (0.6,-0.5) {$(\bphi_x)_i$};
    \node[below] at (-0.6,-0.5) {$(\phi_x)_i$};
    \node[above] at (-0.8,0) {$x$};
\end{tikzpicture}
\end{equation}
where $d_x$ is the quantum dimension of $x\in\simp(\CC)$. For example, group-like invertible lines have quantum dimension 1, so no additional index $i$ is needed. The Even and Odd states in \secref{sec:repd8mps} are not injective on a two-site unit cell but become injective when two unit cells are grouped into a four-site block.

The action tensors $(\phi_x)_i,(\bphi_x)_j$ also satisfy the orthogonality condition,
\begin{equation}\label{eq:actionorthogonality}
    \begin{tikzpicture}
    \tikzset{baseline=(current bounding box.center)}
    \draw[thick, cyan, -<-=0.6] (-1,-0.5) -- (-0.5,-.5);
    \draw[thick, cyan, -<-=0.6] (-0.5,-0.5) -- (0.5,-.5);
    \draw[thick, cyan, -<-=0.6] (0.5,-0.5) -- (1,-.5);
    \draw[thick, black] (-0.5,-0.5) -- (-0.5,0);
    \draw[thick, black] (0.5,-0.5) -- (0.5,0);
    \draw[thick, black,->-=0.6] (-0.5,0) -- (0.5,0);
    \filldraw[fill=white, draw=black, thick] (-0.5,-0.5) circle (0.08);
    \filldraw[fill=black, draw=black, thick] (0.5,-0.5) circle (0.08);
    \node[below] at (-0.5,-0.5) {$(\bphi_x)_i$};
    \node[below] at (0.5,-0.5) {$(\phi_{x'})_j$};
    \node[above] at (0,0) {$x$};
\end{tikzpicture}\ =\ \delta_{i,j}\delta_{x,x'}\begin{tikzpicture}
    \tikzset{baseline=(current bounding box.center)}
    \draw[thick, cyan, -<-=0.6] (-0.5,-0.5) -- (0.5,-.5);
\end{tikzpicture}
\end{equation}
Following from the above conditions, the MPO $\hCO_x$ acting on the $\CC$-symmetric MPS $\ket{A}$ yields,
\begin{equation}
    \hCO_x \ket{A} = d_x \ket{A}.
\end{equation}

For an on-site MPO realization, the symmetry can also be defined on an open chain by factorizing the MPO into left and right tensors and choosing boundary tensors compatible with fusion \cite{lan2024mpo}. Additional boundary degrees of freedom may be needed, and the boundary terms must preserve the resulting symmetry action. Protected edge degeneracy is compatible with an unbroken bulk symmetry. We use periodic boundary conditions to avoid choosing a physical boundary termination, while the action tensors in \eqref{eq:oactonA} retain the boundary symmetry data used in our diagnostics.

The SPT phases with fusion category symmetry $\CC$ are characterized by the $L$-symbols up to gauge transformations \cite{kansei2024nispt}. The $L$-symbols are defined by the following diagrams:
\begin{equation}\label{eq:Lsymbol}
    \begin{tikzpicture}
    \tikzset{baseline=(current bounding box.center)}
    \draw[thick, cyan, -<-=0.6] (-1.5,-0.5) -- (-0.5,-.5);
    \draw[thick, cyan, -<-=0.6] (-0.5,-0.5) -- (0.5,-.5);
    \draw[thick, cyan, -<-=0.6] (0.5,-0.5) -- (1.5,-.5);
    \draw[thick, black] (-0.5,-0.5) -- (-0.5,0);
    \draw[thick, black] (0.5,-0.5) -- (0.5,0.5);
    \draw[thick, black,->-=0.6] (-1.5,0) -- (-0.5,0);
    \draw[thick, black,->-=0.6] (-1.5,0.5) -- (0.5,0.5);
    \filldraw[fill=black, draw=black, thick] (-0.5,-0.5) circle (0.08);
    \filldraw[fill=black, draw=black, thick] (0.5,-0.5) circle (0.08);
    \node[below] at (-0.5,-0.5) {$(\phi_y)_j$};
    \node[below] at (0.5,-0.5) {$(\phi_x)_i$};
    \node[left] at (-1.5,0.5) {$x$};
    \node[left] at (-1.5,0.0) {$y$};
\end{tikzpicture}\ =\ \sum_{z,k,\mu} (L_{xy}^{z})_{(i,j),(k,\mu)}\begin{tikzpicture}
    \tikzset{baseline=(current bounding box.center)}
    \draw[thick, cyan, -<-=0.6] (-1.5,-0.5) -- (0.5,-.5);
    \draw[thick, cyan, -<-=0.6] (0.5,-0.5) -- (1.5,-.5);
    \draw[thick, black] (0.5,-0.5) -- (0.5,0.25);
    \draw[thick, black,->-=0.6] (-1.5,0) -- (-0.5,0);
    \draw[thick, black,->-=0.6] (-1.5,0.5) -- (-0.5,0.5);
    \draw[thick, black] (-0.5,0) -- (-0.5,0.5);
    \draw[thick, black,->-=0.6] (-0.5,0.25) -- (0.5,0.25);
    \filldraw[fill=black, draw=black, thick] (-0.5,0.25) circle (0.08);
    \filldraw[fill=black, draw=black, thick] (0.5,-0.5) circle (0.08);
    \node[above] at (-0.5,0.4) {$(\phi_{xy}^z)_\mu$};
    \node[below] at (0.5,-0.5) {$(\phi_z)_k$};
    \node[left] at (-1.5,0.5) {$x$};
    \node[left] at (-1.5,0.0) {$y$};
    \node[right] at (0.5,0.0) {$z$};
\end{tikzpicture}
\end{equation}
An analogous diagram for the white action tensors $(\bphi_x)_i$ defines $(\bar{L}_{xy}^z)_{(i,j),(k,\mu)}$ \cite{kansei2024nispt}. For anomaly-free group $G$ symmetry, the $L$-symbols reduce to $U(1)$-valued 2-cocycles, whose cohomology classes in $H^2(G,U(1))$ classify the SPT phases. In \appref{app:FLRepD8}, we calculate the $L$-symbols of the three $\Rep(D_8)$ NISPTs using the ZX calculus.

\subsection{Interface algebra for different fusion category symmetric SPTs}\label{sec:interalg}
As for ordinary group SPTs, interfaces between distinct NISPT phases with the same fusion category symmetry support symmetry-protected interface modes. Examples are provided by interfaces between the three inequivalent $\Rep(D_8)$ NISPT phases \cite{Thorngren:2019iar,Sahand2024cluster,kansei2024nispt}. Their protected degeneracy follows from the absence of one-dimensional representations of the interface algebra \cite{kansei2024nispt,choi2024bdytube}.

Diagrammatically, for the two NISPTs which are given by the MPS states with different colors, the symmetry action on the interface (labeled by yellow diamond) between the two MPS states is,
\begin{equation}
    \cdots\begin{tikzpicture}
    \tikzset{baseline=(current bounding box.center)}
    \draw[thick, black, ->-=.3,->-=0.9] (-0.5,0) -- (0.5,0);
    \draw[thick,gray, ->-=0.35,->-=0.95] (0,-0.5) -- (0,0.5);
    \filldraw[fill=gray!30, draw=black, thick] (0,0) circle (0.12);
    \draw[thick, cyan, -<-=.3,-<-=0.9] (-0.5,-0.5) -- (0.5,-0.5);
    \filldraw[fill=blue!30, draw=black, thick] (0,-0.5) circle (0.12);
    \draw[thick, black, ->-=.3,->-=0.9] (0.5,0) -- (1.5,0);
    \draw[thick,gray, ->-=0.35,->-=0.95] (1,-0.5) -- (1,0.5);
    \filldraw[fill=gray!30, draw=black, thick] (1,0) circle (0.12);
    \draw[thick, cyan, -<-=.3,-<-=0.9] (0.5,-0.5) -- (1.5,-0.5);
    \filldraw[fill=blue!30, draw=black, thick] (1,-0.5) circle (0.12);
    \node[above] at (-0.5,0) {$\CO_x$};
    \draw[thick, black, ->-=.3,->-=0.9] (1.5,0) -- (2.5,0);
    \draw[thick,gray, ->-=0.35,->-=0.95] (2,-0.5) -- (2,0.5);
    \filldraw[fill=gray!30, draw=black, thick] (2,0) circle (0.12);
    \draw[thick, orange, -<-=.3,-<-=0.9] (1.5,-0.5) -- (2.5,-0.5);
    \filldraw[fill=yellow!30, draw=black, thick] (2,-0.5) circle (0.12);
    \draw[thick, black, ->-=.3,->-=0.9] (2.5,0) -- (3.5,0);
    \draw[thick,gray, ->-=0.35,->-=0.95] (3,-0.5) -- (3,0.5);
    \filldraw[fill=gray!30, draw=black, thick] (3,0) circle (0.12);
    \draw[thick, orange, -<-=.3,-<-=0.9] (2.5,-0.5) -- (3.5,-0.5);
    \filldraw[fill=yellow!30, draw=black, thick] (3,-0.5) circle (0.12);
    \node[diamond, draw,minimum size=2mm, inner sep=0mm,thick, fill=yellow] at (1.5,-0.5) {};
    \node[above] at (-0.5,-0.5) {$A$};
    \node[above] at (3.5,-0.5) {$B$};
\end{tikzpicture}\cdots\ = \ \sum_{i,i'=1,\cdots d_x} \cdots\begin{tikzpicture}
    \tikzset{baseline=(current bounding box.center)}
    \draw[thick, cyan, -<-=.3,-<-=0.9] (-1.5,0) -- (-0.5,0);
    \draw[thick,gray, ->-=0.7] (-1,0) -- (-1,0.5);
    \filldraw[fill=blue!30, draw=black, thick] (-1,0) circle (0.12);
    \draw[thick, cyan, -<-=.15,-<-=0.7] (-0.5,0) -- (1,0);
    \draw[thick,gray, ->-=0.7] (0,0) -- (0,0.5);
    \filldraw[fill=blue!30, draw=black, thick] (0,0) circle (0.12);
    \node[above] at (-1.5,0) {$A$};
    \draw[thick, orange, -<-=.3,-<-=0.9] (1,0) -- (2.5,0);
    \draw[thick,gray, ->-=0.7] (2,0) -- (2,0.5);
    \filldraw[fill=yellow!30, draw=black, thick] (2,0) circle (0.12);
    \draw[thick, orange, -<-=.3,-<-=0.9] (2.5,0) -- (3.5,0);
    \draw[thick,gray, ->-=0.7] (3,0) -- (3,0.5);
    \filldraw[fill=yellow!30, draw=black, thick] (3,0) circle (0.12);
    \node[above] at (3.5,0) {$B$};
    \draw[thick, black, ->-=0.6] (0.5,0.4) -- (1.5,0.4);
    \draw[thick, black] (0.5,0) -- (0.5,0.4);
    \draw[thick, black] (1.5,0) -- (1.5,0.4);
    \node[diamond, draw,minimum size=2mm, inner sep=0mm,thick, fill=yellow] at (1,0) {};
    \filldraw[fill=white, draw=black, thick] (0.5,0) circle (0.08);
    \filldraw[fill=black, draw=black, thick] (1.5,0) circle (0.08);
    \node[above] at (1,0.4) {$x$};
    \node[below] at (0.5,0) {$(\bphi_x^A)_i$};
    \node[below] at (1.5,0) {$(\phi_x^B)_{i'}$};
\end{tikzpicture}
\end{equation}
{\color{black}We write the corresponding interface-algebra element as}
\begin{equation}\label{eq:interfacealg}
    \CI_{x,(i,i')}^{A|B} \equiv \begin{tikzpicture}
    \tikzset{baseline=(current bounding box.center)}
    \draw[thick, cyan, -<-=.3,-<-=0.9] (-1.5,0) -- (-0.5,0);
    \draw[thick, orange, -<-=.3,-<-=0.9] (0.5,0) -- (1.5,0);
    \draw[thick, black, ->-=0.6] (-1,0.5) -- (1,0.5);
    \draw[thick, black] (-1,0) -- (-1,0.5);
    \draw[thick, black] (1,0) -- (1,0.5);
    \filldraw[fill=white, draw=black, thick] (-1,0) circle (0.08);
    \filldraw[fill=black, draw=black, thick] (1,0) circle (0.08);
    \node[above] at (0,0.5) {$x$};
    \node[below] at (-1,0) {$(\bphi_x^A)_i$};
    \node[below] at (1,0) {$(\phi_x^B)_{i'}$};
\end{tikzpicture}
\end{equation}
{\color{black}For the chosen MPSs $A$ and $B$, this gives a representation of the interface algebra.}

{\color{black}The elements multiply according to}
\begin{align}
    &\CI_{x,(i,i')}^{A|B}\times \CI_{y,(j,j')}^{A|B} \ = \ \begin{tikzpicture}
    \tikzset{baseline=(current bounding box.center)}
    \draw[thick, cyan, -<-=.3,-<-=0.9] (-1.5,0) -- (-0.5,0);
    \draw[thick, orange, -<-=.3,-<-=0.9] (0.5,0) -- (1.5,0);
    \draw[thick, black, ->-=0.6] (-1,0.5) -- (1,0.5);
    \draw[thick, black] (-1,0) -- (-1,0.5);
    \draw[thick, black] (1,0) -- (1,0.5);
    \filldraw[fill=white, draw=black, thick] (-1,0) circle (0.08);
    \filldraw[fill=black, draw=black, thick] (1,0) circle (0.08);
    \node[above] at (0,0.5) {$y$};
    \node[below] at (-1,0) {$(\bphi_y^A)_{j}$};
    \node[below] at (1,0) {$(\phi_y^B)_{j'}$};
    \draw[thick, cyan, -<-=.3,-<-=0.9] (-2.5,0) -- (-1.5,0);
    \draw[thick, orange, -<-=.3,-<-=0.9] (1.5,0) -- (2.5,0);
    \draw[thick, black, ->-=0.6] (-2,1) -- (2,1);
    \draw[thick, black] (-2,0) -- (-2,1);
    \draw[thick, black] (2,0) -- (2,1);
    \filldraw[fill=white, draw=black, thick] (-2,0) circle (0.08);
    \filldraw[fill=black, draw=black, thick] (2,0) circle (0.08);
    \node[above] at (0,1) {$x$};
    \node[below] at (-2,0) {$(\bphi_x^A)_{i}$};
    \node[below] at (2,0) {$(\phi_x^B)_{i'}$};
\end{tikzpicture}\nonumber\\
&=\ \sum_{z,\mu}\sum_{k,k'}({}^A\bar{L}_{xy}^{z})_{(i,j),(k,\mu)}({}^B L_{xy}^{z})_{(i',j'),(k',\mu)} \CI_{z,(k,k')}^{A|B} \label{eq:interalg}
\end{align}
which follows from the definition of the $L$-symbol. The interface algebra is spanned by $\CI_{x,(i,i')}^{A|B}$. For the ordinary $G$-SPT classified by the $H^2(G,U(1))$, the interface algebra of two SPTs labeled by $\psi_1,\psi_2  \in H^2(G,U(1))$ is isomorphic to the twisted group algebra $\IC G^{\psi_1 \psi_2^{-1}}$, whose irreducible representations are the projective representations of $G$ when $\psi_1,\psi_2$ are in different cohomology classes.

If both sides are the same NISPT of $\Rep(H)$, the interface algebra is called the self-interface algebra and is the dual Hopf algebra $H^*$. Here $H$ is reconstructed from the fiber functor of the chosen phase and can depend on that choice. The local Hilbert space is an object in the representation category of the self-interface algebra $H^*$ \cite{lan2024mpo}. For example, the self-interface algebra of at least one $\Rep(G)$ NISPT (there is always a forgetful functor $\Rep(G)\rightarrow \VEC$ \cite{etingof2015tensor}) forms dual of group algebra, $\IC G^*\cong \IC^G$, whose representation category is $\VEC_G$, meaning that the local Hilbert space is $G$-graded Hilbert space. For $\Rep(D_8)$, the self-interface algebras of all 3 NISPTs are $\IC^{D_8}$ (dual to $\IC D_8$) \cite{kansei2024nispt,lan2024mpo}. For $\Rep(D_{16})$, the self-interface algebra of 1 NISPT is $\IC^{D_{16}}$, while the self-interface algebras of the other 2 NISPTs correspond to non-group Hopf algebras \cite{lu2025symset,xiong2020cocycle} \footnote{We thank Conghuan Luo for pointing out the non-group Hopf algebras.}. These examples illustrate differences between NISPTs of different fusion categories \cite{xgw2025nispt}.

More generally, the interface algebra of $\CC$-NISPTs considered here is a special case of the algebra $\CA_{M,N}$ in \cite{kitaev2012gapbdy} (also the ladder category \cite{jones2019laddercat}) with $M,N$ being the module categories over $\CC$ with a single simple object. The connection between the interface algebra and module functors is illustrated in \appref{app:functor}. The self-interface algebra is the dual of the corresponding reconstructed Hopf algebra $H$ (a special case of the strip algebra \cite{clay2024repsol} or annular algebra \cite{frank2023annualalg} with the module category containing a single simple object), since the symmetry is an anomaly free fusion category that admits fiber functor, i.e. $\CC=\Rep(H)$. The dimensions of the irreducible representations of the interface algebra relate to the symmetry protected edge modes on NISPT \cite{kansei2024nispt}, and we will illustrate the relation to the strange correlators and the degeneracy of the entanglement spectrum in the following sections. 

\begingroup
\color{black}
\hypersetup{citecolor=ceruleanblue,linkcolor=ceruleanblue}
Related boundary and strip algebra formulations are discussed in \cite{clay2024bdyalg,copetti2024stripalg1,copetti2024stripalg2,copetti2025stripalg1}.
\par\endgroup

For the ordinary on-site $\IZ_2\times\IZ_2$ symmetry, the product and cluster states belong to distinct SPT phases. Their counterparts in the Hopf-algebra realization of $\Rep(H_8)$, where $H_8$ is the Kac-Paljutkin Hopf algebra, belong to the same phase: their comodule algebras are equivariantly Morita equivalent \cite{tambara2000representations,vangrinsven2025comodule}.

A realization based on Kramers-Wannier duality mixes with translation in its lattice fusion \cite{lu2024latticep,chen2026nissb}. For the interface algebra calculation, we instead use an eight-dimensional $H_8$ qudit at each site, which admits an ``on-site'' non-invertible symmetry MPO of bond dimension two \cite{lu2026hopfising}. On the four-dimensional local subspace supporting these states, the invertible MPOs generate only a $\IZ_2$ action, distinct from the ordinary $\IZ_2\times\IZ_2$ action above.

As shown in \appref{app:h8actions}, the full interface algebra has four one-dimensional irreducible representations and one two-dimensional irreducible representation. The one-dimensional representations allow a nondegenerate symmetry-preserving interface. However, the action tensors of the chosen product and cluster representatives realize the two-dimensional representation in \eqref{eq:h8interfacerepresentations}, and their connected strange correlator constructed from \eqref{eq:strangecond} has long-range order. This signal alone would therefore falsely distinguish the two representatives as different SPT phases.

{\color{black}\hypersetup{citecolor=ceruleanblue,linkcolor=ceruleanblue}The cluster state already has zero correlation length and has MPS bond dimension two. As shown in \appref{app:h8actions}, a depth-two circuit of locally $\Rep(H_8)$-symmetric gates on the original qudits maps it to the product state. The two states therefore admit the same representative of bond dimension one. For this choice the interface matrices are scalar, and the connected strange correlator vanishes.}

\section{Strange correlator for NISPTs in 1+1d}\label{sec:strangecor}
The strange correlator offers a way to distinguish different NISPTs using only local operators on a closed chain. These local operators are charged under the global symmetry, but not all charged operators are suitable for detecting even ordinary SPTs. In this section, we construct strange charged operators for general NISPTs and identify the conditions they satisfy to distinguish a given pair of NISPTs by design. We give the examples of ordinary $\IZ_2\times \IZ_2$ SPTs as well as the NISPTs of $\Rep(D_8)$. We discuss the generalization to higher dimension in the end of this section.

\begingroup
\color{black}
\hypersetup{citecolor=ceruleanblue,linkcolor=ceruleanblue}
Strange correlators have been applied to ordinary SPTs \cite{yizhuang2014strange,meng2015strange1,tanaka2016strange,ksd2016strange1,meng2016strange2,luo2017strange1,ye2022strange1,paganelli2023stranget,qi2023strange1,ye2024strange2}, and related overlap constructions have been used for topological orders and conformal field theories \cite{vanhove2018strange,Ji:2019jhk,frank2020strange,gavin2020strange,symtft2021Sakura,cenke2021strange,robin2022strange,robin2022strange2,sukeno2024stranget,hung2024tnstrange,lingyan2024lattsymtft,shen2025stranget,ingo2024strange}. Extensions to open quantum systems have also been used to detect average SPT order \cite{Lee2022aspt,bi2022strangem,sagar2025strangem}.
\par\endgroup

\subsection{General construction of strange charged operator}\label{sec:scnispt}
{\color{black}To diagnose phases, we first choose representatives $\ket{A},\ket{B}$ with minimal bond dimension and use their action tensors to construct action tensors. The same construction applies directly to two given $\CC$-symmetric MPSs satisfying the assumptions below. In either case the strange correlator is the normalized matrix element}
\begin{equation}
		SC^{A|B}(I,J)\equiv \frac{\bra{B} \bCS_I \CS_J \ket{A}}{\bra{B}\ket{A}} = \cdots \begin{tikzpicture}
		\tikzset{baseline=(current bounding box.center)}
		\draw[thick, cyan, -<-=.3,-<-=0.9] (-1.5,0) -- (-0.5,0);
		\draw[thick,gray, ->-=0.7] (-1,0) -- (-1,0.5);
		\filldraw[fill=blue!30, draw=black, thick] (-1,0) circle (0.12);
		\draw[thick, cyan, -<-=.3,-<-=0.9] (-0.5,0) -- (0.5,0);
		\draw[thick,gray, ->-=0.7] (0,0) -- (0,0.5);
		\filldraw[fill=blue!30, draw=black, thick] (0,0) circle (0.12);
		\draw[thick, cyan, -<-=.2,-<-=0.9] (0.5,0) -- (1.5,0);
		\draw[thick,gray, ->-=0.7] (1,0) -- (1,0.5);
		\filldraw[fill=blue!30, draw=black, thick] (1,0) circle (0.12);
		\node[above] at (-1.5,0) {$A$};
		\draw[thick, orange, ->-=.3,->-=0.9] (-1.5,1) -- (-0.5,1);
		\draw[thick,gray, ->-=0.7] (-1,.5) -- (-1,1);
		\filldraw[fill=yellow!30, draw=black, thick] (-1,1) circle (0.12);
		\draw[thick, orange, ->-=.3,->-=0.9] (-0.5,1) -- (0.5,1);
		\draw[thick,gray, ->-=0.7] (0,0.5) -- (0,1);
		\filldraw[fill=yellow!30, draw=black, thick] (0,1) circle (0.12);
		\draw[thick, orange, ->-=.2,->-=0.9] (0.5,1) -- (1.5,1);
		\draw[thick,gray, ->-=0.7] (1,0.5) -- (1,1);
		\filldraw[fill=yellow!30, draw=black, thick] (1,1) circle (0.12);
		\node[below] at (-1.5,1) {$B$};
		\node[draw, rectangle,minimum width=2mm, minimum height=2mm, inner sep=0mm,thick, fill=dgreen] at (0,0.5) {};
		\node[left] at (0,0.5) {$\bCS_I$};
		\node[above] at (0,1) {$B^{\otimes n}$};
		\node[below] at (0,0) {$A^{\otimes n}$};
		\draw[thick, cyan, -<-=.3,-<-=0.9] (1.5,0) -- (2.5,0);
		\draw[thick,gray, ->-=0.7] (2,0) -- (2,0.5);
		\filldraw[fill=blue!30, draw=black, thick] (2,0) circle (0.12);
		\draw[thick, cyan, -<-=.3,-<-=0.9] (2.5,0) -- (3.5,0);
		\draw[thick,gray, ->-=0.7] (3,0) -- (3,0.5);
		\filldraw[fill=blue!30, draw=black, thick] (3,0) circle (0.12);
		\draw[thick, cyan, -<-=.2,-<-=0.9] (3.5,0) -- (4.5,0);
		\draw[thick,gray, ->-=0.7] (4,0) -- (4,0.5);
		\filldraw[fill=blue!30, draw=black, thick] (4,0) circle (0.12);
		\draw[thick, orange, ->-=.3,->-=0.9] (1.5,1) -- (2.5,1);
		\draw[thick,gray, ->-=0.7] (2,.5) -- (2,1);
		\filldraw[fill=yellow!30, draw=black, thick] (2,1) circle (0.12);
		\draw[thick, orange, ->-=.3,->-=0.9] (2.5,1) -- (3.5,1);
		\draw[thick,gray, ->-=0.7] (3,0.5) -- (3,1);
		\filldraw[fill=yellow!30, draw=black, thick] (3,1) circle (0.12);
		\draw[thick, orange, ->-=.2,->-=0.9] (3.5,1) -- (4.5,1);
		\draw[thick,gray, ->-=0.7] (4,0.5) -- (4,1);
		\filldraw[fill=yellow!30, draw=black, thick] (4,1) circle (0.12);
		\node[] at (1.5,0.5) {$\cdots$};
		\node[draw, rectangle,minimum width=2mm, minimum height=2mm, inner sep=0mm,thick, fill=dgreen] at (3,0.5) {};
		\node[left] at (3,0.5) {$\CS_J$};
		\node[above] at (3,1) {$B^{\otimes n}$};
		\node[below] at (3,0) {$A^{\otimes n}$};
	\end{tikzpicture}\cdots /\bra{B}\ket{A}
\end{equation}
The normalized ratio is considered for nonzero overlap, even when both numerator and denominator tend to zero in the thermodynamic limit. If the fixed point representatives have zero overlap, we can evaluate the strange correlator for nearby states with nonzero overlap obtained by a small symmetry-preserving deformation \cite{yizhuang2014strange}. We reconstruct the local insertions from their blocked MPS tensors and keep the deformation fixed while taking the long-distance limit. For the Tri--Even pair in \secref{sec:repd8strange}, this procedure gives a nonzero connected strange correlator independent of separation. The original MPS tensors need not be injective: choose a common finite $n$ such that $A^{\otimes n}$ and $B^{\otimes n}$ are both injective. The local operator $\CS_I$ acts on the corresponding finite block $I$. The blocking convention for the $\Rep(D_8)$ examples is specified in \secref{sec:repd8mps}.

Contracting the action tensors of the two MPSs gives the following conditions on the local insertions.
\begin{equation}
	\begin{tikzpicture}
		\tikzset{baseline=(current bounding box.center)}
		\draw[thick, cyan, -<-=.3,-<-=0.9] (-0.5,0) -- (0.5,0);
		\draw[thick,gray, ->-=0.7] (0,0) -- (0,0.5);
		\filldraw[fill=blue!30, draw=black, thick] (0,0) circle (0.12);
		\node[below] at (0,0) {$A^{\otimes n}$};
		\draw[thick, orange, ->-=.3,->-=0.9] (-0.5,1) -- (0.5,1);
		\draw[thick,gray, ->-=0.7] (0,0.5) -- (0,1);
		\filldraw[fill=yellow!30, draw=black, thick] (0,1) circle (0.12);
		\node[above] at (0,1) {$B^{\otimes n}$};
		\node[draw, rectangle,minimum width=2mm, minimum height=2mm, inner sep=0mm,thick, fill=dgreen] at (0,0.5) {};
		\node[right] at (0,0.5) {$\CS_{I,(x,y),(i,i')}^{A|B,(z,k)}$};
	\end{tikzpicture} \ = \ \begin{tikzpicture}
		\tikzset{baseline=(current bounding box.center)}
		\draw[thick, cyan, -<-=.2,-<-=0.6] (-0.5,0) -- (1.2,0);
		\draw[thick, orange, ->-=.2,->-=0.6] (-0.5,1) -- (1.2,1);
		\draw[thick, black] (0,0) -- (0,0.3);
		\draw[thick, black] (0,0.7) -- (0,1);
		\draw[thick, black,->-=0.6] (0,0.3) -- (0.4,0.3);
		\draw[thick, black,->-=0.6] (0.4,0.7) -- (0.,0.7);
		\draw[thick, black,->-=0.7] (0.4,0.5) -- (0.8,0.5);
		\draw[thick, black] (0.8,0.5) -- (0.8,0);
		\draw[thick, black] (0.4,0.3) -- (0.4,0.7);
		\filldraw[fill=white, draw=black, thick] (0,0) circle (0.08);
		\filldraw[fill=black, draw=black, thick] (0,1) circle (0.08);
		\filldraw[fill=black, draw=black, thick] (0.4,0.5) circle (0.08);
		\filldraw[fill=black, draw=black, thick] (0.8,0) circle (0.08);
		\node[above] at (0,1) {$(\phi^B_{\bar{y}})_{i'}$};
		\node[below] at (-0.1,0) {$(\bphi^A_{x})_{i}$};
		\node[left] at (0,0.3) {$x$};
		\node[left] at (0.4,0.5) {$\mu$};
		\node[left] at (0,0.7) {$\bar{y}$};
		\node[right] at (0.8,0.3) {$z$};
		\node[below] at (1,0) {$(\phi^A_{z})_{k}$};
	\end{tikzpicture},\quad
	\begin{tikzpicture}
		\tikzset{baseline=(current bounding box.center)}
		\draw[thick, cyan, -<-=.3,-<-=0.9] (-0.5,0) -- (0.5,0);
		\draw[thick,gray, ->-=0.7] (0,0) -- (0,0.5);
		\filldraw[fill=blue!30, draw=black, thick] (0,0) circle (0.12);
		\node[below] at (0,0) {$A^{\otimes n}$};
		\draw[thick, orange, ->-=.3,->-=0.9] (-0.5,1) -- (0.5,1);
		\draw[thick,gray, ->-=0.7] (0,0.5) -- (0,1);
		\filldraw[fill=yellow!30, draw=black, thick] (0,1) circle (0.12);
		\node[above] at (0,1) {$B^{\otimes n}$};
		\node[draw, rectangle,minimum width=2mm, minimum height=2mm, inner sep=0mm,thick, fill=dgreen] at (0,0.5) {};
		\node[right] at (0,0.5) {$\bCS_{I,(x,y),(i,i')}^{A|B,(z,k)}$};
	\end{tikzpicture}\ = \ \begin{tikzpicture}
		\tikzset{baseline=(current bounding box.center)}
		\draw[thick, cyan, -<-=.2,-<-=0.6] (-0.5,0) -- (1.2,0);
		\draw[thick, orange, ->-=.6,->-=0.9] (-0.5,1) -- (1.2,1);
		\draw[thick, black] (0.8,0) -- (0.8,0.3);
		\draw[thick, black] (0.8,0.7) -- (0.8,1);
		\draw[thick, black,->-=0.6] (0.4,0.3) -- (0.8,0.3);
		\draw[thick, black,->-=0.6] (0.8,0.7) -- (0.4,0.7);
		\draw[thick, black,->-=0.7] (0,0.5) -- (0.4,0.5);
		\draw[thick, black] (0.0,0.5) -- (0.0,0);
		\draw[thick, black] (0.4,0.3) -- (0.4,0.7);
		\filldraw[fill=white, draw=black, thick] (0,0) circle (0.08);
		\filldraw[fill=white, draw=black, thick] (0.8,1) circle (0.08);
		\filldraw[fill=white, draw=black, thick] (0.4,0.5) circle (0.08);
		\filldraw[fill=black, draw=black, thick] (0.8,0) circle (0.08);
		\node[above] at (0.8,1) {$(\bphi^B_{\bar{y}})_{i'}$};
		\node[below] at (-0.1,0) {$(\bphi^A_{z})_{k}$};
		\node[right] at (0.8,0.3) {$x$};
		\node[right] at (0.4,0.5) {$\mu$};
		\node[right] at (0.8,0.7) {$\bar{y}$};
		\node[left] at (0.0,0.3) {$z$};
		\node[below] at (1,0) {$(\phi^A_{x})_{i}$};
	\end{tikzpicture}
\end{equation}
where $\bar{y} \in \simp(\CC)$ is the orientation reversed object of $y$. {\color{black}For each fusion outcome $z$, we fix the fusion channel $\mu$ shown in the diagrams and suppress its label on the local operators.} Since both $A^{\otimes n}$ and $B^{\otimes n}$ are injective, a solution always exists for the above equations. For states obtained from $\ketx{A},\ketx{B}$ by {\color{black}finite-depth circuits of locally symmetric gates}, the local operators can be reconstructed from the same equations using their blocked MPS tensors.

An interesting special case of the above construction is taking $y=x$, since the identity $1$ is inside the multiplication of $x\times \bar{x}$, we can choose $z=1$, then the local strange charged operators should satisfy,
\begin{equation}\label{eq:strangecond}
	\begin{tikzpicture}
		\tikzset{baseline=(current bounding box.center)}
		\draw[thick, cyan, -<-=.3,-<-=0.9] (-0.5,0) -- (0.5,0);
		\draw[thick,gray, ->-=0.7] (0,0) -- (0,0.5);
		\filldraw[fill=blue!30, draw=black, thick] (0,0) circle (0.12);
		\node[below] at (0,0) {$A^{\otimes n}$};
		\draw[thick, orange, ->-=.3,->-=0.9] (-0.5,1) -- (0.5,1);
		\draw[thick,gray, ->-=0.7] (0,0.5) -- (0,1);
		\filldraw[fill=yellow!30, draw=black, thick] (0,1) circle (0.12);
		\node[above] at (0,1) {$B^{\otimes n}$};
		\node[draw, rectangle,minimum width=2mm, minimum height=2mm, inner sep=0mm,thick, fill=dgreen] at (0,0.5) {};
		\node[right] at (0,0.5) {$\CS_{I,x,(i,i')}^{A|B}$};
	\end{tikzpicture}\ = \ \begin{tikzpicture}
		\tikzset{baseline=(current bounding box.center)}
		\draw[thick, cyan, -<-=.3,-<-=0.9] (-0.5,0) -- (0.5,0);
		\draw[thick, orange, ->-=.3,->-=0.9] (-0.5,1) -- (0.5,1);
		\draw[thick, black] (0,0) -- (0,0.3);
		\draw[thick, black] (0,0.7) -- (0,1);
		\draw[thick, black,->-=0.6] (0,0.3) -- (0.4,0.3);
		\draw[thick, black,->-=0.6] (0.4,0.7) -- (0.,0.7);
		\draw[thick, black] (0.4,0.3) -- (0.4,0.7);
		\filldraw[fill=white, draw=black, thick] (0,0) circle (0.08);
		\filldraw[fill=black, draw=black, thick] (0,1) circle (0.08);
		\node[above] at (0,1) {$(\phi^B_{x})_{i'}$};
		\node[below] at (-0.1,0) {$(\bphi^A_{x})_{i}$};
		\node[left] at (0,0.3) {$x$};
		\node[left] at (0,0.7) {$x$};
	\end{tikzpicture},\quad \begin{tikzpicture}
		\tikzset{baseline=(current bounding box.center)}
		\draw[thick, cyan, -<-=.3,-<-=0.9] (-0.5,0) -- (0.5,0);
		\draw[thick,gray, ->-=0.7] (0,0) -- (0,0.5);
		\filldraw[fill=blue!30, draw=black, thick] (0,0) circle (0.12);
		\node[below] at (0,0) {$A^{\otimes n}$};
		\draw[thick, orange, ->-=.3,->-=0.9] (-0.5,1) -- (0.5,1);
		\draw[thick,gray, ->-=0.7] (0,0.5) -- (0,1);
		\filldraw[fill=yellow!30, draw=black, thick] (0,1) circle (0.12);
		\node[above] at (0,1) {$B^{\otimes n}$};
		\node[draw, rectangle,minimum width=2mm, minimum height=2mm, inner sep=0mm,thick, fill=dgreen] at (0,0.5) {};
		\node[right] at (0,0.5) {$\bCS_{I,x,(i,i')}^{A|B}$};
	\end{tikzpicture} \ = \ \begin{tikzpicture}
		\tikzset{baseline=(current bounding box.center)}
		\draw[thick, cyan, -<-=.3,-<-=0.9] (-0.5,0) -- (0.5,0);
		\draw[thick, orange, ->-=.3,->-=0.9] (-0.5,1) -- (0.5,1);
		\draw[thick, black] (0,0) -- (0,0.3);
		\draw[thick, black] (0,0.7) -- (0,1);
		\draw[thick, black,->-=0.6] (-0.4,0.3) -- (0,0.3);
		\draw[thick, black,->-=0.6] (0.0,0.7) -- (-0.4,0.7);
		\draw[thick, black] (-0.4,0.3) -- (-0.4,0.7);
		\filldraw[fill=white, draw=black, thick] (0,1) circle (0.08);
		\filldraw[fill=black, draw=black, thick] (0,0) circle (0.08);
		\node[above] at (0,1) {$(\bphi^B_{x})_{i'}$};
		\node[right] at (0,0.3) {$x$};
		\node[right] at (0,0.7) {$x$};
		\node[below] at (0,0) {$(\phi^A_{x})_{i}$};
	\end{tikzpicture}
\end{equation}
which we dub as the \textit{strange charged operator condition}. The right hand sides are nothing but the elements of the interface algebras $\CI_{x,(i,i')}^{A|B}$ and $\CI_{x,(i',i)}^{B|A}$. When all irreducible representations of the interface algebra have dimension greater than one, the interface necessarily carries a nontrivial multiplet. If a one-dimensional representation exists, a nondegenerate interface is allowed, but higher-dimensional sectors can still occur, as in the $\Rep(H_8)$ example in \eqref{eq:h8mixedinterfacedecomposition}. Since the blocked MPS tensors of $A$ and $B$ are injective, such local operators exist \footnote{We assume for some $n$, such that both $A^{\otimes n}$ and $B^{\otimes n}$ are injective.  We label the virtual indices by Greek letters and the physical indices by capital letters, $(A^{\otimes n})^I_{\alpha,\beta},(B^{* \otimes n})_{J}^{\gamma,\delta}$. The right hand side is some general 4-leg tensor which can be decomposed as $\sum_i \widetilde{v}^i_{\alpha,\beta} \widetilde{u}_i^{\gamma,\delta}$. Because $A^{\otimes n}$ and $B^{\otimes n}$ are injective, there exist $p_I^i, q_i^J$, such that $\sum_I  (A^{\otimes n})^I_{\alpha,\beta} p_I^i=\widetilde{v}^i_{\alpha,\beta}$ and $\sum_J (B^{* \otimes n})_{J}^{\gamma,\delta} q_i^J = \widetilde{u}_i^{\gamma,\delta}$. And the local operator is given by $\sum_i p^i_I q_i^J$. We thank Tomo Soejima for discussing this point.}. Intuitively, the strange charged operators are constructed by the following procedure, first inserting an identity line in between MPS $\bra{B}$ and $\ket{A}$, then making a bubble of $x,\bar{x}$, and finally fusing the line with the states $\bra{B},\ket{A}$.
\begin{align}\label{eq:strangecharge}
	&\begin{tikzpicture}
		\tikzset{baseline=(current bounding box.center)}
		\draw[thick, cyan, -<-=.1,-<-=0.9] (-1.5,0) -- (1.5,0);
		\draw[thick, orange, ->-=.1,->-=0.9] (-1.5,1) -- (1.5,1);
		\draw[thick,gray, ->-=0.3,->-=0.9] (1,0) -- (1,1);
		\draw[thick,gray, ->-=0.3,->-=0.9] (0.2,0) -- (0.2,1);
		\draw[thick,gray, ->-=0.3,->-=0.9] (-0.2,0) -- (-0.2,1);
		\draw[thick,gray, ->-=0.3,->-=0.9] (-1,0) -- (-1,1);
		\node[left] at (-1.5,0) {$A$};
		\node[left] at (-1.5,1) {$B$};
	\end{tikzpicture} =\begin{tikzpicture}
		\tikzset{baseline=(current bounding box.center)}
		\draw[thick, cyan, -<-=.1,-<-=0.9] (-1.5,0) -- (1.5,0);
		\draw[thick, orange, ->-=.1,->-=0.9] (-1.5,1) -- (1.5,1);
		\draw[thick,gray, ->-=0.3,->-=0.9] (1,0) -- (1,1);
		\draw[thick,gray, ->-=0.3,->-=0.9] (0.2,0) -- (0.2,1);
		\draw[thick,gray, ->-=0.3,->-=0.9] (-0.2,0) -- (-0.2,1);
		\draw[thick,gray, ->-=0.3,->-=0.9] (-1,0) -- (-1,1);
		\draw[thick, black, ->-=.1,->-=0.9,dashed] (-1.5,0.5) -- (1.5,0.5);
		\node[left] at (-1.5,0) {$A$};
		\node[left] at (-1.5,1) {$B$};
		\node[left] at (-1.5,0.5) {$1$};
	\end{tikzpicture} = \frac{1}{d_x} \begin{tikzpicture}
		\tikzset{baseline=(current bounding box.center)}
		\draw[thick, cyan, -<-=.1,-<-=0.9] (-1.5,0) -- (1.5,0);
		\draw[thick, orange, ->-=.1,->-=0.9] (-1.5,1) -- (1.5,1);
		\draw[thick,gray, ->-=0.3,->-=0.9] (1,0) -- (1,1);
		\draw[thick,gray, ->-=0.3,->-=0.9] (0.2,0) -- (0.2,1);
		\draw[thick,gray, ->-=0.3,->-=0.9] (-0.2,0) -- (-0.2,1);
		\draw[thick,gray, ->-=0.3,->-=0.9] (-1,0) -- (-1,1);
		\draw[thick, black, ->-=0.4,dashed] (-1.5,0.5) -- (-0.8,0.5);
		\draw[thick, black, ->-=0.8,dashed] (0.8,0.5) -- (1.5,0.5);
		\draw[thick, black, ->-=0.6] (-0.8,0.7) -- (0.8,0.7);
		\draw[thick, black, ->-=0.6] (-0.8,0.3) -- (0.8,0.3);
		\draw[thick, black] (-0.8,0.3) -- (-0.8,0.7);
		\draw[thick, black] (0.8,0.3) -- (0.8,0.7);
		\filldraw[fill=white, draw=black, thick] (-0.8,0.5) circle (0.08);
		\filldraw[fill=black, draw=black, thick] (0.8,0.5) circle (0.08);
		\node[left] at (-1.5,0) {$A$};
		\node[left] at (-1.5,1) {$B$};
		\node[left] at (-1.5,0.5) {$1$};
		\node[above] at (-0.5,0.6) {$\bar{x}$};
		\node[below] at (-0.5,0.35) {$x$};
	\end{tikzpicture} \\
	&= \frac{1}{d_x} \begin{tikzpicture}
		\tikzset{baseline=(current bounding box.center)}
		\draw[thick, cyan, -<-=.1,-<-=0.9] (-1.5,0) -- (1.5,0);
		\draw[thick, orange, ->-=.1,->-=0.9] (-1.5,1) -- (1.5,1);
		\draw[thick,gray, ->-=0.3,->-=0.9] (1,0) -- (1,1);
		\draw[thick,gray, ->-=0.3,->-=0.9] (0.2,0) -- (0.2,1);
		\draw[thick,gray, ->-=0.3,->-=0.9] (-0.2,0) -- (-0.2,1);
		\draw[thick,gray, ->-=0.3,->-=0.9] (-1,0) -- (-1,1);
		\draw[thick, black, ->-=0.4,dashed] (-1.5,0.5) -- (-0.8,0.5);
		\draw[thick, black, ->-=0.8,dashed] (0.8,0.5) -- (1.5,0.5);
		\draw[thick, black, -<-=0.6] (-0.8,0.7) -- (0.8,0.7);
		\draw[thick, black, ->-=0.6] (-0.8,0.3) -- (0.8,0.3);
		\draw[thick, black] (-0.8,0.3) -- (-0.8,0.7);
		\draw[thick, black] (0.8,0.3) -- (0.8,0.7);
		\filldraw[fill=white, draw=black, thick] (-0.8,0.5) circle (0.08);
		\filldraw[fill=black, draw=black, thick] (0.8,0.5) circle (0.08);
		\node[left] at (-1.5,0) {$A$};
		\node[left] at (-1.5,1) {$B$};
		\node[left] at (-1.5,0.5) {$1$};
		\node[above] at (-0.5,0.6) {$x$};
		\node[below] at (-0.5,0.35) {$x$};
	\end{tikzpicture} =\frac{1}{d_x}  \sum_{i,i'}\begin{tikzpicture}
		\tikzset{baseline=(current bounding box.center)}
		\draw[thick, cyan, -<-=.1,-<-=0.9] (-1.5,0) -- (1.5,0);
		\draw[thick, orange, ->-=.1,->-=0.9] (-1.5,1) -- (1.5,1);
		\draw[thick,gray, ->-=0.3,->-=0.9] (1,0) -- (1,1);
		\draw[thick,gray, ->-=0.3,->-=0.9] (0.2,0) -- (0.2,1);
		\draw[thick,gray, ->-=0.3,->-=0.9] (-0.2,0) -- (-0.2,1);
		\draw[thick,gray, ->-=0.3,->-=0.9] (-1,0) -- (-1,1);
		\draw[thick, black, -<-=0.6] (-0.8,0.7) -- (-0.4,0.7);
		\draw[thick, black, -<-=0.6] (0.4,0.7) -- (0.8,0.7);
		\draw[thick, black, ->-=0.6] (-0.8,0.3) -- (-0.4,0.3);
		\draw[thick, black, ->-=0.6] (0.4,0.3) -- (0.8,0.3);
		\draw[thick, black] (-0.4,0.7) -- (-0.4,1);
		\draw[thick, black] (0.4,0.7) -- (0.4,1);
		\draw[thick, black] (-0.4,0.3) -- (-0.4,0);
		\draw[thick, black] (0.4,0.3) -- (0.4,0);
		\draw[thick, black] (-0.8,0.3) -- (-0.8,0.7);
		\draw[thick, black] (0.8,0.3) -- (0.8,0.7);
		\filldraw[fill=white, draw=black, thick] (0.4,0) circle (0.08);
		\filldraw[fill=black, draw=black, thick] (-0.4,0) circle (0.08);
		\filldraw[fill=black, draw=black, thick] (0.4,1) circle (0.08);
		\filldraw[fill=white, draw=black, thick] (-0.4,1) circle (0.08);
		\node[left] at (-1.5,0) {$A$};
		\node[left] at (-1.5,1) {$B$};
		\node[below] at (-0.5,0) {$(\phi^A_{x})_{i}$};
		\node[below] at (0.5,0) {$(\bphi^A_{x})_{i}$};
		\node[above] at (-0.5,1) {$(\bphi^B_{x})_{i'}$};
		\node[above] at (0.5,1) {$(\phi^B_{x})_{i'}$};
	\end{tikzpicture}\mathrel{\propto}\frac{1}{d_x}  \sum_{i,i'}\begin{tikzpicture}
		\tikzset{baseline=(current bounding box.center)}
		\draw[thick, cyan, -<-=.1,-<-=0.9] (-1.5,0) -- (1.5,0);
		\draw[thick, orange, ->-=.1,->-=0.9] (-1.5,1) -- (1.5,1);
		\draw[thick,gray, ->-=0.3,->-=0.9] (1,0) -- (1,1);
		\draw[thick,gray, ->-=0.3,->-=0.9] (0.2,0) -- (0.2,1);
		\draw[thick,gray, ->-=0.3,->-=0.9] (-0.2,0) -- (-0.2,1);
		\draw[thick,gray, ->-=0.3,->-=0.9] (-1,0) -- (-1,1);
		\draw[thick,gray, ->-=0.3,->-=0.9] (-0.6,0) -- (-0.6,1);
		\draw[thick,gray, ->-=0.3,->-=0.9] (0.6,0) -- (0.6,1);
		\node[draw, rectangle,minimum width=2mm, minimum height=2mm, inner sep=0mm,thick, fill=dgreen] at (0.6,0.5) {};
		\node[draw, rectangle,minimum width=2mm, minimum height=2mm, inner sep=0mm,thick, fill=dgreen] at (-0.6,0.5) {};
		\node[left] at (-1.5,0) {$A$};
		\node[left] at (-1.5,1) {$B$};
		\node[below] at (-0.8,0) {$\bCS_{I,x,(i,i')}^{A|B}$};
		\node[below] at (0.8,0) {$\CS_{J,x,(i,i')}^{A|B}$};
	\end{tikzpicture}
\end{align}
where $d_x$ is the quantum dimension of symmetry line operator $x$, and from the third diagram we use the orientation reversal to reverse the arrow and change $\bar{x}\rightarrow x$. When the mixed transfer matrix of a block is a nonzero scalar multiple of the identity, replacing two such blocks by the interface matrices changes the normalization by a factor independent of their separation. In this case the contraction above gives
\begin{equation}
	SC_x^{A|B}(I,J)\equiv \sum_{1\le i,i'\le d_x} \frac{\bra{B}\bCS_{I,x,(i,i')}^{A|B}\CS_{J,x,(i,i')}^{A|B}\ket{A}}{\bra{B}\ket{A}} \mathrel{\propto} d_x,\quad \text{as $|I-J|\rightarrow \infty$}
\end{equation}
The proportionality factor depends on the normalization of the tensors and insertions. For the Tri state and the deformed Even state $\ket{\mathrm{Even}_\epsilon}$ of \secref{sec:repd8strange}, the one-point functions of the reconstructed $\CD$ insertions vanish. With the normalization used there, the summed correlator is therefore connected and equals
\begin{equation}
    \label{eq:trievenconnected}SC_{\CD}^{\mathrm{Tri}|\mathrm{Even}_\epsilon}(I,J)=2,\qquad \text{for disjoint eight-site blocks $I,J$.}
\end{equation}
Keeping the original operators fixed does not guarantee a nonzero value throughout the phases. Exponential decay for an unmatched pair requires a selection rule or a transfer-matrix calculation that excludes a nonzero long-distance contribution; it does not follow from the phase labels alone.

If $x$ is group-like element, then there is no summation as the quantum dimension of group-like element is $1$ and reduces to the strange correlator of ordinary SPTs \cite{yizhuang2014strange}.

\subsection{Group case \texorpdfstring{$\IZ_2\times \IZ_2$}{Z2 x Z2}}\label{sec:scz2z2}
In this section, we show that the strange charged operator constructed using the procedure in the previous section matches the known charged operator in the group case. We use ZX calculus to evaluate the tensor contractions and obtain explicit operators. The diagrammatic rules are summarized in \appref{app:qdzx}; see \cite{nat2024zxtensor} for their application to generalized symmetries.

The physical local Hilbert space $\IC^2\otimes \IC^2$ contains two qubits, and the symmetry operator as a MPO is,
\begin{equation}
    \begin{tikzpicture}
    \tikzset{baseline=(current bounding box.center)}
    \draw[thick, black, ->-=.3,->-=0.9] (-0.5,0) -- (0.5,0);
    \draw[thick,gray, ->-=0.25,->-=0.95] (0,-0.5) -- (0,0.5);
    \filldraw[fill=gray!30, draw=black, thick] (0,0) circle (0.12);
    \node[left] at (-0.5,0) {$\CO_{(a,b)}$};
\end{tikzpicture}\ =\  \begin{tikzpicture}
\tikzset{baseline=(current bounding box.center)}
    \begin{pgfonlayer}{nodelayer}
		\node [style=red dot] (0) at (-0.5, 0) {$a$};
		\node [style=red dot] (1) at (0, 0) {$b$};
		\node [style=none] (2) at (-0.5, 0.5) {};
            \node [style=none] (3) at (-0.5, -0.5) {};
		\node [style=none] (4) at (0, 0.5) {};	
		\node [style=none] (5) at (0, -0.5) {};
	\end{pgfonlayer}
	\begin{pgfonlayer}{edgelayer}
		\draw (2.center) to (0);
		\draw (0) to (3.center);
		\draw (1) to (4.center);
		\draw (1) to (5.center);
	\end{pgfonlayer}
\end{tikzpicture} = X^a \otimes X^b
\end{equation}
where $a,b=0,1$ in $X^{a,b}$ denotes the identity matrix $I$ and Pauli $X$ matrix respectively, while in the ZX calculus diagram $a,b\sim a\pi,b\pi$ \footnote{Letters in the ZX diagrams are always understood as multiplied by $\pi$.}, which follows from the definition of the red spider. The vertical legs acts on the physical local Hilbert space $\IC^2\otimes \IC^2$, with qubits on even and odd sites for example. Since the MPO has bond dimension 1, there is no horizontal virtual leg.

The two symmetric gapped phases with a unique ground state of $\IZ_2\times \IZ_2$ symmetry are the trivial product state and the cluster state \footnote{The factor $2^{-1}$ in the product state is due to the normalization of the ZX calculus, $\begin{tikzpicture}
		\begin{pgfonlayer}{nodelayer}
			\node [style=tiny green dot] (0) at (0, 0) {};
			\node [style=none] (1) at (-0.5,0) {};
		\end{pgfonlayer}
		\begin{pgfonlayer}{edgelayer}
			\draw (1.center) to (0);
		\end{pgfonlayer}
	\end{tikzpicture}\ =\ \sqrt{2}\ket{+}$, $\begin{tikzpicture}
		\begin{pgfonlayer}{nodelayer}
			\node [style=tiny red dot] (0) at (0, 0) {};
			\node [style=none] (1) at (-0.5,0) {};
		\end{pgfonlayer}
		\begin{pgfonlayer}{edgelayer}
			\draw (1.center) to (0);
		\end{pgfonlayer}
	\end{tikzpicture}\ =\ \sqrt{2}\ket{0}$.},
\begin{equation}
\begin{tikzpicture}
    \tikzset{baseline=(current bounding box.center)}
    \draw[thick, cyan, -<-=.3,-<-=0.9] (-0.5,0) -- (0.5,0);
    \draw[thick,gray, ->-=0.7] (0,0) -- (0,0.5);
    \filldraw[fill=blue!30, draw=black, thick] (0,0) circle (0.12);
    \node[above] at (-0.5,0) {Tri};
\end{tikzpicture}  = 2^{-1}
 \begin{tikzpicture}
\tikzset{baseline=(current bounding box.center)}
    \begin{pgfonlayer}{nodelayer}
		\node [style=green dot] (0) at (-0.5, 0) {};
            \node [style=none] (1) at (-0.5, 0.5) {};
		\node [style=green dot] (2) at (0, 0) {};
		\node [style=none] (3) at (0, 0.5) {};
	\end{pgfonlayer}
	\begin{pgfonlayer}{edgelayer}
		\draw (1.center) to (0);
		\draw (2) to (3.center);
	\end{pgfonlayer}
\end{tikzpicture} ,\quad   \begin{tikzpicture}
    \tikzset{baseline=(current bounding box.center)}
    \draw[thick, orange, -<-=.3,-<-=0.9] (-0.5,0) -- (0.5,0);
    \draw[thick,gray, ->-=0.7] (0,0) -- (0,0.5);
    \filldraw[fill=yellow!30, draw=black, thick] (0,0) circle (0.12);
    \node[above] at (-0.5,0) {SPT};
\end{tikzpicture} \ = \
 \begin{tikzpicture}
 \tikzset{baseline=(current bounding box.center)}
	\begin{pgfonlayer}{nodelayer}
		\node [style=had] (0) at (-0.5, 2.5) {};
		\node [style=green dot] (1) at (-1, 2.5) {};
		\node [style=green dot] (2) at (0, 2.5) {};
		\node [style=none] (3) at (-1, 3) {};
		\node [style=none] (4) at (0, 3) {};
		\node [style=none] (5) at (1, 2.5) {};
		\node [style=had] (7) at (0.5, 2.5) {};
		\node [style=none] (16) at (-1.5, 2.5) {};
	\end{pgfonlayer}
	\begin{pgfonlayer}{edgelayer}
		\draw (1) to (0);
		\draw (0) to (2);
		\draw (3.center) to (1);
		\draw (4.center) to (2);
		\draw (2) to (7);
		\draw (7) to (5);
		\draw (16.center) to (1);
	\end{pgfonlayer}
\end{tikzpicture}
\end{equation}
where the yellow box is the Hadamard gate $\frac{1}{\sqrt{2}}\left(\begin{smallmatrix}
	1 & 1 \\1 & -1
\end{smallmatrix}\right)$. The split and fusion action tensors for the trivial product state is 1, while for the cluster state,
\begin{equation}
     \begin{tikzpicture}
 \tikzset{baseline=(current bounding box.center)}
	\begin{pgfonlayer}{nodelayer}
		\node [style=had] (0) at (-0.5, 2.5) {};
		\node [style=green dot] (1) at (-1, 2.5) {};
		\node [style=green dot] (2) at (0, 2.5) {};
		\node [style=none] (3) at (-1, 3) {};
		\node [style=none] (4) at (0, 3) {};
		\node [style=none] (5) at (1, 2.5) {};
		\node [style=had] (7) at (0.5, 2.5) {};
		\node [style=none] (16) at (-1.5, 2.5) {};
            \node[style=red dot] (8) at (-1,3) {$a$};
            \node[style=none] (9) at (-1,3.5) {};
            \node[style=red dot] (10) at (-0,3) {$b$};
            \node[style=none] (11) at (0,3.5) {};
	\end{pgfonlayer}
	\begin{pgfonlayer}{edgelayer}
		\draw (1) to (0);
		\draw (0) to (2);
		\draw (3.center) to (1);
		\draw (4.center) to (2);
		\draw (2) to (7);
		\draw (7) to (5);
		\draw (16.center) to (1);
            \draw (9.center) to (8);
            \draw (11.center) to (10);
	\end{pgfonlayer}
\end{tikzpicture} = \begin{tikzpicture}
 \tikzset{baseline=(current bounding box.center)}
	\begin{pgfonlayer}{nodelayer}
		\node [style=had] (0) at (-0.5, 2.5) {};
		\node [style=green dot] (1) at (-1, 2.5) {};
		\node [style=green dot] (2) at (0, 2.5) {};
		\node [style=none] (3) at (-1, 3) {};
		\node [style=none] (4) at (0, 3) {};
		\node [style=none] (5) at (2, 2.5) {};
		\node [style=had] (7) at (0.5, 2.5) {};
		\node [style=none] (16) at (-2.5, 2.5) {};
            \node[style=red dot] (8) at (-2,2.5) {$a$};
            \node[style=green dot] (9) at (-1.5,2.5) {$b$};
            \node[style=red dot] (10) at (1.5,2.5) {$a$};
            \node[style=green dot] (11) at (1,2.5) {$b$};
	\end{pgfonlayer}
	\begin{pgfonlayer}{edgelayer}
		\draw (1) to (0);
		\draw (0) to (2);
		\draw (3.center) to (1);
		\draw (4.center) to (2);
		\draw (2) to (7);
		\draw (7) to (5.center);
		\draw (16.center) to (1);
	\end{pgfonlayer}
\end{tikzpicture}
\end{equation}
Therefore, the fusion and split action tensor for the cluster state are,
\begin{equation}
    \begin{tikzpicture}
    \tikzset{baseline=(current bounding box.center)}
    \draw[thick, orange, -<-=.3,-<-=0.9] (-0.5,0) -- (0.5,0);
    \draw[thick, black] (0.0,0.0) -- (0.0,0.5);
    \draw[thick, black,->-=0.6] (-0.5,0.5) -- (0.0,0.5);
    \filldraw[fill=black, draw=black, thick] (0,0) circle (0.08);
    \node[above] at (-0.5,0) {SPT};
    \node[above] at (-0.5,0.5) {${(a,b)}$};
\end{tikzpicture}\ = \ \begin{tikzpicture}
 \tikzset{baseline=(current bounding box.center)}
	\begin{pgfonlayer}{nodelayer}
		\node[style=none] (0) at (0,0) {};
            \node[style=red dot] (1) at (0.5,0) {$a$};
            \node[style=green dot] (2) at (1,0) {$b$};
            \node[style=none] (3) at (1.5,0) {};
	\end{pgfonlayer}
	\begin{pgfonlayer}{edgelayer}
		\draw (0.center) to (3.center);
	\end{pgfonlayer}
\end{tikzpicture},\quad     \begin{tikzpicture}
    \tikzset{baseline=(current bounding box.center)}
    \draw[thick, orange, -<-=.3,-<-=0.9] (-0.5,0) -- (0.5,0);
    \draw[thick, black] (0.0,0.0) -- (0.0,0.5);
    \draw[thick, black,->-=0.6] (0,0.5) -- (0.5,0.5);
    \filldraw[fill=white, draw=black, thick] (0,0) circle (0.08);
    \node[above] at (-0.5,0) {SPT};
    \node[above] at (0.5,0.5) {${(a,b)}$};
\end{tikzpicture}\ = \ \begin{tikzpicture}
 \tikzset{baseline=(current bounding box.center)}
	\begin{pgfonlayer}{nodelayer}
		\node[style=none] (0) at (0,0) {};
            \node[style=red dot] (1) at (1,0) {$a$};
            \node[style=green dot] (2) at (0.5,0) {$b$};
            \node[style=none] (3) at (1.5,0) {};
	\end{pgfonlayer}
	\begin{pgfonlayer}{edgelayer}
		\draw (0.center) to (3.center);
	\end{pgfonlayer}
\end{tikzpicture}
\end{equation}
From the action tensors, one can check that the $L$-symbol is,
\begin{equation}
    {}^\text{SPT}L_{(a_1,b_1),(a_2,b_2)}^{(a_1+a_2,b_1+b_2)} = (-1)^{a_1b_2}
\end{equation}
which is the non-trivial 2-cocycle in $H^2(\IZ_2 \times \IZ_2,U(1))$. Next, following the previous procedure, we want to find local operators such that,
\begin{equation}
    \begin{tikzpicture}
 \tikzset{baseline=(current bounding box.center)}
	\begin{pgfonlayer}{nodelayer}
		\node[style=none] (0) at (0.5,1) {};
            \node[style=had] (1) at (2.5,1) {};
            \node[style=green dot] (2) at (1,1) {};
            \node[style=had] (3) at (1.5,1) {};
            \node[style=green dot] (4) at (2,1) {};
            \node[style=none] (5) at (3,1) {};
            \node[style=green dot] (6) at (1,0) {};
            \node[style=green dot] (7) at (2,0) {};
	\end{pgfonlayer}
	\begin{pgfonlayer}{edgelayer}
		\draw (0.center) to (5.center);
            \draw (2.center) to (6.center);
            \draw (4.center) to (7.center);
	\end{pgfonlayer}
    \node[draw, rectangle,minimum width=12mm, minimum height=2mm, inner sep=0mm,thick, fill=dgreen] at (1.5,0.5) {};
    \node[right] at (2.2,0.5) {$\CS_{(I_1,I_2),(a,b)}^{\text{Tri}|\text{SPT}}$};
    \node[below] at (1,-0.2) {$I_1$};
    \node[left] at (1,0) {$2^{-1}$};
    \node[below] at (2,-0.2) {$I_2$};
\end{tikzpicture} = \begin{tikzpicture}
    \tikzset{baseline=(current bounding box.center)}
    \draw[thick, cyan, -<-=.3,-<-=0.9] (-0.5,0) -- (0.5,0);
    \draw[thick, orange, ->-=.3,->-=0.9] (-0.5,1) -- (0.5,1);
    \draw[thick, black] (0,0) -- (0,0.3);
    \draw[thick, black] (0,0.7) -- (0,1);
    \draw[thick, black,->-=0.6] (0,0.3) -- (0.4,0.3);
    \draw[thick, black,->-=0.6] (0.4,0.7) -- (0.,0.7);
    \draw[thick, black] (0.4,0.3) -- (0.4,0.7);
    \filldraw[fill=white, draw=black, thick] (0,0) circle (0.08);
    \filldraw[fill=black, draw=black, thick] (0,1) circle (0.08);
    \node[above] at (0,1) {$\phi^{\text{SPT}}_{(a,b)}$};
    \node[below] at (-0.1,0) {$\bphi^{\text{Tri}}_{(a,b)}$};
    \node[left] at (0,0.7) {${(a,b)}$};
\end{tikzpicture}= \begin{tikzpicture}
 \tikzset{baseline=(current bounding box.center)}
	\begin{pgfonlayer}{nodelayer}
		\node[style=none] (0) at (0,0) {};
            \node[style=red dot] (1) at (1,0) {$a$};
            \node[style=green dot] (2) at (0.5,0) {$b$};
            \node[style=none] (3) at (1.5,0) {};
	\end{pgfonlayer}
	\begin{pgfonlayer}{edgelayer}
		\draw (0.center) to (3.center);
	\end{pgfonlayer}
\end{tikzpicture} \Rightarrow \CS_{(I_1,I_2),(a,b)}^{\text{Tri}|\text{SPT}} =  2Z^b_{I_1}Z^a_{I_2}
\end{equation}
where $I_1,I_2$ are the lattice indices. Similarly, one can find $\bCS_{(I_1,I_2),(a,b)}^{\text{Tri}|\text{SPT}} = 2(-1)^{ab} Z^b_{I_1}Z^a_{I_2}$. This aligns with the strange correlator in \cite{yizhuang2014strange}. 

\subsection{Non-invertible SPTs of \texorpdfstring{$\Rep D_8$}{Rep(D8)}}\label{sec:screpd8}
\begin{figure}[htbp]
    \centering
    \includegraphics[width=0.3\linewidth]{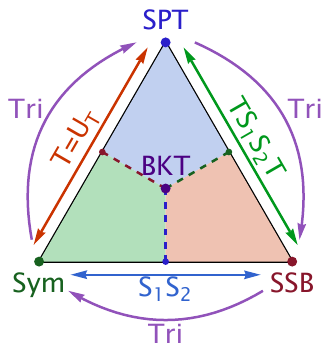}
    \caption{Phase diagram of $\IZ_2\times \IZ_2$ symmetric phases from \cite{lu2024latticep}. The three corners are the symmetric gapped phases, where Sym stands for symmetric trivial gapped phase, i.e. product state, SSB stands for the $\IZ_2\times \IZ_2$ SSB phase, and SPT is the cluster state as the non-trivial $\IZ_2\times \IZ_2$ SPT. Interpolating the each two gapped phases will have gapless critical theories which are labeled by dashed lines. The 3 critical theories intersects at the Berezinskii-Kosterlitz-Thouless (BKT) transition point.}
    \label{fig:z2z2tri}
\end{figure}

\begingroup
\color{black}
\hypersetup{citecolor=ceruleanblue,linkcolor=ceruleanblue}
The $\IZ_2\times\IZ_2$ cluster state is invariant under Kramers-Wannier duality (the bottom blue arrow $S_1S_2$ in \figref{fig:z2z2tri}), while the product state is mapped to a symmetry-breaking state \cite{Sahand2024cluster,lu2024latticep}. This duality MPO has bond dimension four and does not satisfy the on-site condition. Conjugating it by the SPT entangler $U_T=\prod_i\cz_{i,i+1}$ gives the Kennedy-Tasaki transformation (the right green arrow $TS_1S_2T$). Here, $\cz_{i,i+1}$ is the controlled-$Z$ gate on neighboring sites. The transformed symmetry admits a bond-dimension-two on-site MPO and leaves the product state invariant \cite{jose2023mpoclas,lan2024mpo,sahand2025gaugempo}.
\par\endgroup

Besides the trivial product state, two other $\Rep(D_8)$ NISPTs were identified in continuum field theory \cite{Thorngren:2019iar}, consistent with the mathematical classification \cite{tambara2000representations}. The three NISPTs were first realized on the lattice via the Kennedy-Tasaki or twisted gauging maps, relating them to SSB phases of the dual symmetry \cite{Sahand2024cluster}; later constructions came from the algebraic perspective \cite{lan2024mpo,alison2024repd8s}. In the following, we will use the states constructed in \cite{Sahand2024cluster} with an additional SPT entangler, such that the MPOs are ``on-site'' and each unit cell contains two qubits \cite{jose2023mpoclas,kansei2024nispt,lan2024mpo}.

\subsubsection{MPO of \texorpdfstring{$\Rep(D_8)$}{Rep(D8)} fusion category symmetry}
We consider the ``on-site'' realization of the $\TY(\IZ_2\times \IZ_2,\chi_{\text{off-diag}},+1)\cong \Rep(D_8)$ fusion category symmetry, such that the trivial product state is the symmetric state.\footnote{\color{black}$\chi_{\text{off-diag}}$ is the symmetric, nondegenerate bicharacter $\chi_{\text{off-diag}}((a_1,b_1),(a_2,b_2))=(-1)^{a_1b_2+a_2b_1}$ on $\IZ_2\times\IZ_2$.} The fusion category contains $\IZ_2\times \IZ_2$ and the Kennedy-Tasaki transformation $\CD$ as the simple lines. In the MPO presentation,
\begin{equation}
    \CO_{(a,b)}=\begin{tikzpicture}
\tikzset{baseline=(current bounding box.center)}
    \begin{pgfonlayer}{nodelayer}
		\node [style=red dot] (0) at (-0.5, 0) {$a$};
		\node [style=red dot] (1) at (0, 0) {$b$};
		\node [style=none] (2) at (-0.5, 0.5) {};
            \node [style=none] (3) at (-0.5, -0.5) {};
		\node [style=none] (4) at (0, 0.5) {};	
		\node [style=none] (5) at (0, -0.5) {};
	\end{pgfonlayer}
	\begin{pgfonlayer}{edgelayer}
		\draw (2.center) to (0);
		\draw (0) to (3.center);
		\draw (1) to (4.center);
		\draw (1) to (5.center);
	\end{pgfonlayer}
\end{tikzpicture},\quad \CO_\CD =2 \begin{tikzpicture}
\tikzset{baseline=(current bounding box.center)}
    \begin{pgfonlayer}{nodelayer}
    \node[style=none] (0) at (0,0) {};
    \node[style=had] (1) at (0.5,0) {};
    \node[style=green dot] (2) at (1,0) {};
    \node[style=red dot] (3) at (1.5,-0.5) {};
    \node[style=none] (4) at (1.5,0.5) {};
    \node[style=none] (5) at (1.5,-1) {};
    \node[style=none] (6) at (3.5,0) {};
    \node[style=had] (7) at (2,0) {};
    \node[style=green dot] (8) at (2.5,0) {};
    \node[style=red dot] (9) at (3,-0.5) {};
    \node[style=none] (10) at (3,-1) {};
    \node[style=none] (11) at (3,0.5) {};
	\end{pgfonlayer}
	\begin{pgfonlayer}{edgelayer}
		\draw (6.center) to (0.center);
            \draw (2) to (3);
            \draw (8) to (9);
            \draw (4.center) to (5.center);
            \draw (10.center) to (11.center);
	\end{pgfonlayer}
\end{tikzpicture}
\end{equation}
where
\begin{equation}\frac{1}{\sqrt{2}}
    \begin{tikzpicture}
\tikzset{baseline=(current bounding box.center)}
    \begin{pgfonlayer}{nodelayer}
    \node[style=red dot] (0) at (0,0) {$i_1$};
    \node[style=had] (1) at (0.5,0) {};
    \node[style=green dot] (2) at (1,0) {};
    \node[style=red dot] (3) at (1.5,-0.5) {};
    \node[style=none] (4) at (1.5,0.5) {};
    \node[style=none] (5) at (1.5,-1) {};
    \node[style=red dot] (6) at (2,0) {$i_2$};
	\end{pgfonlayer}
	\begin{pgfonlayer}{edgelayer}
		\draw (6.center) to (0.center);
            \draw (2) to (3);
            \draw (4.center) to (5.center);
	\end{pgfonlayer}
\end{tikzpicture} =\frac{1}{\sqrt{2}}\begin{pmatrix}
    I & X\\
    I & -X
\end{pmatrix}_{(i_1,i_2)}.
\end{equation}
The fusion rule of the above MPOs are,
\begin{align}
	&\hCO_{(a_1,b_1)}\times \hCO{(a_2,b_2)} = \hCO{(a_1+a_2,b_1+b_2)},\\
	&\hCO{(a_1,b_1)}\times \hCO_\CD = \hCO_\CD \times \hCO_{(a_1,b_1)} = \hCO_\CD, \hCO_\CD\times \hCO_\CD = \sum_{a_1,b_1=0,1} \hCO_{(a_1,b_1)}.
\end{align}
where $a_i,b_i,i=1,2$ are $\IZ_2$ valued. These MPOs satisfy the on-site condition \cite{jose2023mpoclas,lan2024mpo,sahand2025gaugempo}. The fusion junctions for the MPOs are
\begin{align}
&\begin{tikzpicture}
\tikzset{baseline=(current bounding box.center)}
    \draw[thick, black, ->-=.7] (-0.5,0.25) -- (0,0.25);
    \draw[thick, black, ->-=.7] (-0.5,-0.25) -- (0,-0.25);
    \draw[thick, black] (0,0.25) -- (0,-0.25);
    \draw[thick, black, ->-=.7] (0,0) -- (0.5,0);
    \filldraw[fill=black, draw=black, thick] (0,0) circle (0.05);
    \node[left] at (-0.5,0.25) {${(a_1,b_1)}$};
    \node[left] at (-0.5,-0.25) {${(a_2,b_2)}$};
    \node[right] at (0.5,0) {${(a_1+a_2,b_1+b_2)}$};
\end{tikzpicture}= (-1)^{a_1 b_2},\quad  \begin{tikzpicture}
\tikzset{baseline=(current bounding box.center)}
    \draw[thick, black, ->-=.7] (-0.5,0.25) -- (0,0.25);
    \draw[thick, black, ->-=.7] (-0.5,-0.25) -- (0,-0.25);
    \draw[thick, black] (0,0.25) -- (0,-0.25);
    \draw[thick, black, ->-=.7] (0,0) -- (0.5,0);
    \filldraw[fill=black, draw=black, thick] (0,0) circle (0.05);
    \node[left] at (-0.5,0.25) {${\CD}$};
    \node[left] at (-0.5,-0.25) {${\CD}$};
    \node[right] at (0.5,0) {${(a,b)}$};
\end{tikzpicture}=\frac{1}{\sqrt{2}} \begin{tikzpicture}
 \tikzset{baseline=(current bounding box.center)}
	\begin{pgfonlayer}{nodelayer}
		\node[style=none] (0) at (0,1) {};
            \node[style=none] (1) at (0.5,1) {};
            \node[style=none] (2) at (0.5,0) {};
            \node[style=none] (3) at (0,0) {};
            \node[style=green dot] (4) at (0.5,0.25) {$a$};
            \node[style=red dot] (5) at (0.5,0.75) {$b$};
	\end{pgfonlayer}
	\begin{pgfonlayer}{edgelayer}
		\draw (0.center) to (1.center);
        \draw (1.center) to (2.center);
        \draw (2.center) to (3.center);
	\end{pgfonlayer}
\end{tikzpicture}\\
&\begin{tikzpicture}
\tikzset{baseline=(current bounding box.center)}
    \draw[thick, black, ->-=.7] (-0.5,0.25) -- (0,0.25);
    \draw[thick, black, ->-=.7] (-0.5,-0.25) -- (0,-0.25);
    \draw[thick, black] (0,0.25) -- (0,-0.25);
    \draw[thick, black, ->-=.7] (0,0) -- (0.5,0);
    \filldraw[fill=black, draw=black, thick] (0,0) circle (0.05);
    \node[left] at (-0.5,0.25) {${(a,b)}$};
    \node[left] at (-0.5,-0.25) {${\CD}$};
    \node[right] at (0.5,0) {${\CD}$};
\end{tikzpicture}= \begin{tikzpicture}
 \tikzset{baseline=(current bounding box.center)}
	\begin{pgfonlayer}{nodelayer}
		\node[style=none] (0) at (0,0) {};
            \node[style=green dot] (1) at (0.5,0) {$a$};
            \node[style=red dot] (2) at (1,0) {$b$};
            \node[style=none] (3) at (1.5,0) {};
	\end{pgfonlayer}
	\begin{pgfonlayer}{edgelayer}
		\draw (0.center) to (3.center);
	\end{pgfonlayer}
\end{tikzpicture}, \quad \begin{tikzpicture}
\tikzset{baseline=(current bounding box.center)}
    \draw[thick, black, ->-=.7] (-0.5,0.25) -- (0,0.25);
    \draw[thick, black, ->-=.7] (-0.5,-0.25) -- (0,-0.25);
    \draw[thick, black] (0,0.25) -- (0,-0.25);
    \draw[thick, black, ->-=.7] (0,0) -- (0.5,0);
    \filldraw[fill=black, draw=black, thick] (0,0) circle (0.05);
    \node[left] at (-0.5,-0.25) {${(a,b)}$};
    \node[left] at (-0.5,0.25) {${\CD}$};
    \node[right] at (0.5,0) {${\CD}$};
\end{tikzpicture}= \begin{tikzpicture}
 \tikzset{baseline=(current bounding box.center)}
	\begin{pgfonlayer}{nodelayer}
		\node[style=none] (0) at (0,0) {};
            \node[style=green dot] (1) at (1,0) {$a$};
            \node[style=red dot] (2) at (0.5,0) {$b$};
            \node[style=none] (3) at (1.5,0) {};
	\end{pgfonlayer}
	\begin{pgfonlayer}{edgelayer}
		\draw (0.center) to (3.center);
	\end{pgfonlayer}
\end{tikzpicture}
\end{align}
Following \eqref{eq:Fsymbol}, one can check these fusion junctions give the correct $F$-symbols of the fusion category $\Rep(D_8)$ as shown in \appref{app:FLRepD8}, also see \cite{sahand2025gaugempo}. We derive all the fusion and splitting junctions in \appref{app:repd81} and $F$-symbols in \appref{app:FLRepD8}.

\subsubsection{MPSs of three \texorpdfstring{$\Rep(D_8)$}{Rep(D8)} SPTs}\label{sec:repd8mps}
Besides the product state, there are two other $\Rep(D_8)$-invariant states, and any pair among the three hosts non-trivial edge modes when imposing the $\Rep(D_8)$ symmetry, i.e. distinct $\Rep(D_8)$ NISPTs \cite{Sahand2024cluster}. These three states are obtained by acting the SPT entangler $U_T=\prod_i \cz_{i,i+1}$ on the states constructed in \cite{Sahand2024cluster}, the cluster state becomes product state, while the other ``Even'' and ``Odd'' states have the following form,
\begin{equation}\label{eq:repd8sptsmps}
 \begin{tikzpicture}
    \tikzset{baseline=(current bounding box.center)}
    \draw[thick, cyan, -<-=.3,-<-=0.9] (-0.5,0) -- (0.5,0);
    \draw[thick,gray, ->-=0.7] (0,0) -- (0,0.5);
    \filldraw[fill=blue!30, draw=black, thick] (0,0) circle (0.12);
    \node[above] at (-0.5,0) {Tri};
\end{tikzpicture} \ = \
\frac{1}{2}\;
 \begin{tikzpicture}
\tikzset{baseline=(current bounding box.center)}
    \begin{pgfonlayer}{nodelayer}
		\node [style=green dot] (0) at (-0.5, 0) {};
            \node [style=none] (1) at (-0.5, 0.5) {};
		\node [style=green dot] (2) at (0, 0) {};
		\node [style=none] (3) at (0, 0.5) {};
	\end{pgfonlayer}
	\begin{pgfonlayer}{edgelayer}
		\draw (1.center) to (0);
		\draw (2) to (3.center);
	\end{pgfonlayer}
\end{tikzpicture},\quad
\begin{tikzpicture}
    \tikzset{baseline=(current bounding box.center)}
    \draw[thick, orange, -<-=.3,-<-=0.9] (-0.5,0) -- (0.5,0);
    \draw[thick,gray, ->-=0.7] (0,0) -- (0,0.5);
    \filldraw[fill=yellow!30, draw=black, thick] (0,0) circle (0.12);
    \node[above] at (-0.5,0) {Even};
\end{tikzpicture} \ = \
\frac{1}{\sqrt{2}}\;
 \begin{tikzpicture}
\tikzset{baseline=(current bounding box.center)}
    \begin{pgfonlayer}{nodelayer}
            \node [style=none] (0) at (0, 0) {};
		\node [style=green dot] (1) at (0.5, 0) {$\pi$};
            \node[style=had] (2) at (1,0) {};
            \node [style=green dot] (3) at (1.5, 0.25) {$\pi$};
            \node [style=none] (4) at (2, 0) {};
            \node [style=none] (5) at (0.5, 1) {};
            \node [style=none] (6) at (1.5, 1) {};
	\end{pgfonlayer}
	\begin{pgfonlayer}{edgelayer}
		\draw (4.center) to (0.center);
        \draw (1) to (5);
        \draw (3) to (6);
	\end{pgfonlayer}
\end{tikzpicture},\quad
\begin{tikzpicture}
    \tikzset{baseline=(current bounding box.center)}
    \draw[thick, purple, -<-=.3,-<-=0.9] (-0.5,0) -- (0.5,0);
    \draw[thick,gray, ->-=0.7] (0,0) -- (0,0.5);
    \filldraw[fill=violet!30, draw=black, thick] (0,0) circle (0.12);
    \node[above] at (-0.5,0) {Odd};
\end{tikzpicture} \ = \
\frac{1}{\sqrt{2}}\;
 \begin{tikzpicture}
\tikzset{baseline=(current bounding box.center)}
    \begin{pgfonlayer}{nodelayer}
            \node [style=none] (0) at (0, 0) {};
		\node [style=green dot] (1) at (0.5, 0.25) {$\pi$};
            \node[style=had] (2) at (1,0) {};
            \node [style=green dot] (3) at (1.5, 00) {$\pi$};
            \node [style=none] (4) at (2, 0) {};
            \node [style=none] (5) at (0.5, 1) {};
            \node [style=none] (6) at (1.5, 1) {};
	\end{pgfonlayer}
	\begin{pgfonlayer}{edgelayer}
		\draw (4.center) to (0.center);
        \draw (1) to (5);
        \draw (3) to (6);
	\end{pgfonlayer}
\end{tikzpicture}
\end{equation}
The Even and Odd MPSs become injective after grouping four physical sites. With the tensor phases shown here, the $\CD$ action on a four-site block has an overall minus sign. Eight-site blocking removes this scalar phase and gives \eqref{eq:oactonA}; we therefore evaluate the correlators on periodic chains whose lengths are multiples of eight. The four-site diagrams below specify the local tensor conventions, and the same action tensors apply after eight-site blocking. The fusion action tensors are
\begin{equation}
    \begin{tikzpicture}
    \tikzset{baseline=(current bounding box.center)}
    \draw[thick, cyan, -<-=.3,-<-=0.9] (-0.5,0) -- (0.5,0);
    \draw[thick, black] (0.0,0.0) -- (0.0,0.5);
    \draw[thick, black,->-=0.6] (-0.5,0.5) -- (0.0,0.5);
    \filldraw[fill=black, draw=black, thick] (0,0) circle (0.08);
    \node[above] at (-0.5,0) {Tri};
    \node[above] at (-0.5,0.5) {${(a,b)}$};
\end{tikzpicture}\ = \ 1,\quad \begin{tikzpicture}
    \tikzset{baseline=(current bounding box.center)}
    \draw[thick, orange, -<-=.3,-<-=0.9] (-0.5,0) -- (0.5,0);
    \draw[thick, black] (0.0,0.0) -- (0.0,0.5);
    \draw[thick, black,->-=0.6] (-0.5,0.5) -- (0.0,0.5);
    \filldraw[fill=black, draw=black, thick] (0,0) circle (0.08);
    \node[above] at (-0.5,0) {Even};
    \node[above] at (-0.5,0.5) {${(a,b)}$};
\end{tikzpicture}\ = \
\begin{tikzpicture}
 \tikzset{baseline=(current bounding box.center)}
	\begin{pgfonlayer}{nodelayer}
		\node[style=none] (0) at (0,0) {};
            \node[style=red dot] (1) at (0.5,0) {$a$};
            \node[style=green dot] (2) at (1,0) {$a$};
            \node[style=none] (3) at (1.5,0) {};
	\end{pgfonlayer}
	\begin{pgfonlayer}{edgelayer}
		\draw (0.center) to (3.center);
	\end{pgfonlayer}
\end{tikzpicture},\quad \begin{tikzpicture}
    \tikzset{baseline=(current bounding box.center)}
    \draw[thick, purple, -<-=.3,-<-=0.9] (-0.5,0) -- (0.5,0);
    \draw[thick, black] (0.0,0.0) -- (0.0,0.5);
    \draw[thick, black,->-=0.6] (-0.5,0.5) -- (0.0,0.5);
    \filldraw[fill=black, draw=black, thick] (0,0) circle (0.08);
    \node[above] at (-0.5,0) {Odd};
    \node[above] at (-0.5,0.5) {${(a,b)}$};
\end{tikzpicture}\ = \
\begin{tikzpicture}
 \tikzset{baseline=(current bounding box.center)}
	\begin{pgfonlayer}{nodelayer}
		\node[style=none] (0) at (0,0) {};
            \node[style=green dot] (1) at (0.5,0) {$b$};
            \node[style=red dot] (2) at (1,0) {$b$};
            \node[style=none] (3) at (1.5,0) {};
	\end{pgfonlayer}
	\begin{pgfonlayer}{edgelayer}
		\draw (0.center) to (3.center);
	\end{pgfonlayer}
\end{tikzpicture}
\end{equation}

\begin{equation}
    \begin{tikzpicture}
    \tikzset{baseline=(current bounding box.center)}
    \draw[thick, cyan, -<-=.3,-<-=0.9] (-0.5,0) -- (0.5,0);
    \draw[thick, black] (0.0,0.0) -- (0.0,0.5);
    \draw[thick, black,->-=0.6] (-0.5,0.5) -- (0.0,0.5);
    \filldraw[fill=black, draw=black, thick] (0,0) circle (0.08);
    \node[above] at (-0.5,0) {Tri};
    \node[above] at (-0.5,0.5) {${\CD}$};
    \node[below] at (0,0) {$i$};
\end{tikzpicture}\ = \ \frac{1}{\sqrt{2}}\begin{tikzpicture}
 \tikzset{baseline=(current bounding box.center)}
	\begin{pgfonlayer}{nodelayer}
		\node[style=none] (0) at (0,0) {};
            \node[style=red dot] (1) at (0.5,0) {$i$};
	\end{pgfonlayer}
	\begin{pgfonlayer}{edgelayer}
		\draw (0.center) to (1.center);
	\end{pgfonlayer}
\end{tikzpicture},\quad \begin{tikzpicture}
    \tikzset{baseline=(current bounding box.center)}
    \draw[thick, orange, -<-=.3,-<-=0.9] (-0.5,0) -- (0.5,0);
    \draw[thick, black] (0.0,0.0) -- (0.0,0.5);
    \draw[thick, black,->-=0.6] (-0.5,0.5) -- (0.0,0.5);
    \filldraw[fill=black, draw=black, thick] (0,0) circle (0.08);
    \node[above] at (-0.5,0) {Even};
    \node[above] at (-0.5,0.5) {${\CD}$};
    \node[below] at (0,0) {$i$};
\end{tikzpicture}\ = \
\frac{1}{\sqrt{2}}
\begin{tikzpicture}
 \tikzset{baseline=(current bounding box.center)}
	\begin{pgfonlayer}{nodelayer}
		\node[style=none] (0) at (0,0) {};
            \node[style=red dot] (1) at (0.5,0) {$i$};
            \node[style=green dot] (2) at (1,0) {$\bar{i}$};
            \node[style=none] (3) at (1.5,0) {};
            \node[style=green dot] (4) at (0.5,0.5) {$i$};
            \node[style=none] (5) at (0,0.5) {};
	\end{pgfonlayer}
	\begin{pgfonlayer}{edgelayer}
		\draw (0.center) to (3.center);
        \draw (4.center) to (5.center);
	\end{pgfonlayer}
\end{tikzpicture},\quad \begin{tikzpicture}
    \tikzset{baseline=(current bounding box.center)}
    \draw[thick, purple, -<-=.3,-<-=0.9] (-0.5,0) -- (0.5,0);
    \draw[thick, black] (0.0,0.0) -- (0.0,0.5);
    \draw[thick, black,->-=0.6] (-0.5,0.5) -- (0.0,0.5);
    \filldraw[fill=black, draw=black, thick] (0,0) circle (0.08);
    \node[above] at (-0.5,0) {Odd};
    \node[above] at (-0.5,0.5) {${\CD}$};
    \node[below] at (0,0) {$i$};
\end{tikzpicture}\ = \
\frac{1}{\sqrt{2}}
\begin{tikzpicture}
 \tikzset{baseline=(current bounding box.center)}
	\begin{pgfonlayer}{nodelayer}
		\node[style=none] (0) at (0,0) {};
            \node[style=green dot] (1) at (0.5,0) {$i$};
            \node[style=red dot] (2) at (1,0) {$\bar{i}$};
            \node[style=none] (3) at (1.5,0) {};
            \node[style=red dot] (4) at (0.5,0.5) {$\bar{i}$};
            \node[style=none] (5) at (0,0.5) {};
	\end{pgfonlayer}
	\begin{pgfonlayer}{edgelayer}
		\draw (0.center) to (3.center);
        \draw (4.center) to (5.center);
	\end{pgfonlayer}
\end{tikzpicture}
\end{equation}
The splitting action tensors are derived analogously in \appref{app:repd82}. One can check the corresponding $L$-symbols for each state are gauge inequivalent as shown in \appref{app:FLRepD8}, meaning that they are distinct NISPT phases \cite{kansei2024nispt}.

The full interface algebra between distinct $\Rep(D_8)$ NISPTs is eight-dimensional and isomorphic to $M_2(\IC)\oplus M_2(\IC)$, so it has two inequivalent two-dimensional irreducible representations \cite{kansei2024nispt}. For Tri and Even, the action-tensor contraction realizes the first representation:
\begin{equation}
    \CI^{\text{Tri|Even}}_{(a,b)} = \begin{tikzpicture}
 \tikzset{baseline=(current bounding box.center)}
	\begin{pgfonlayer}{nodelayer}
		\node[style=none] (0) at (0,0) {};
            \node[style=green dot] (1) at (0.5,0) {$a$};
            \node[style=red dot] (2) at (1,0) {$a$};
            \node[style=none] (3) at (1.5,0) {};
	\end{pgfonlayer}
	\begin{pgfonlayer}{edgelayer}
		\draw (0.center) to (3.center);
	\end{pgfonlayer}
\end{tikzpicture},\quad \CI^{\text{Tri|Even}}_{\CD,(i,j)} = 2^{-1}\begin{tikzpicture}
 \tikzset{baseline=(current bounding box.center)}
	\begin{pgfonlayer}{nodelayer}
		\node[style=none] (0) at (0,0) {};
            \node[style=green dot] (1) at (0.5,0) {$\bar{i}$};
            \node[style=red dot] (2) at (1,0) {$i$};
            \node[style=none] (3) at (1.5,0) {};
            \node[style=green dot] (4) at (0.5,-0.5) {$i$};
            \node[style=red dot] (5) at (1,-0.5) {$j$};
	\end{pgfonlayer}
	\begin{pgfonlayer}{edgelayer}
		\draw (0.center) to (3.center);
        \draw (4.center) to (5.center);
	\end{pgfonlayer}
\end{tikzpicture}
\end{equation}
The table gives both gauge-inequivalent representations for the interface algebra. To obtain the second representation, multiply the Tri action tensors $\bphi^{\text{Tri}}_{(a,b)}$ by $(-1)^b$ and exchange $(\bphi^{\text{Tri}}_{\CD})_0$ and $(\bphi^{\text{Tri}}_{\CD})_1$, applying the same change to their dual tensors. This preserves the $L$-symbols, thus, the multiplication rule is unchanged. Conjugating the resulting interface matrices by $(\widetilde{I}+i\widetilde{Y})/\sqrt{2}$ gives the second row of the table.
\begin{center}
\begin{tabular}{|c|c|c|c|c|c|}
    \hline
    \gape{$\CI^{\text{Tri|Even}}_{x,(i,j)}$} & $(0,0)$ & $(0,1)$ & $(1,0)$ & $(1,1)$ & $\CD$ \\
    \hline
    First representation & $\widetilde{I}$ & $\widetilde{I}$ & $i\widetilde{Y}$ & $i\widetilde{Y}$ & \gape{$\frac{1}{\sqrt{2}}\left(\begin{array}{cc}-\widetilde{X} & \widetilde{X} \\ \widetilde{Z} & \widetilde{Z}\end{array}\right)_{i,j}$} \\
    \hline
    Second representation & $\widetilde{I}$ & $-\widetilde{I}$ & $i\widetilde{Y}$ & $-i\widetilde{Y}$ & \gape{$\frac{1}{\sqrt{2}}\left(\begin{array}{cc}-\widetilde{X} & -\widetilde{X} \\ -\widetilde{Z} & \widetilde{Z}\end{array}\right)_{i,j}$} \\
    \hline
\end{tabular}
\end{center}
The central element $\CI^{\text{Tri|Even}}_{(0,1)}$ acts as $+\widetilde{I}$ in the first representation and $-\widetilde{I}$ in the second, distinguishing the two representations. Each row spans $M_2(\IC)$, while their direct sum represents the full interface algebra faithfully. Both representations have dimension two, so the minimum protected interface degeneracy is two; symmetry does not require the corresponding interface multiplets to have the same energy. If the non-invertible symmetry $\CD$ is removed, the remaining invertible generators can be simultaneously diagonalized into one-dimensional representations, and the distinction between these NISPTs is lost.

\subsubsection{Strange correlator for non-invertible \texorpdfstring{$\Rep(D_8)$}{Rep(D8)} SPTs}\label{sec:repd8strange}
With all these ingredients, we can construct the strange charged operator following the general procedure in \secref{sec:scnispt}. Let's demonstrate using the Tri and Even states. First, we construct the $\CS_{I,(a,b)}^{\text{Tri|Even}}$,
\begin{equation}
 \begin{tikzpicture}
 \tikzset{baseline=(current bounding box.center)}
	\begin{pgfonlayer}{nodelayer}
		\node[style=none] (0) at (1,0) {};
            \node[style=none] (1) at (4.5,0) {};
            \node[style=had] (2) at (3.5,0) {};
            \node[style=green dot] (3) at (1.5,0) {$\pi$};
            \node[style=had] (4) at (2,0) {};
            \node[style=green dot] (5) at (3,0) {$\pi$};
            \node[style=green dot] (6) at (4,-0.5) {$\pi$};
            \node[style=green dot] (7) at (4,-1) {$\pi$};
            \node[style=green dot] (8) at (4,-1.5) {};
            \node[style=green dot] (9) at (2.5,-0.5) {$\pi$};
            \node[style=green dot] (10) at (2.5,-1) {$\pi$};
            \node[style=green dot] (11) at (2.5,-1.5) {};
            \node[style=green dot] (12) at (1.5,-1) {$\bar{a}$};
            \node[style=green dot] (13) at (1.5,-1.5) {};
            \node[style=green dot] (14) at (3,-1) {$\bar{a}$};
            \node[style=green dot] (15) at (3,-1.5) {};
	\end{pgfonlayer}
	\begin{pgfonlayer}{edgelayer}
		\draw (0.center) to (1.center);
            \draw (6) to (8);
            \draw (9) to (11);
            \draw (3) to (13);
            \draw (5) to (15);
	\end{pgfonlayer}
    \draw[dashed] (1,-0.75) -- (4.5,-0.75);
    \draw[dashed] (1,-1.25) -- (4.5,-1.25);
    \node[right] at (4.5,-1) {$\CS_{I,(a,b)}^{\text{Tri|Even}}$};
    \node[right] at (4.5,0) {$\bra{\text{Even}}$};
    \node[right] at (4.5,-1.5) {$\ket{\text{Tri}}$};
    \node[left] at (1,0) {$2^{-1}$};
    \node[left] at (1,-1.5) {$2^{-2}$};
    \node[left] at (1,-1.0) {$2$};
\end{tikzpicture}   \ = \ \begin{tikzpicture}
 \tikzset{baseline=(current bounding box.center)}
	\begin{pgfonlayer}{nodelayer}
		\node[style=none] (0) at (0,0) {};
            \node[style=green dot] (1) at (0.5,0) {$a$};
            \node[style=red dot] (2) at (1,0) {$a$};
            \node[style=none] (3) at (1.5,0) {};
	\end{pgfonlayer}
	\begin{pgfonlayer}{edgelayer}
		\draw (0.center) to (3.center);
	\end{pgfonlayer}
\end{tikzpicture} = \CI^{\text{Tri|Even}}_{(a,b)}.
\end{equation}
Therefore, $\CS_{I,(a,b)}^{\text{Tri|Even}} = {2} Z^{\bar{a}}_{I_1} Z_{I_2} Z^{\bar{a}}_{I_3} Z_{I_4}$. For $\CS_{I,\CD,(i,j)}^{\text{Tri|Even}}$,
\begin{equation}
 \begin{tikzpicture}
 \tikzset{baseline=(current bounding box.center)}
	\begin{pgfonlayer}{nodelayer}
		\node[style=none] (0) at (1,0) {};
            \node[style=none] (1) at (4.5,0) {};
            \node[style=had] (2) at (3.5,0) {};
            \node[style=green dot] (3) at (1.5,0) {$\pi$};
            \node[style=had] (4) at (2,0) {};
            \node[style=green dot] (5) at (3,0) {$\pi$};
            \node[style=green dot] (6) at (4,-0.5) {$\pi$};
            \node[style=green dot] (7) at (4,-1.5) {$\pi$};
            \node[style=green dot] (8) at (4,-2.5) {};
            \node[style=green dot] (9) at (2.5,-0.5) {$\pi$};
            \node[style=green dot] (10) at (2.5,-1.0) {$\bar{j}$};
            \node[style=had] (16) at (2.5,-1.5) {};
            \node[style=green dot] (10) at (2.5,-2.0) {$i$};
            \node[style=green dot] (11) at (2.5,-2.5) {};
            \node[style=green dot] (12) at (1.5,-1.5) {$j$};
            \node[style=green dot] (13) at (1.5,-2.5) {};
            \node[style=green dot] (14) at (3,-1.5) {$\bar{j}$};
            \node[style=green dot] (15) at (3,-2.5) {};
	\end{pgfonlayer}
	\begin{pgfonlayer}{edgelayer}
		\draw (0.center) to (1.center);
            \draw (6) to (8);
            \draw (9) to (11);
            \draw (3) to (13);
            \draw (5) to (15);
	\end{pgfonlayer}
    \draw[dashed] (1,-0.75) -- (4.5,-0.75);
    \draw[dashed] (1,-2.25) -- (4.5,-2.25);
    \node[right] at (4.5,-1.5) {$\CS_{I,\CD,(i,j)}^{\text{Tri|Even}}$};
    \node[right] at (4.5,0) {$\bra{\text{Even}}$};
    \node[right] at (4.5,-2.5) {$\ket{\text{Tri}}$};
    \node[left] at (1,0) {$2^{-1}$};
    \node[left] at (1,-2.5) {$2^{-2}$};
    \node[left] at (1,-1.5) {$2$};
\end{tikzpicture}   \ = \ 2^{-1} \begin{tikzpicture}
 \tikzset{baseline=(current bounding box.center)}
	\begin{pgfonlayer}{nodelayer}
		\node[style=none] (0) at (0,0) {};
            \node[style=green dot] (1) at (0.5,0) {$\bar{j}$};
            \node[style=red dot] (2) at (1,0) {$j$};
            \node[style=none] (3) at (1.5,0) {};
            \node[style=green dot] (4) at (0.5,-0.5) {$j$};
            \node[style=red dot] (5) at (1,-0.5) {$i$};
	\end{pgfonlayer}
	\begin{pgfonlayer}{edgelayer}
		\draw (0.center) to (3.center);
        \draw (4.center) to (5.center);
	\end{pgfonlayer}
\end{tikzpicture} = \CI^{\text{Tri|Even}}_{\CD,(i,j)}.
\end{equation}
Therefore,
\begin{equation}
\CS_{(I_1,I_2,I_3,I_4),\CD,(i,j)}^{\text{Tri|Even}} = {2} Z_{I_1}^j Z^{\bar{j}}_{I_2} H_{I_2} Z^i_{I_2} Z^{\bar{j}}_{I_3} Z_{I_4}.
\end{equation}
where $I_1,\cdots,I_4$ are the lattice indices and $H$ is the Hadamard matrix. The operators $\bCS^{\text{Tri|Even}}_{I,x}$ are obtained analogously. The indices here use the action-tensor bases in the diagrams; their conversion to the matrix-table basis is given in \appref{app:FLRepD8}. The displayed $\CD$ operators are Hermitian when $i=\bar j$.

The Tri and Even representatives have zero overlap. We therefore evaluate the normalized strange correlator after a small symmetry-preserving deformation of the Even state, with the local insertions reconstructed from its blocked MPS. The deformation is kept nonzero while forming the ratio and taking the long-distance limit \cite{yizhuang2014strange}.\footnote{For $L$ divisible by eight, one choice is $\ket{\mathrm{Even}_\epsilon}=\prod_{j=0}^{L/4-1}\exp[\tfrac{i\epsilon}{2}Z_{4j+1}(1+X_{4j+2})Z_{4j+3}]\ket{\mathrm{Even}}$. Each gate commutes with the symmetry MPOs. The mixed transfer matrix on eight sites is $(\sin^2\epsilon/4)\,\widetilde{I}$, so $\langle\mathrm{Even}_\epsilon|\mathrm{Tri}\rangle=2(\sin^2\epsilon/4)^{L/8}\ne0$ for sufficiently small nonzero $\epsilon$. Normalizing each insertion so that its mixed transfer matrix is $(\sin^2\epsilon/4)$ times the corresponding interface matrix gives \eqref{eq:trievenconnected}.} The four-site expressions above determine the interface action used to reconstruct the insertions on eight-site blocks.

\subsection{Strange correlator in higher dimension}\label{sec:schigher}
\begingroup
\color{black}
\hypersetup{citecolor=ceruleanblue,linkcolor=ceruleanblue}
In $n+1$ spacetime dimensions, finite non-invertible symmetries are described by fusion $n$-categories, which include symmetry operators of different codimensions and their junction and braiding data \cite{kong2015higher,kong2017higher,kong2018higher,kong2020higher,kong2021higher,kong2022higher,kong2024higher,Douglas:2018fusion2,pearson2024gapped21,sakura2025gapped21,wen2025gapless21,sakura2025gapless21}.
\par\endgroup

Taking the $2+1d$ as an example, the symmetry operators contain surface operators for the $0$-form symmetry and line operators for the $1$-form symmetry. The picture of \eqref{eq:strangecharge} can be generalized to higher dimension, with both $0$-form and $1$-form symmetry being taken into account. It is known that there is no pure $1$-form SPT in $2+1d$ \cite{anton2013higherspt}, so the $0$-form symmetry is inevitable for the fusion 2-category symmetric SPTs.

\begingroup
\color{black}
\hypersetup{citecolor=ceruleanblue,linkcolor=ceruleanblue}
{\color{black}For a $0$-form symmetry in $2+1$d, expanding a symmetry surface between the two states gives a strange charged operator on the boundary loop, by the same reasoning as in \eqref{eq:strangecond}.}
\par\endgroup
\begin{equation}
\includegraphics[width=0.9\textwidth]{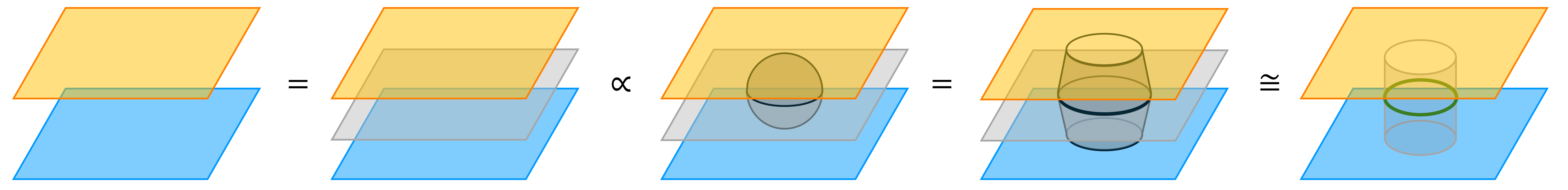}
\end{equation}

{\color{black}In the $\IZ_2$ CZX model, $U_{CZX,\partial A}=(\prod_{i\in\partial A}\cz_{i,i+1})(\prod_{i\in\partial A}X_i)$ acts on the effective boundary spins \cite{chen2011z2spt}. We normalize $\bra{\text{$\IZ_2$ SPT}}U_{CZX,\partial A}\ket{\text{Trivial}}$ by the overlap $\langle\text{$\IZ_2$ SPT}\vert\text{Trivial}\rangle$. If the fixed-point states have zero overlap, applying a symmetry-preserving finite-depth circuit to one of them may give a nonzero overlap. If the fixed-point tensor identities allow the loop to shrink freely, this ratio is independent of its length. Away from the fixed point, it depends on the perimeter.}

{\color{black}To understand the local strange charged operators, stretch the loop into a narrow segment from $i$ to $j$. Its two sides fuse to the identity in the interior, leaving $Z_iZ_j$ at the two ends up to an overall phase. For the CZX plaquette state with the product reference $(\ket{0000}+\ket{1111})/\sqrt{2}$ on each site, the connected $Z_iZ_j$ strange correlator is $1$. When the overlap describes a critical theory, this two-point function can decay as a power of $|i-j|$, as observed for a deformed $\IZ_2$ SPT \cite{yizhuang2014strange}.}

For $1$-form symmetry, such construction will give strange charged local operators. The detailed construction for strange correlators of higher dimension NISPTs will be left for the future study.

\section{String order parameters for NISPTs in 1+1d}\label{sec:stringopn}
\begingroup
\color{black}
\hypersetup{citecolor=ceruleanblue,linkcolor=ceruleanblue}
At a fixed point, a truncated symmetry MPO can be absorbed into the state except for action tensors at the two boundaries:
\par\endgroup
\begin{equation}
	\begin{tikzpicture}
		\tikzset{baseline=(current bounding box.center)}
		\draw[thick, black] (-1,0.5) -- (1,0.5);
		\draw[thick, cyan] (-1.5,0) -- (1.5,0);
	\end{tikzpicture}=\begin{tikzpicture}
		\tikzset{baseline=(current bounding box.center)}
		\draw[thick, black] (-1,0.5) -- (-0.75,0.5);
		\draw[thick, black] (0.75,0.5) -- (1,0.5);
		\draw[thick, black] (0.75,0.5) -- (0.75,0);
		\draw[thick, black] (-0.75,0.5) -- (-0.75,0);
		\draw[thick, cyan] (-1.5,0) -- (1.5,0);
		\filldraw[fill=white, draw=black, thick] (0.75,0) circle (0.04);
		\filldraw[fill=black, draw=black, thick] (-0.75,0) circle (0.04);
	\end{tikzpicture}
\end{equation}
\begingroup
\color{black}
\hypersetup{citecolor=ceruleanblue,linkcolor=ceruleanblue}
Local charge decoration operators reproduce these action tensors:
\par\endgroup
\begin{equation}
	\begin{tikzpicture}
		\tikzset{baseline=(current bounding box.center)}
		\draw[thick, black] (0,0.3) -- (0.25,0.3);
		\draw[thick, gray!50] (0,0) -- (0,0.6);
		\draw[thick, cyan] (-0.5,0) -- (0.5,0);
		\node[draw, regular polygon, regular polygon sides=3, minimum size=3mm, inner sep=0mm, thick,fill=blue!30, shape border rotate=-30] at (0,0.3) {};
	\end{tikzpicture} = \begin{tikzpicture}
		\tikzset{baseline=(current bounding box.center)}
		\draw[thick, black] (0.25,0.3) -- (0.5,0.3);
		\draw[thick, black] (0.25,0) -- (0.25,0.3);
		\draw[thick, cyan] (-0.5,0) -- (0.5,0);
		\draw[thick, gray!50] (0,0) -- (0,0.5);
		\filldraw[fill=white, draw=black, thick] (0.25,0) circle (0.04);
	\end{tikzpicture},\quad 
	\begin{tikzpicture}
		\tikzset{baseline=(current bounding box.center)}
		\draw[thick, black] (0,0.3) -- (-0.25,0.3);
		\draw[thick, gray!50] (0,0) -- (0,0.6);
		\draw[thick, cyan] (-0.5,0) -- (0.5,0);
		\node[draw, regular polygon, regular polygon sides=3, minimum size=3mm, inner sep=0mm, thick,fill=blue!30, shape border rotate=30] at (0,0.3) {};
	\end{tikzpicture} = \begin{tikzpicture}
		\tikzset{baseline=(current bounding box.center)}
		\draw[thick, black] (-0.25,0.3) -- (-0.5,0.3);
		\draw[thick, black] (-0.25,0) -- (-0.25,0.3);
		\draw[thick, cyan] (-0.5,0) -- (0.5,0);
		\draw[thick, gray!50] (0,0) -- (0,0.5);
		\filldraw[fill=black, draw=black, thick] (-0.25,0) circle (0.04);
	\end{tikzpicture}
\end{equation}
\begingroup
\color{black}
\hypersetup{citecolor=ceruleanblue,linkcolor=ceruleanblue}
Contracting the decoration operators with the truncated MPO gives the string order parameter:
\par\endgroup
\begin{equation}
	\begin{tikzpicture}
		\tikzset{baseline=(current bounding box.center)}
		\draw[thick, black] (-1,0.5) -- (1,0.5);
		\draw[thick, cyan] (-1.5,0) -- (1.5,0);
		\draw[thick, gray!50] (-1,0) -- (-1,0.8);
		\draw[thick, gray!50] (1,0) -- (1,0.8);
		\node[draw, regular polygon, regular polygon sides=3, minimum size=3mm, inner sep=0mm, thick,fill=blue!30, shape border rotate=-30] at (-1,0.5) {};
		\node[draw, regular polygon, regular polygon sides=3, minimum size=3mm, inner sep=0mm, thick,fill=blue!30, shape border rotate=30] at (1,0.5) {};
	\end{tikzpicture}=\begin{tikzpicture}
		\tikzset{baseline=(current bounding box.center)}
		\draw[thick, black] (-1,0.5) -- (-0.75,0.5);
		\draw[thick, black] (0.75,0.5) -- (1,0.5);
		\draw[thick, black] (0.75,0.5) -- (0.75,0);
		\draw[thick, black] (-0.75,0.5) -- (-0.75,0);
		\draw[thick, black] (-1,0.5) -- (-1,0);
		\draw[thick, black] (1,0.5) -- (1,0);
		\draw[thick, gray!50] (1.3,0.8) -- (1.3,0);
		\draw[thick, gray!50] (-1.3,0.8) -- (-1.3,0);
		\draw[thick, cyan] (-1.5,0) -- (1.5,0);
		\filldraw[fill=white, draw=black, thick] (0.75,0) circle (0.04);
		\filldraw[fill=black, draw=black, thick] (-0.75,0) circle (0.04);
		\filldraw[fill=white, draw=black, thick] (-1,0) circle (0.04);
		\filldraw[fill=black, draw=black, thick] (1,0) circle (0.04);
		\node at (0,0.8) {};
	\end{tikzpicture}
	=\begin{tikzpicture}
		\tikzset{baseline=(current bounding box.center)}
		\node at (0,0.8) {};
		\draw[thick, cyan] (-1.5,0) -- (1.5,0);
	\end{tikzpicture}
\end{equation}
\begingroup
\color{black}
\hypersetup{citecolor=ceruleanblue,linkcolor=ceruleanblue}
The gray vertical legs denote physical indices. The two action-tensor indices are contracted consistently, as made explicit below in \eqref{eq:stringop}.
\par\endgroup

\begingroup
\color{black}
\hypersetup{citecolor=ceruleanblue,linkcolor=ceruleanblue}
To write this construction explicitly for an MPS in $1+1$d, we introduce the \textit{charge decoration operator}, which satisfies
\par\endgroup
\begin{equation}\label{eq:stringcond}
\begin{tikzpicture}
    \tikzset{baseline=(current bounding box.center)}
    \draw[thick, black, ->-=.8] (0,0) -- (0.5,0);
    \draw[thick,gray, ->-=0.35,->-=0.95] (0,-0.5) -- (0,0.5);
    \node[draw, regular polygon, regular polygon sides=3, minimum size=3mm, inner sep=0mm, thick,fill=blue!30, shape border rotate=-30] at (0,0) {};
    \draw[thick, cyan, -<-=.3,-<-=0.9] (-0.5,-0.5) -- (0.5,-0.5);
    \filldraw[fill=blue!30, draw=black, thick] (0,-0.5) circle (0.12);
    \node[left] at (0,0) {${}^A(\bar{\CE}_x)_i$};
    \node[below] at (0,-0.6) {$A^{\otimes n}$};
\end{tikzpicture} \ =\ \begin{tikzpicture}
    \tikzset{baseline=(current bounding box.center)}
    \draw[thick,gray, ->-=0.8] (0,-0.5) -- (0,0.5);
    \draw[thick, cyan, -<-=.3,-<-=0.9] (-0.5,-0.5) -- (0.5,-0.5);
    \draw[thick, cyan, -<-=0.6] (0.6,-0.5) -- (1,-.5);
    \draw[thick, black] (0.6,-0.5) -- (0.6,0);
    \draw[thick, black,->-=0.6] (0.6,0) -- (1,0);
    \filldraw[fill=blue!30, draw=black, thick] (0,-0.5) circle (0.12);
    \filldraw[fill=white, draw=black, thick] (0.6,-0.5) circle (0.08);
    \node[above] at (-0.4,-0.5) {$A^{\otimes n}$};
    \node[below] at (0.6,-0.5) {$(\bphi_x)_i$};
    \node[above] at (0.8,0) {$x$};
\end{tikzpicture},\quad \begin{tikzpicture}
    \tikzset{baseline=(current bounding box.center)}
    \draw[thick, black, ->-=.6] (-0.5,0) -- (0,0);
    \draw[thick,gray, ->-=0.35,->-=0.95] (0,-0.5) -- (0,0.5);
    \node[draw, regular polygon, regular polygon sides=3, minimum size=3mm, inner sep=0mm, thick,fill=blue!30, shape border rotate=30] at (0,0) {};
    \draw[thick, cyan, -<-=.3,-<-=0.9] (-0.5,-0.5) -- (0.5,-0.5);
    \filldraw[fill=blue!30, draw=black, thick] (0,-0.5) circle (0.12);
    \node[right] at (00,0) {${}^A(\CE_x)_i$};
    \node[below] at (0,-0.6) {$A^{\otimes n}$};
\end{tikzpicture} \ =\ \begin{tikzpicture}
    \tikzset{baseline=(current bounding box.center)}
    \draw[thick,gray, ->-=0.6] (0,-0.5) -- (0,0.5);
    \draw[thick, cyan, -<-=.3,-<-=0.9] (-0.5,-0.5) -- (0.5,-0.5);
    \draw[thick, cyan, -<-=0.6] (-1,-0.5) -- (-0.6,-.5);
    \draw[thick, black] (-0.6,-0.5) -- (-0.6,0);
    \draw[thick, black,->-=0.6] (-1,0) -- (-0.6,0);
    \filldraw[fill=blue!30, draw=black, thick] (0,-0.5) circle (0.12);
    \filldraw[fill=black, draw=black, thick] (-0.6,-0.5) circle (0.08);
    \node[above] at (0.4,-0.5) {$A^{\otimes n}$};
    \node[below] at (-0.6,-0.5) {$(\phi_x)_i$};
    \node[above] at (-0.8,0) {$x$};
\end{tikzpicture}
\end{equation}
where the charge decoration operator acts on a finite block chosen so that $A^{\otimes n}$ is injective. Then the string order parameter for NISPT $A$ is given by,
\begin{align}\label{eq:stringop}
    &({}^A\hCO^\text{string}_{x})_{i;I,J} \equiv\begin{tikzpicture}
    	\tikzset{baseline=(current bounding box.center)}
    	\draw[thick, black, ->-=.3,->-=0.9] (-1.5,0) -- (-0.5,0);
    	\draw[thick,gray, ->-=0.25,->-=0.95] (-1,-0.5) -- (-1,0.5);
    	\filldraw[fill=gray!30, draw=black, thick] (-1,0) circle (0.12);
    	\draw[thick, black, ->-=.3,->-=0.9] (-0.5,0) -- (0.5,0);
    	\draw[thick,gray, ->-=0.25,->-=0.95] (0,-0.5) -- (0,0.5);
    	\filldraw[fill=gray!30, draw=black, thick] (0,0) circle (0.12);
    	\draw[thick, black, ->-=.3,->-=0.9] (1,0) -- (2,0);
    	\draw[thick,gray, ->-=0.25,->-=0.95] (1.5,-0.5) -- (1.5,0.5);
    	\filldraw[fill=gray!30, draw=black, thick] (1.5,0) circle (0.12);
    	\node[above] at (-1.5,0) {$\CO_x$};
    	\node[] at (0.75,0) {\scriptsize$\cdots$};
    	\draw[thick, black, ->-=.8] (-2,0) -- (-1.5,0);
    	\draw[thick,gray, ->-=0.35,->-=0.95] (-2,-0.5) -- (-2,0.5);
    	\node[draw, regular polygon, regular polygon sides=3, minimum size=3mm, inner sep=0mm, thick,fill=blue!30, shape border rotate=-30] at (-2,0) {};
    	\node[left] at (-2,0) {${}^A(\bar{\CE}_{{x}})_i$};
    	\draw[thick, black, ->-=.6] (2,0) -- (2.5,0);
    	\draw[thick,gray, ->-=0.35,->-=0.95] (2.5,-0.5) -- (2.5,0.5);
    	\node[draw, regular polygon, regular polygon sides=3, minimum size=3mm, inner sep=0mm, thick,fill=blue!30, shape border rotate=30] at (2.5,0) {};
    	\node[right] at (2.5,0) {${}^A(\CE_{{x}})_i$};
    \end{tikzpicture}\nonumber\\
    &={}^A(\bar{\CE}_x)_{i;I-n+1,\cdots,I}\CO_{x;I+1}\cdots\CO_{x;J-1} {}^A(\CE_x)_{i;J,\cdots,J+n-1} ,\quad i=1,\cdots,d_x
\end{align}
where we restore the lattice indices $I,J$. The local dressing operators absorb the action tensors at the two boundaries of the truncated MPO. For normalized $\ket{A}$ and disjoint dressing blocks, \eqref{eq:stringcond}, \eqref{eq:oactonA}, and \eqref{eq:actionorthogonality} give
\begin{equation}
    \bra{A}({}^A\hCO^\text{string}_{x})_{i;I,J}\ket{A}=1.
\end{equation}
A string constructed for one phase need not vanish in another phase. Its long-distance behavior depends on the compatibility of the local dressing operators with the symmetry data of the state. Exponential decay follows when a selection rule excludes a nonzero long-distance contribution. For example, the sign string in \secref{sec:s3diagnostics} has expectation value one in both the $\IZ_2$ and $S_3$ fixed point states, whereas the three observables selected there distinguish the three fixed point states. When the state is deformed within its phase, the local dressing operators may need to be adapted.

The action-tensor orthogonality in \eqref{eq:actionorthogonality} implies that choosing indices $i$ and $j$ for the two local dressing operators gives a factor $\delta_{ij}$ when the string acts on $\ket{A}$. This explains the common index $i$ in \eqref{eq:stringop}.

\subsection{Warm-up: string order parameter for \texorpdfstring{$\IZ_2\times \IZ_2 $}{Z2 x Z2} SPT}
As an illustrative example, let us begin by constructing the string order parameter for the $\mathbb{Z}_2\times\mathbb{Z}_2$ symmetry group. There are two distinct SPT phases: the trivial product state and the non-trivial cluster state.

For the trivial product state phase, the charge decoration operator remains trivial,
\begin{equation}
	{}^{\text{Tri}}\bar{\CE}_{{(a,b)}}={}^{\text{Tri}}\CE_{{(a,b)}}=1
\end{equation}
So the string order parameter for the product state is the truncated symmetry operator without any charge decoration,
\begin{equation}
\begin{tikzpicture}[scale=0.7]
    \tikzset{baseline=(current bounding box.center)}
	\begin{pgfonlayer}{nodelayer}
		\node [style=none] (3) at (6.5, 1.25) {};
		\node [style=none] (4) at (6.5, -0.25) {};
		\node [style=none] (5) at (5.75, -0.25) {};
		\node [style=none] (6) at (5.75, 1.25) {};
		\node [style=none] (10) at (0.5, 1.25) {};
		\node [style=none] (11) at (0.5, -0.25) {};
		\node [style=none] (12) at (1.25, 1.25) {};
		\node [style=none] (13) at (1.25, -0.25) {};
		\node [style=red dot] (14) at (2, 0.5) {$a$};
		\node [style=red dot] (15) at (2.75, 0.5) {$b$};
		\node [style=red dot] (16) at (5, 0.5) {$b$};
		\node [style=red dot] (17) at (4.25, 0.5) {$a$};
		\node [style=none] (18) at (3.5, 0.5) {$\dots$};
		\node [style=none] (19) at (2, 1.25) {};
		\node [style=none] (20) at (2, -0.25) {};
		\node [style=none] (21) at (2.75, -0.25) {};
		\node [style=none] (22) at (2.75, 1.25) {};
		\node [style=none] (23) at (4.25, 1.25) {};
		\node [style=none] (24) at (4.25, -0.25) {};
		\node [style=none] (25) at (5, -0.25) {};
		\node [style=none] (26) at (5, 1.25) {};
	\end{pgfonlayer}
	\begin{pgfonlayer}{edgelayer}
		\draw (19.center) to (14);
		\draw (14) to (20.center);
		\draw (22.center) to (15);
		\draw (15) to (21.center);
		\draw (23.center) to (17);
		\draw (17) to (24.center);
		\draw (26.center) to (16);
		\draw (16) to (25.center);
		\draw (10.center) to (11.center);
		\draw (12.center) to (13.center);
		\draw (6.center) to (5.center);
		\draw (3.center) to (4.center);
	\end{pgfonlayer}
\end{tikzpicture} =\prod_{i=I}^{J} X^a_{2i} X^b_{2i+1}
\end{equation}

In contrast, for the cluster state phase (the non-trivial SPT phase), we derive the charge decoration operator,
\begin{equation}
\begin{tikzpicture}
    \tikzset{baseline=(current bounding box.center)}
    \draw[thick, black, ->-=.8] (0,0) -- (0.5,0);
    \draw[thick,gray, ->-=0.35,->-=0.95] (0,-0.5) -- (0,0.5);
    \node[draw, regular polygon, regular polygon sides=3, minimum size=3mm, inner sep=0mm, thick,fill=yellow!30, shape border rotate=-30] at (0,0) {};
    \draw[thick, orange, -<-=.3,-<-=0.9] (-0.5,-0.5) -- (0.5,-0.5);
    \filldraw[fill=yellow!30, draw=black, thick] (0,-0.5) circle (0.12);
    \node[left] at (0,0) {${}^\text{SPT}\bar{\CE}_{{(a,b)}}$};
\end{tikzpicture}=\begin{tikzpicture}
    \tikzset{baseline=(current bounding box.center)}
    \draw[thick,gray, ->-=0.8] (0,-0.5) -- (0,0.5);
    \draw[thick, orange, -<-=.3,-<-=0.9] (-0.5,-0.5) -- (0.5,-0.5);
    \draw[thick, orange, -<-=0.6] (0.6,-0.5) -- (1,-.5);
    \draw[thick, black] (0.6,-0.5) -- (0.6,0);
    \draw[thick, black,->-=0.6] (0.6,0) -- (1,0);
    \filldraw[fill=yellow!30, draw=black, thick] (0,-0.5) circle (0.12);
    \filldraw[fill=white, draw=black, thick] (0.6,-0.5) circle (0.08);
    \node[above] at (-0.8,-0.5) {$\text{SPT}$};
    \node[below] at (0.6,-0.5) {};
    \node[above] at (0.8,0) {${(a,b)}$};
\end{tikzpicture}=
\begin{tikzpicture}
    \tikzset{baseline=(current bounding box.center)}
	\begin{pgfonlayer}{nodelayer}
		\node [style=green dot] (0) at (-1, 0) {};
		\node [style=had] (1) at (-0.5, 0) {};
		\node [style=green dot] (2) at (0, 0) {};
		\node [style=had] (3) at (0.5, 0) {};
		\node [style=red dot] (4) at (1.5, 0) {$a$};
		\node [style=green dot] (5) at (1, 0) {$b$};
		\node [style=none] (6) at (-1.5, 0) {};
		\node [style=none] (7) at (2, 0) {};
		\node [style=none] (8) at (-1, 0.75) {};
		\node [style=none] (9) at (0, 0.75) {};
	\end{pgfonlayer}
	\begin{pgfonlayer}{edgelayer}
		\draw (6.center) to (0);
		\draw (0) to (1);
		\draw (1) to (2);
		\draw (2) to (3);
		\draw (8.center) to (0);
		\draw (9.center) to (2);
		\draw (3) to (5);
		\draw (5) to (4);
		\draw (4) to (7.center);
	\end{pgfonlayer}
\end{tikzpicture}\Rightarrow\begin{tikzpicture}
    \tikzset{baseline=(current bounding box.center)}
    \draw[thick, black, ->-=.8] (0,0) -- (0.5,0);
    \draw[thick,gray, ->-=0.35,->-=0.95] (0,-0.5) -- (0,0.5);
    \node[draw, regular polygon, regular polygon sides=3, minimum size=3mm, inner sep=0mm, thick,fill=yellow!30, shape border rotate=-30] at (0,0) {};
    \node[left] at (0,0) {${}^\text{SPT}\bar{\CE}_{{(a,b)}}$};
\end{tikzpicture}=
\begin{tikzpicture}
    \tikzset{baseline=(current bounding box.center)}
	\begin{pgfonlayer}{nodelayer}
		\node [style=green dot] (0) at (0, 0.5) {$b$};
		\node [style=red dot] (1) at (0.5, 0.75) {$b$};
		\node [style=green dot] (2) at (0.5, 0.25) {$a$};
		\node [style=none] (3) at (0, 1) {};
		\node [style=none] (4) at (0, 0) {};
		\node [style=none] (5) at (0.5, 0) {};
		\node [style=none] (6) at (0.5, 1) {};
	\end{pgfonlayer}
	\begin{pgfonlayer}{edgelayer}
		\draw (3.center) to (0);
		\draw (0) to (4.center);
		\draw (6.center) to (1);
		\draw (1) to (2);
		\draw (2) to (5.center);
	\end{pgfonlayer}
\end{tikzpicture}
\end{equation}
Similarly, we can derive,
\begin{equation}
\begin{tikzpicture}
    \tikzset{baseline=(current bounding box.center)}
    \draw[thick, black, ->-=.6] (-0.5,0) -- (0,0);
    \draw[thick,gray, ->-=0.35,->-=0.95] (0,-0.5) -- (0,0.5);
    \node[draw, regular polygon, regular polygon sides=3, minimum size=3mm, inner sep=0mm, thick,fill=yellow!30, shape border rotate=30] at (0,0) {};
    \node[right] at (00,0) {${}^\text{SPT}\CE_{{(a,b)}}$};
\end{tikzpicture} =
\begin{tikzpicture}
    \tikzset{baseline=(current bounding box.center)}
	\begin{pgfonlayer}{nodelayer}
		\node [style=green dot] (0) at (0.5, 0.5) {$a$};
		\node [style=green dot] (1) at (0, 0.75) {$b$};
		\node [style=red dot] (2) at (0, 0.25) {$a$};
		\node [style=none] (3) at (0.5, 1) {};
		\node [style=none] (4) at (0.5, 0) {};
		\node [style=none] (5) at (0, 0) {};
		\node [style=none] (6) at (0, 1) {};
	\end{pgfonlayer}
	\begin{pgfonlayer}{edgelayer}
		\draw (3.center) to (0);
		\draw (0) to (4.center);
		\draw (6.center) to (1);
		\draw (1) to (2);
		\draw (2) to (5.center);
	\end{pgfonlayer}
\end{tikzpicture}
\end{equation}
So the string order parameter for the cluster state is,
\begin{equation}
\begin{tikzpicture}[tikzfig]
    \tikzset{baseline=(current bounding box.center)}
	\begin{pgfonlayer}{nodelayer}
		\node [style=green dot] (0) at (3, 0.5) {$a$};
		\node [style=green dot] (1) at (2.25, 1) {$b$};
		\node [style=red dot] (2) at (2.25, 0) {$a$};
		\node [style=none] (3) at (3, 1.75) {};
		\node [style=none] (4) at (3, -0.75) {};
		\node [style=none] (5) at (2.25, -0.75) {};
		\node [style=none] (6) at (2.25, 1.75) {};
		\node [style=green dot] (7) at (-4, 0.5) {$b$};
		\node [style=green dot] (8) at (-3.25, 0) {$a$};
		\node [style=red dot] (9) at (-3.25, 1) {$b$};
		\node [style=none] (10) at (-4, 1.75) {};
		\node [style=none] (11) at (-4, -0.75) {};
		\node [style=none] (12) at (-3.25, 1.75) {};
		\node [style=none] (13) at (-3.25, -0.75) {};
		\node [style=red dot] (14) at (-2.5, 0.5) {$a$};
		\node [style=red dot] (15) at (-1.75, 0.5) {$b$};
		\node [style=red dot] (16) at (1.5, 0.5) {$b$};
		\node [style=red dot] (17) at (0.75, 0.5) {$a$};
		\node [style=none] (18) at (-0.5, 0.5) {$\dots$};
		\node [style=none] (19) at (-2.5, 1.75) {};
		\node [style=none] (20) at (-2.5, -0.75) {};
		\node [style=none] (21) at (-1.75, -0.75) {};
		\node [style=none] (22) at (-1.75, 1.75) {};
		\node [style=none] (23) at (0.75, 1.75) {};
		\node [style=none] (24) at (0.75, -0.75) {};
		\node [style=none] (25) at (1.5, -0.75) {};
		\node [style=none] (26) at (1.5, 1.75) {};
	\end{pgfonlayer}
	\begin{pgfonlayer}{edgelayer}
		\draw (3.center) to (0);
		\draw (0) to (4.center);
		\draw (6.center) to (1);
		\draw (1) to (2);
		\draw (2) to (5.center);
		\draw (10.center) to (7);
		\draw (7) to (11.center);
		\draw (13.center) to (8);
		\draw (8) to (9);
		\draw (9) to (12.center);
		\draw (19.center) to (14);
		\draw (14) to (20.center);
		\draw (22.center) to (15);
		\draw (15) to (21.center);
		\draw (23.center) to (17);
		\draw (17) to (24.center);
		\draw (26.center) to (16);
		\draw (16) to (25.center);
	\end{pgfonlayer}
\end{tikzpicture} = Z^b_{2I-2}(X^b Z^a)_{2I-1}\left(\prod_{i=I}^{J} X^a_{2i} X^b_{2i+1}\right) (Z^b X^a)_{2J+2} Z^a_{2J+3}
\end{equation}
which aligns with the ordinary string order parameter for the cluster state of $\IZ_2\times \IZ_2$. In particular, the truncated $\IZ_2$ string is decorated with the charge of the other $\IZ_2$. 

\subsection{String order parameter for three \texorpdfstring{$\Rep(D_8)$}{Rep(D8)} NISPTs}
For the string order parameters of three $\Rep(D_8)$ NISPTs, we summarize the results as follows. The MPSs and their colors are given in \eqref{eq:repd8sptsmps}. For the $\CD$ strings below, we choose $J-I+1\in4\IZ$, so the intervening MPO segment consists of eight-site blocks and its scalar phase cancels.

For the Trivial state,
\begin{equation}
,\quad i=0,1
\end{equation}
\begin{samepage}
{\color{black}In the ZX convention of \appref{app:qdzx}, a controlled-$X$ gate is $\sqrt{2}$ times the diagram of a green spider joined to a red spider. Each two-site $\CD$ MPO cell contains two such gates and contributes $(\sqrt{2})^2=2$. With $J-I+1$ cells and a factor of $1/\sqrt{2}$ from each boundary vector, the overall normalization is $2^{J-I+1}/2=2^{J-I}$.} The corresponding string order parameter for the Tri state is
\begin{equation}
    ({}^\text{Tri}\hCO^\text{string}_{\CD})_{i;I,J}=\textcolor{black}{2^{J-I}}%
,\quad i=0,1
\end{equation}
\end{samepage}

For the Even state, recall that the MPS will be injective when grouping two unit-cells, so the charge decoration operator supports on 4 sites,

\begin{equation}
%
,\quad i=0,1
$}
\end{equation}
For example, the string order parameter of non-invertible symmetry $\CD$ is given by,
\begin{align}
    &({}^\text{Even}\hCO^\text{string}_{\CD})_{i;I,J}=\textcolor{black}{-2^{J-I}}%
,\quad i=0,1 \\
&={-}Z^{\bar{i}}_{2I-4} X^{\bar{i}}_{2I-2}Z^{i}_{2I-2} \bra{i}_x(\CO_\CD)_{2I,2I+1}\cdots (\CO_\CD)_{2J,2J+1} \ket{i}_x Z^{\bar{i}}_{2J+2}X^i_{2J+2}Z^i_{2J+4}
\end{align}
Here $\ket{i}_x$ are the normalized eigenstates of Pauli $X$ on the MPO bond. The truncated non-invertible symmetry MPO is dressed by the $\IZ_2^\text{even}$ charges for the Even NISPT. 

For the Odd state, as for the Even state, blocking two unit cells makes the MPS injective, and the charge decoration operator acts on 4 sites,
\begin{equation}
%
,\quad i=0,1
\end{equation}
The truncated non-invertible symmetry MPO is dressed by the $\IZ_2^\text{odd}$ charges for the Odd NISPT,
\begin{align}
    \label{eq:repd8oddstring}
	&({}^\text{Odd}\hCO^\text{string}_{\CD})_{i;I,J}=\textcolor{black}{-2^{J-I}}%
,\quad i=0,1 \\
	&={-}Z^{\bar{i}}_{2I-3} X^{\bar{i}}_{2I-1}Z^{i}_{2I-1} \bra{\bar{i}}_z(\CO_\CD)_{2I,2I+1}\cdots (\CO_\CD)_{2J,2J+1} \ket{\bar{i}}_z Z^{\bar{i}}_{2J+3}X^i_{2J+3}Z^i_{2J+5}
\end{align}
Here $\ket{i}_z$ are the normalized eigenstates of Pauli $Z$ on the MPO bond. It is clear that there are charge decorations of the string order parameter for Even and Odd NISPTs compared with the Tri NISPT of $\Rep(D_8)$. The non-invertible symmetry action on the edges of all the NISPT states are correlated as a general consequences from \eqref{eq:stringop}.

\section{Intrinsically mixed \texorpdfstring{$G\times\Rep(G)$}{G x Rep(G)} SPTs and their interface algebra}
\label{sec:grepgspt}

The $G\times\Rep(G)$ SPTs provide a concrete class of NISPTs with simple lattice constructions. A basic example is the group-based cluster state, which generalizes the ordinary cluster state from qubits to group-valued spins. It can be prepared from a product state by a finite depth circuit of controlled group multiplication gates, which generates entanglement between neighboring spins \cite{brell2015gcluster,nat2023gcluster}. These generalized cluster states are also motivated by applications in quantum information, particularly measurement-based quantum computation \cite{brell2015gcluster,nat2023gcluster}. Their entanglement spectra and gapless extensions have been studied in \cite{xu2026entanglementspectra}.

These SPTs also admit a bulk-boundary description as gapped physical boundaries of the $G\times G$ quantum double model, with a fixed reference boundary specifying the $G\times\Rep(G)$ symmetry \cite{nat2023gcluster,youxuan2026mixed}. For the group-based cluster state, unfolding this construction gives a thin strip of the $G$ quantum double model with rough and smooth boundaries \cite{nat2023gcluster}.

In this section, we apply the general construction of strange correlators and string order parameters to the fusion category symmetry
\begin{equation}
    \CC_G\equiv \VEC_G\boxtimes\Rep(G),
\end{equation}
which describes the $G\times\Rep(G)$ symmetry. We first specify the family of intrinsically mixed NISPTs and their $L$-symbols, then use these data to derive the interface algebra. For $G=S_3$, we compute the interface multiplets and show how the same data determine strange correlators, string order parameters and entanglement spectrum degeneracies.

\subsection{Fiber functors and \texorpdfstring{$L$}{L}-symbols}
\label{sec:grepgfiber}

Throughout this section, $G$ is a finite group with identity element $e$. A simple symmetry line of $\CC_G$ is labeled by $(g,\rho)$, where $g\in G$ and $\rho\in\simp(\Rep(G))$. We denote the representation space of $\rho$ by $V_\rho$ and choose an orthonormal basis in which $\rho(g)$ is unitary. The fusion rule and quantum dimension are
\begin{equation}
    (g,\rho)\otimes(h,\sigma)=(gh,\rho\otimes\sigma),
    \qquad d_{(g,\rho)}=d_\rho\equiv\dim(V_\rho).
\end{equation}
Here $h\in G$, $\sigma\in\simp(\Rep(G))$, and $\rho\otimes\sigma$ is generally reducible. The $G$ symmetry lines are labeled by $(g,\mathbf{1})$, while the $\Rep(G)$ symmetry lines are labeled by $(e,\rho)$. Here $\mathbf{1}$ is the trivial one-dimensional representation.

We consider phases that reduce to the trivial ordinary SPT when only $G$ is kept, and to the phase defined by the canonical forgetful functor when only $\Rep(G)$ is kept. These are the restrictions of the product reference phase; the non-trivial phases with the same restrictions are intrinsically mixed, since their protection requires both symmetry factors. Including the product reference phase, this family is classified by group homomorphisms $\varphi:G\rightarrow G$ up to conjugation \cite{youxuan2026mixed}. A homomorphism preserves multiplication, $\varphi(gh)=\varphi(g)\varphi(h)$. We denote the corresponding phase by $[\varphi]$ and its fixed point state by $\ketx{A_\varphi}$. Two homomorphisms $\varphi$ and $\psi$ label the same phase precisely when
\begin{equation}
    [\varphi]=[\psi]
    \quad\Longleftrightarrow\quad
    \varphi(g)=x\psi(g)x^{-1}
    \quad\text{for all }g\in G,
    \label{eq:grepgclass}
\end{equation}
for a single $x\in G$ independent of $g$. This conjugation is implemented by compatible changes of basis $\rho(x)$ on the representation spaces, giving equivalent fiber functors.

The fiber functor for phase $[\varphi]$ assigns $V_\rho$ to $(g,\rho)$. Its monoidal structure, which specifies how these vector spaces combine under fusion, is given by \cite{youxuan2026mixed}
\begin{align}
    J^\varphi_{(g,\rho),(h,\sigma)}
    &:V_\rho\otimes V_\sigma\longrightarrow
    V_\rho\otimes V_\sigma,\nonumber\\
    J^\varphi_{(g,\rho),(h,\sigma)}
    &=\dsi_{d_\rho}\otimes\sigma\big(\varphi(g)\big).
    \label{eq:grepgJ}
\end{align}
Here $\dsi_{d_\rho}$ is the identity matrix on $V_\rho$. The vector space on the right corresponds to the fused line $(gh,\rho\otimes\sigma)$. For fusion within either $\VEC_G$ or $\Rep(G)$, $J^\varphi$ is the identity map, so it acts non-trivially only when both symmetry factors are involved. The choice $\varphi(g)=e$ for every $g$ gives the product reference state, while $\varphi(g)=g$ gives the group-based cluster state \cite{nat2023gcluster,youxuan2026mixed}.

We choose a unitary Clebsch--Gordan change of basis for $\rho\otimes\sigma$. For the fusion channel $(g,\rho)\otimes(h,\sigma)\rightarrow(gh,\tau)$, the $L$-symbol in \eqref{eq:Lsymbol} is
\begin{equation}
    \left({}^{\varphi}L_{(g,\rho),(h,\sigma)}^{(gh,\tau)}\right)_{(i,j),(k,\mu)}
    =
    \sum_{j'=1}^{d_\sigma}
    \left\langle\tau,k,\mu\,\middle|\,\rho,i;\sigma,j'\right\rangle
    \sigma\big(\varphi(g)\big)_{j'j}.
    \label{eq:grepgL}
\end{equation}
The bracket is the Clebsch--Gordan coefficient between the tensor product basis labeled by $(i,j')$ and the irreducible basis labeled by $(k,\mu)$. Here $\tau\in\simp(\Rep(G))$ occurs in $\rho\otimes\sigma$, and $\mu$ runs over its fusion multiplicity. The indices $i=1,\ldots,d_\rho$, $j,j'=1,\ldots,d_\sigma$, and $k=1,\ldots,d_\tau$ refer to the bases of $V_\rho$, $V_\sigma$, and $V_\tau$, respectively. Therefore, the phase-dependent part of the $L$-symbol is the matrix $\sigma(\varphi(g))$, which describes the mixed fusion of the $G$ symmetry line $(g,\mathbf{1})$ and the $\Rep(G)$ symmetry line $(e,\sigma)$. These $L$-symbols determine the interface algebra in the following subsection.

\subsection{Interface algebra}
\label{sec:grepginterface}

We consider the interface $\varphi|\psi$ between the ket state $\ketx{A_\varphi}$ below and the bra state $\brax{A_\psi}$ above, following the ordering in \eqref{eq:interfacealg}. The interface algebra $\mathfrak{I}_{\varphi|\psi}$ is spanned by $\CI^{\varphi|\psi}_{(g,\rho),(i,j)}$. The indices $i,j=1,\ldots,d_\rho$ refer to the action tensors of $A_\varphi$ and $A_\psi$, respectively. To compute the multiplication rule, we use the Fourier transform for finite groups to write these operators in a basis labeled by group elements. For $h,g\in G$, we define
\begin{align}
    E^{\varphi|\psi}_{h,g}
    &\equiv
    \frac{1}{|G|}
    \sum_{\rho\in\simp(\Rep(G))}d_\rho
    \sum_{i,j=1}^{d_\rho}
    \rho(h)_{ji}\,
    \CI^{\varphi|\psi}_{(g,\rho),(i,j)},\nonumber\\
    \CI^{\varphi|\psi}_{(g,\rho),(i,j)}
    &=\sum_{h\in G}\rho(h^{-1})_{ij}
    E^{\varphi|\psi}_{h,g}.
    \label{eq:grepgPW}
\end{align}
The inverse transformation follows from the orthogonality of representation matrix elements. Substituting \eqref{eq:grepgL} into \eqref{eq:interalg} and using the Clebsch--Gordan completeness relation gives
\begin{equation}
    E^{\varphi|\psi}_{h,g}E^{\varphi|\psi}_{h',g'}
    =
    \delta_{h,\psi(g)h'\varphi(g)^{-1}}\,
    E^{\varphi|\psi}_{h,gg'}.
    \label{eq:grepgcrossedproduct}
\end{equation}
The identity element is $\sum_{h\in G}E^{\varphi|\psi}_{h,e}$. As an associative algebra, the interface algebra is the crossed product $\mathfrak{I}_{\varphi|\psi}\cong\IC^G\rtimes\IC[G]$, where $g\in G$ acts on $h\in G$ by $h\mapsto\psi(g)h\varphi(g)^{-1}$. The basis has $|G|^2$ elements, so $\dim\mathfrak{I}_{\varphi|\psi}=|G|^2$.

The irreducible representations are determined by the orbits of this action and the representations of their stabilizers \cite{koornwinder1997double}. For each orbit $\Omega\subset G$, choose $h\in\Omega$. The subgroup that leaves $h$ fixed is its stabilizer, $G_h=\{g\in G\mid\psi(g)h\varphi(g)^{-1}=h\}$. Each orbit contributes a block $M_{|\Omega|}(\IC[G_h])$, where $M_n(\cdot)$ denotes $n\times n$ matrices with entries in the indicated algebra. Decomposing the stabilizer group algebra gives
\begin{equation}
    \mathfrak{I}_{\varphi|\psi}
    \cong
    \bigoplus_{\Omega}
    \bigoplus_{\lambda\in\simp(\Rep(G_h))}
    M_{|\Omega|\dim(\lambda)}(\IC).
    \label{eq:grepgwedderburn}
\end{equation}
Here $\lambda$ runs over the irreducible representations of $G_h$. Thus, an irreducible representation of the interface algebra is labeled by $(\Omega,\lambda)$ and has dimension $|\Omega|\dim(\lambda)$. Since every $G_h$ has a trivial one-dimensional representation, the interface algebra admits a one-dimensional representation if and only if some orbit has $|\Omega|=1$. Such a fixed point satisfies $\varphi(g)=h^{-1}\psi(g)h$ for all $g\in G$, which is the condition for $[\varphi]=[\psi]$ in \eqref{eq:grepgclass}. Therefore, when $[\varphi]\ne[\psi]$, all irreducible representations of the interface algebra have dimensions greater than one. The interface then carries symmetry protected degenerate edge modes, in agreement with \secref{sec:interalg} \cite{kansei2024nispt}.

\subsection{The \texorpdfstring{$S_3\times\Rep(S_3)$}{S3 x Rep(S3)} example}
\label{sec:s3reps3}

\begingroup
We realize $S_3\times\Rep(S_3)$ symmetry on a periodic chain of $2N$ sites, each carrying a six-dimensional Hilbert space with basis $\ket{g}$, $g\in S_3$. The $S_3$ group is presented as $\langle r,s\mid r^3=s^2=e,\ srs=r^{-1}\rangle$. Following \cite{nat2023gcluster}, define $\overrightarrow{X}_g\ket{h}=\ket{gh}$ and $\overleftarrow{X}_g\ket{h}=\ket{hg^{-1}}$, and the matrix-valued operator $\overline{Z}_\rho=\sum_{h\in S_3}\rho(h)\otimes\ket{h}\bra{h}$, where $\rho(h)$ is the representation matrix of $h$. The symmetry operators associated with the two factors act on the odd and even sublattices, respectively:
\begin{equation}
    \overleftarrow{A}_g=\prod_{j=1}^{N}\overleftarrow{X}_g^{(2j-1)},
    \qquad
    \widehat{B}_\rho=\operatorname{Tr}\!\left[\overline{Z}_\rho^{(2)}\overline{Z}_\rho^{(4)}\cdots\overline{Z}_\rho^{(2N)}\right],
    \label{eq:s3chainsym}
\end{equation}
The matrices are multiplied in site order $2,4,\ldots,2N$, and the trace is taken over the $d_\rho$-dimensional MPO bond space $V_\rho$. $\overleftarrow{A}_g$ generates the $G$ symmetry, while $\widehat{B}_\rho$ generates the $\Rep(G)$ symmetry. The combined symmetry MPO is $\hCO_{(g,\rho)}=\overleftarrow{A}_g\widehat{B}_\rho$. The Hamiltonian of the trivial product state and its ground state wavefunction are
\begin{equation}
\begin{aligned}
    H_0&=-\frac{1}{6}\sum_{j=1}^{N}\left[\sum_{g\in S_3}\overrightarrow{X}_g^{(2j-1)}+\sum_{\rho\in\simp(\Rep(S_3))}d_\rho\operatorname{Tr}\overline{Z}_\rho^{(2j)}\right],\\
    \ketx{A_0}&=\bigotimes_{j=1}^{N}\ket{\mathbf{1}}_{2j-1}\ket{e}_{2j},
\end{aligned}
    \label{eq:s3productham}
\end{equation}
where $\ket{\mathbf{1}}=6^{-1/2}\sum_{g\in S_3}\ket{g}$. Character orthogonality identifies the even-site projector with $\ket{e}\bra{e}$.
\par\endgroup

\begin{samepage}
We label the three phases by $a=0,\IZ_2,S_3$, according to the trivial, order-two, or full image of $\varphi_a$. All three have the same $S_3\times\Rep(S_3)$ symmetry. The homomorphism classes have the following representatives \cite{nat2023gcluster,youxuan2026mixed}:
\begingroup
\renewcommand{\arraystretch}{1.25}
\setlength{\arraycolsep}{7pt}
\begin{equation}
\begin{array}{|c|c|c|c|}
    \hline
    a & \varphi_a(r) & \varphi_a(s) & \text{phase}\\
    \hline
    0 & e & e & \text{product state}\\
    \hline
    \IZ_2 & e & s & \text{$\IZ_2\times \IZ_2$ cluster-like state}\\
    \hline
    S_3 & r & s & \text{group-based cluster state}\\
    \hline
\end{array}.
    \label{eq:s3endmaps}
\end{equation}
\endgroup
\end{samepage}
In particular, $\varphi_0(g)=e$ and $\varphi_{S_3}(g)=g$. The map $\varphi_{\IZ_2}$ sends even permutations to $e$ and odd permutations to $s$. It is the quotient by the alternating subgroup $A_3=\langle r\rangle$, followed by the inclusion of $\IZ_2\cong\langle s\rangle$ into $S_3$. Its kernel is $A_3$, so it is not an automorphism.

For $S_3$, the kernel of a homomorphism must be one of the normal subgroups $S_3,A_3,\{e\}$, giving a trivial map, an order-two image, or an automorphism, respectively. All transpositions are conjugate, so replacing $s$ by another transposition in $\varphi_{\IZ_2}$ gives the same phase by \eqref{eq:grepgclass}. Every automorphism of $S_3$ is inner and hence equivalent to $\varphi_{S_3}$. The three representatives have different kernels and therefore cannot be related by conjugation.

\begingroup
For $a=0,\IZ_2,S_3$, write $\ketx{A_a}=\ketx{A_{\varphi_a}}$. Define the controlled multiplication gates $C\overrightarrow{X}_\varphi\ket{g,h}=\ket{g,\varphi(g)h}$ and $C\overleftarrow{X}_\varphi\ket{g,h}=\ket{g,h\varphi(g)^{-1}}$, with the first site as control and the second as target. These gates form the SPT entangler that prepares $\ketx{A_\varphi}$ from the product state,
\begin{equation}
    U_\varphi=\prod_{j=1}^{N}C\overrightarrow{X}_\varphi^{(2j-1,2j)}C\overleftarrow{X}_\varphi^{(2j+1,2j)},
    \qquad
    \ketx{A_\varphi}=U_\varphi\ketx{A_0},
    \label{eq:s3entangler}
\end{equation}
with site indices understood modulo $2N$. Conjugating \eqref{eq:s3productham} by this two-layer circuit gives
\begin{align}
    H_\varphi&=U_\varphi H_0U_\varphi^\dagger\nonumber\\
    &=-\frac{1}{6}\sum_{j=1}^{N}\Bigg[\sum_{g\in S_3}\overleftarrow{X}_{\varphi(g)}^{(2j-2)}\overrightarrow{X}_g^{(2j-1)}\overrightarrow{X}_{\varphi(g)}^{(2j)}\nonumber\\
    &\hspace{38mm}+\sum_{\rho\in\simp(\Rep(S_3))}d_\rho\operatorname{Tr}\!\left[\left(\overline{Z}_{\rho\circ\varphi}^{(2j-1)}\right)^\dagger\overline{Z}_\rho^{(2j)}\overline{Z}_{\rho\circ\varphi}^{(2j+1)}\right]\Bigg].
    \label{eq:s3sptHamiltonian}
\end{align}
Here $\overline{Z}_{\rho\circ\varphi}$ is defined by replacing $\rho(h)$ with $\rho(\varphi(h))$ in $\overline{Z}_\rho$. The second term enforces $h=\varphi(g)\varphi(k)^{-1}$, where $h$ labels an even-site spin and $g,k$ label its neighboring odd-site spins. Each $H_\varphi$ is a commuting-projector Hamiltonian with unique symmetric ground state $\ketx{A_\varphi}$ on the ring and the same symmetry operators \eqref{eq:s3chainsym}.

Setting $\varphi=\varphi_{S_3}$ gives the $S_3$ controlled-$X$ gates and recovers the group-based cluster Hamiltonian of \cite{nat2023gcluster}. For $\varphi_{\IZ_2}$, identify $\ket{r^m s^n}=\ket{m}\otimes\ket{n}$ with $m=0,1,2$ and $n=0,1$, so that $\varphi_{\IZ_2}(r^m s^n)=s^n$. On the even-site subspace spanned by $\ket{e},\ket{s}$, both controlled gates reduce to controlled-$X$ on the parity qubits $n$. Thus, the $\IZ_2\times\IZ_2$ cluster-like state $\ketx{A_{\IZ_2}}$ consists of a cluster state on the parity qubits, with the qutrits remaining in a uniform product state on odd sites and in $\ket{0}$ on even sites. A Hadamard change of basis on the even parity qubits gives the usual $\prod_i\cz_{i,i+1}$ preparation of the cluster state in \secref{sec:scz2z2}. The group multiplication in \eqref{eq:s3entangler} extends this transformation to the full local Hilbert space and ensures that $H_{\varphi_{\IZ_2}}$ respects the full $\Rep(S_3)$ symmetry.
\par\endgroup

We now compare these states using the interface algebra in \secref{sec:grepginterface}, writing $a|b$ for the interface between phases $a,b\in\{0,\IZ_2,S_3\}$. Inversion $h\mapsto h^{-1}$ relates the group actions for $a|b$ and $b|a$, so reversed interfaces have the same irreducible dimensions. Applying \eqref{eq:grepgwedderburn} gives the following orbit data and matrix block decompositions, where $M_n$ denotes $M_n(\IC)$:
\begin{center}
\small
\renewcommand{\arraystretch}{1.25}
\setlength{\tabcolsep}{4pt}
\begin{tabular}{|c|c|c|c|}
\hline
interface &
$(|\Omega|,G_h)$ &
$\mathfrak{I}_{a|b}$ &
irreducible dimensions\\
\hline
$0|0$ &
$6\times(1,S_3)$ &
$\IC^{\oplus 12}\oplus M_2^{\oplus 6}$ &
$1^{\times 12},\,2^{\times 6}$\\
\hline
$0|\IZ_2,\ \IZ_2|0$ &
$3\times(2,\IZ_3)$ &
$M_2^{\oplus 9}$ &
$2^{\times 9}$\\
\hline
$0|S_3,\ S_3|0$ &
$(6,\{e\})$ &
$M_6$ &
$6$\\
\hline
$\IZ_2|\IZ_2$ &
$2\times(1,S_3),\ 2\times(2,\IZ_3)$ &
$\IC^{\oplus 4}\oplus M_2^{\oplus 8}$ &
$1^{\times 4},\,2^{\times 8}$\\
\hline
$\IZ_2|S_3,\ S_3|\IZ_2$ &
$2\times(3,\IZ_2)$ &
$M_3^{\oplus 4}$ &
$3^{\times 4}$\\
\hline
$S_3|S_3$ &
$(1,S_3),\ (2,\IZ_3),\ (3,\IZ_2)$ &
$\IC^{\oplus 2}\oplus M_2^{\oplus 4}\oplus M_3^{\oplus 2}$ &
$1^{\times 2},\,2^{\times 4},\,3^{\times 2}$\\
\hline
\end{tabular}
\end{center}
For the $0|\IZ_2$ interface, even symmetry elements $g$ leave the group label $h$ unchanged, while odd $g$ send $h$ to $sh$. The six labels therefore split into three pairs, $\{e,s\}$, $\{r,r^2s\}$, and $\{r^2,rs\}$, each forming one orbit. The stabilizer of each label is $\{e,r,r^2\}\cong\IZ_3$.

For $0|S_3$, the action is $h\mapsto gh$. Any label can be mapped to any other, and only $g=e$ leaves a label unchanged. Thus, there is one orbit containing all six labels, with trivial stabilizer.

For $\IZ_2|S_3$, the action $h\mapsto gh\varphi_{\IZ_2}(g)^{-1}$ preserves permutation parity. The even labels $\{e,r,r^2\}$ form one orbit and the odd labels $\{s,rs,r^2s\}$ form the other. Choosing $h=e$ and $h=s$ as representatives, each is left unchanged only by $g=e,s$, giving a $\IZ_2$ stabilizer for each orbit.

For the $S_3|S_3$ interface, $\varphi_{S_3}(g)=g$, so the action is conjugation and its orbits are the three conjugacy classes of $S_3$. The multiplication rule \eqref{eq:grepgcrossedproduct} then identifies the interface algebra with the underlying associative algebra of the Drinfeld double \cite{kitaev2003anyons}:
\begin{equation}
    \mathfrak{I}_{S_3|S_3}
    \cong \IC^{S_3}\rtimes\IC[S_3]
    \cong D(S_3).
\end{equation}
Its eight irreducible representations have dimensions $1,1,2,2,2,2,3,3$. The coproduct and braiding require additional data and are not part of this identification.

For each interface, the squared irreducible dimensions sum to $36$, the dimension of the interface algebra. The irreducible representations describe the symmetry protected interface multiplets, and a microscopic interface can contain one or several such multiplets. Denoting the smallest allowed multiplet dimension by $d_{\min}(a|b)$, we find
\begin{equation}
\begin{aligned}
    d_{\min}(0|\IZ_2)&=d_{\min}(\IZ_2|0)=2,\\
    d_{\min}(\IZ_2|S_3)&=d_{\min}(S_3|\IZ_2)=3,\\
    d_{\min}(0|S_3)&=d_{\min}(S_3|0)=6.
\end{aligned}
    \label{eq:s3interfacedeg}
\end{equation}

\subsection{Strange correlator, string order parameter and entanglement spectrum}
\label{sec:s3diagnostics}

\paragraph{Strange correlator}
Following \secref{sec:scnispt}, we construct local strange charged operators from the interface elements of $a|b$, which are contractions of the action tensors of $A_a$ and $A_b$. For the fixed point tensors used here, the factors from the unit cells between $I$ and $J$ cancel in the normalized ratio, so it is independent of their separation. For pairs with nonzero overlap, summing over the two action-tensor indices gives $SC_{(g,\rho)}^{a|b}(I,J)\propto d_\rho$, with a nonzero factor set by the tensor and operator normalizations. Here $d_\rho$ is $1$ or $2$ for $S_3$. The phase dependence enters through the strange charged operators, which are constructed for the pair $a|b$.

When $a\ne b$, the interface algebra has no one-dimensional representations, so suitable strange charged operators probe protected interface multiplets. The operators must be chosen to distinguish the phases of interest.

\paragraph{String order parameter}
For phase $a$, we construct the string order parameter as in \eqref{eq:stringop}, using local dressing operators that satisfy \eqref{eq:stringcond} for $A_a$. The two ends have correlated action-tensor indices, as in the general construction. For a truncated $\Rep(S_3)$ string labeled by $\rho$, the symmetry representations of the local dressing operators are determined by $g\mapsto\rho(\varphi_a(g))$, as follows from \eqref{eq:grepgJ}. In the group-based cluster phase $S_3$, this recovers the transformation properties of the local dressing operators in the group-based cluster string order parameters \cite{nat2023gcluster}. Writing $\mathbf{sgn}$ for the sign representation and $\mathbf{2}$ for the two-dimensional irreducible representation, the decomposition of $\rho\circ\varphi_a$ is
\begingroup
\renewcommand{\arraystretch}{1.25}
\setlength{\arraycolsep}{7pt}
\begin{equation}
\begin{array}{|c|c|c|c|}
    \hline
    a\backslash\rho & \mathbf{1} & \mathbf{sgn} & \mathbf{2}\\
    \hline
    0 & \mathbf{1} & \mathbf{1} & \mathbf{1}\oplus\mathbf{1}\\
    \hline
    \IZ_2 & \mathbf{1} & \mathbf{sgn} & \mathbf{1}\oplus\mathbf{sgn}\\
    \hline
    S_3 & \mathbf{1} & \mathbf{sgn} & \mathbf{2}\\
    \hline
\end{array}.
    \label{eq:s3pullback}
\end{equation}
\endgroup
\begingroup
For the lattice realization in \eqref{eq:s3chainsym}, take distinct odd sites $I<J$ and truncate the $\Rep(S_3)$ MPO to the even sites between them. The local dressing operators are the $i$-th row of $(\overline{Z}_{\rho\circ\varphi_a}^{(I)})^\dagger$ on the left and the $i$-th column of $\overline{Z}_{\rho\circ\varphi_a}^{(J)}$ on the right. Thus, \eqref{eq:stringop} becomes
\begin{equation}
    ({}^{A_a}\hCO^{\text{string}}_{(e,\rho)})_{i;I,J}
    =\left[
    (\overline{Z}_{\rho\circ\varphi_a}^{(I)})^\dagger
    \left(\prod_{\substack{I<k<J\\k\text{ even}}}\overline{Z}_\rho^{(k)}\right)
    \overline{Z}_{\rho\circ\varphi_a}^{(J)}
    \right]_{ii}.
    \label{eq:s3dressedstring}
\end{equation}
The product is ordered along the chain and contracts the MPO representation indices. On $\ketx{A_a}$, the fixed point constraint below \eqref{eq:s3sptHamiltonian} makes the even-site factors cancel successively, and the local dressing operators make the matrix in brackets the identity. Each diagonal component therefore has expectation value $1$.

For a fixed symmetry MPO, the long-distance string expectation value vanishes unless the symmetry representation of the local dressing operators appears in the corresponding entry of \eqref{eq:s3pullback}. An analogous condition holds for ordinary SPTs \cite{pollmann2012stringo}. For a $\mathbf{sgn}$ string, for example, the local dressing operators transform in the sign representation in both phases $\IZ_2$ and $S_3$, so their symmetry representations alone cannot distinguish these phases. For phase $0$, choose $\rho=\mathbf{sgn}$; both local dressing operators are the identity. For phase $\IZ_2$, choose $\rho=\mathbf{2}$ in a basis with $\rho(s)=\operatorname{diag}(1,-1)$. Then $\overline{Z}_{\mathbf{2}\circ\varphi_{\IZ_2}}=\operatorname{diag}(\dsi,\overline{Z}_{\mathbf{sgn}})$, so taking $i=2$ selects the local operator transforming in the sign representation and dresses the $22$ component of the MPO with $\overline{Z}_{\mathbf{sgn}}$ at both ends. This local operator acts as $\overline{Z}_{\mathbf{sgn}}\ket{r^m s^n}=(-1)^n\ket{r^m s^n}$. For phase $S_3$, choose $\rho=\mathbf{2}$ and average the $i=1,2$ components of \eqref{eq:s3dressedstring}; the local dressing operators are the doublet matrix elements of $\overline{Z}_{\mathbf{2}}^\dagger$ and $\overline{Z}_{\mathbf{2}}$. These three choices have expectation value $1$ in their corresponding fixed point states and $0$ in the other two. The same symmetry constraint forces exponential decay in the other two gapped phases away from the fixed points.
\par\endgroup

Contracting the local dressing operators of phases $a$ and $b$ in the identity channel gives the corresponding strange charged operator, as shown in \eqref{eq:stringdecorstrange}. Both constructions therefore use the same phase-dependent action tensors.

\paragraph{Entanglement spectrum}
The interface algebra also constrains the entanglement spectrum. For the lattice realization in \eqref{eq:s3chainsym}, $\ketx{A_0}$ provides the symmetric product reference. Dividing an infinite chain into two halves, the symmetry MPO can be truncated at the cut, and the left Schmidt states in phase $a$ transform under $\mathfrak{I}_{a|0}$ \cite{lan2024mpo}. The reduced density matrix commutes with this action, so each of its eigenspaces decomposes into irreducible interface multiplets. This is the same symmetry argument that enforces entanglement degeneracies in ordinary SPTs \cite{pollmann2010entanglement}.

From \secref{sec:s3reps3}, all irreducible representations of $\mathfrak{I}_{\IZ_2|0}$ have dimension $2$, while $\mathfrak{I}_{S_3|0}$ has a single irreducible representation of dimension $6$. Therefore, every nonzero eigenvalue of the reduced density matrix has a multiplicity that is a multiple of $2$ in phase $\IZ_2$ and a multiple of $6$ in the group-based cluster phase $S_3$. The product phase has no symmetry-enforced degeneracy. In particular, at the group-based cluster fixed point $\ketx{A_{S_3}}$, the reduced density matrix has six nonzero eigenvalues, all equal to $1/6$ \cite{nat2023gcluster,xu2026entanglementspectra}.

\section{General $\Rep(H)$ SPTs from Drinfeld twists of a Hopf algebra}\label{sec:hopftwist}
The $G\times\Rep(G)$ NISPTs studied in \secref{sec:grepgspt} are labelled by group homomorphisms $\varphi:G\to G$ up to conjugation. For a general anomaly-free symmetry $\Rep(H)$, the NISPTs are labelled by gauge-equivalence classes of unitary Drinfeld twists of $H$, which is the generalization of the 2-cocycle twist of the ordinary group SPTs. In this section, we use the Drinfeld twist to construct a fixed point MPS, its action tensors and its string order parameters.

\subsection{From the twist to a fixed point state}
We consider a periodic chain of $N$ qudits of dimension $\dim H$, where $H$ is a finite-dimensional Hopf $C^*$-algebra. The local Hilbert space is the vector space $H$ with inner product $\langle h|k\rangle=\lambda(h^*k)$, where $\lambda\in H^*$ is the normalized Haar trace and $h^*$ is the adjoint of $h$ in $H$. We write $m$, $\Delta$, $\epsilon$ and $S$ for the multiplication, coproduct, counit and antipode, respectively, and $1$ for the unit element. We write $f\circ g$ for composition, with $g$ acting first. Juxtaposition and $\cdot$ denote algebra or matrix multiplication. See \cite{lu2026hopfising} for more details on Hopf algebra and the construction of physical models in $1+1$d.

For an irreducible representation $\rho$ of $H$, let $\rho_{ab}(h)=\langle a|\rho(h)|b\rangle$ in an orthonormal basis of $V_\rho$. The local symmetry tensor applies the coproduct to the physical input and evaluates $\rho_{ab}$ on the second leg \cite[Eqs.~(3.2) and~(3.28)]{lu2026hopfising}. Thus $(\CO_\rho)_{ab}=(\mathrm{id}\otimes\rho_{ab})\circ\Delta$ sends $\ket{k}$ to $\sum_{(k)}\ket{k_{(1)}}\,\rho_{ab}(k_{(2)})$, where $\Delta(k)=\sum_{(k)}k_{(1)}\otimes k_{(2)}$. Its matrix element between the input $\ket{k}$ and output $\ket{h}$ is
\tikzset{
    hopf mpo diagram/.style={
        tensor/.style={draw,thick,fill=white,minimum width=0.66cm,minimum height=0.40cm,inner sep=2.5pt},
        mpo/.style={thick,black,->-=0.55},
        physical/.style={thick,gray,->-=0.55},
        index/.style={font=\scriptsize,text=black,inner sep=2pt}},
    pics/hopf coproduct mpo/.style={code={
            \node[tensor,minimum width=0.76cm] (-cop) at (0,0.78) {$\Delta$};
            \node[tensor] (-rep) at (0.82,1.60) {$\rho$};
            \draw[thick,gray] ([xshift=-0.18cm]-cop.north)
                .. controls +(-0.06,0.27) and +(0,-0.24) .. (-0.80,1.52);
            \draw[physical] (-0.80,1.52) -- (-0.80,2.50) node[above,text=black] {$h$};
            \node[index,left] at (-0.40,1.19) {$k_{(1)}$};
            \draw[thick,black,->-=0.59] ([xshift=0.18cm]-cop.north)
                .. controls +(0.08,0.24) and +(0,-0.21) .. (-rep.south);
            \node[index,right] at (0.66,1.05) {$k_{(2)}$};
            \draw[mpo] (0.06,1.60) node[left] {$a$} -- (-rep.west);
            \draw[mpo] (-rep.east) -- (1.58,1.60) node[right] {$b$};
    }}}
\begingroup
\tikzexternaldisable
\begin{equation}
    (\CO_\rho^{h,k})_{ab}
    =\begin{tikzpicture}[baseline=(current bounding box.center),hopf mpo diagram]
        \node[tensor] (op) at (0,1.60) {$\CO_\rho$};
        \draw[physical] (0,0.16) node[below,text=black] {$k$} -- (op.south);
        \draw[physical] (op.north) -- (0,2.50) node[above,text=black] {$h$};
        \draw[mpo] (-0.96,1.60) node[left] {$a$} -- (op.west);
        \draw[mpo] (op.east) -- (0.96,1.60) node[right] {$b$};
        \node at (1.84,1.39) {$=$};
        \begin{scope}[xshift=3.46cm]
            \pic (local) at (0,0) {hopf coproduct mpo};
            \draw[physical] (0,0.16) node[below,text=black] {$k$} -- (local-cop.south);
        \end{scope}
    \end{tikzpicture}
    =\sum_{(k)}\langle h\mid k_{(1)}\rangle\,\rho_{ab}(k_{(2)}).
    \label{eq:hopflocalmpo}
\end{equation}
\endgroup
Contracting the indices $a,b$ around the ring gives the symmetry operator $\hCO_\rho$. By coassociativity of $\Delta$, stacking two such MPOs gives $\hCO_\rho\circ\hCO_\sigma=\hCO_{\rho\otimes\sigma}$, where $h\in H$ acts on $V_\rho\otimes V_\sigma$ by $\big[(\rho\otimes\sigma)\circ\Delta\big](h)$. Since $\Delta(1)=1\otimes1$, the local tensor satisfies $(\CO_\rho)_{ab}\ket{1}=\delta_{ab}\ket{1}$. Closing the MPO on the product state $\ket{1}^{\otimes N}$ gives the symmetry eigenvalue $d_\rho=\dim V_\rho$.

For the product state, the action tensors of $\hCO_\rho$ and $\hCO_\sigma$ combine by the ordinary tensor product. This gives the canonical forgetful functor on $\Rep(H)$. To construct other SPT phases with the same symmetry MPOs, we use a normalized unitary Drinfeld twist $J\in H\otimes H$ satisfying $(\epsilon\otimes\mathrm{id})(J)=(\mathrm{id}\otimes\epsilon)(J)=1$ and the associativity condition, 
\begin{equation}
    (\Delta\otimes\mathrm{id})(J)\cdot(J\otimes1)=(\mathrm{id}\otimes\Delta)(J)\cdot(1\otimes J).
    \label{eq:hopftwisteq}
\end{equation}
For the phase specified by $J$, the $L$-symbol defined in \eqref{eq:Lsymbol} for the fusion channel $\rho\otimes\sigma\to\tau$ is
\begin{equation}
    (L_{\rho\sigma}^{\tau})_{(i,j),(k,\mu)}
    =\sum_{i'=1}^{d_\rho}\sum_{j'=1}^{d_\sigma}
    \big\langle\tau,k,\mu\,\big|\,\rho,i';\sigma,j'\big\rangle
    \big[(\rho\otimes\sigma)(J)\big]_{(i',j'),(i,j)} .
    \label{eq:hopftwistactionfusion}
\end{equation}
The bracket is the Clebsch--Gordan coefficient projecting the product basis vector $\ket{\rho,i';\sigma,j'}$ onto the $k$th basis vector in the $\mu$th copy of $V_\tau$. The second factor is the matrix element of the twist in the product basis of $V_\rho\otimes V_\sigma$. For unitary $u\in H$ with $\epsilon(u)=1$, the gauge-equivalent twists $J$ and $\Delta(u)J(u^*\otimes u^*)$ describe the same NISPT phase \cite{aljadeff2002twisting,etingof2015tensor}. 

For example, take $H=\IC^G\otimes\IC[G]$ for the symmetry in \secref{sec:grepgspt}. With $\delta_g$ the function supported at $g$ with value one, the twist $J=\sum_g(\delta_g\otimes 1)\otimes(1\otimes\varphi(g))$ acts on the pair $(g,\rho),(h,\sigma)$ as $\dsi_{d_\rho}\otimes(\sigma\circ\varphi)(g)$, so \eqref{eq:hopftwistactionfusion} reduces to \eqref{eq:grepgL}.

To construct an MPS for the phase specified by $J$, we first choose its virtual bond space. The twist defines an associative product on $H^*$ by $(f\star_J g)(h)=(f\otimes g)(\Delta(h)J)$, with unit $\epsilon$ \cite{aljadeff2002twisting}. Each $f\in H^*$ acts on $H^*$ by left multiplication, $g\mapsto f\star_J g$. We use one irreducible representation of the twisted algebra $(H^*,\star_J)$ on the MPS bond. Its representation space $W\subseteq H^*$ is invariant under left multiplication, so the bond matrices act by $\widetilde{\pi}(f)w=f\star_J w$ for $w\in W$. The twist $J$ fixes the SPT phase, while the bond dimension $\dim W$ can depend on the choice of irreducible representation. For ordinary group symmetry, $H=\IC^G$, this gives the familiar projective representation on the MPS bond.

We collect the matrices $\widetilde{\pi}(f)$ in a tensor $T\in\operatorname{End}(W)\otimes H$ defined by $(\mathrm{id}\otimes f)(T)=\widetilde{\pi}(f)$ for every $f\in H^*$. Let $\widetilde{I}$ denote the identity on $W$. Successive left multiplications satisfy $\widetilde{\pi}(f)\circ\widetilde{\pi}(g)=\widetilde{\pi}(f\star_J g)$, which gives
\begingroup
\tikzexternaldisable
\begin{equation}
\begin{gathered}
    (\mathrm{id}\otimes\Delta)(T)\cdot(\widetilde{I}\otimes J)=T_{12}T_{13}.\\[4pt]
    \begin{tikzpicture}[baseline=(current bounding box.center),
        tensor/.style={draw,thick,fill=white,minimum height=0.39cm,minimum width=0.61cm,inner sep=2.3pt},
        bond/.style={thick,cyan,-<-=0.5},
        hopf/.style={thick,black,->-=0.55}]
        \begin{scope}
            \node[tensor] (t) at (-0.66,0) {$T$};
            \draw[bond] (-1.53,0) node[below,text=black] {$\scriptstyle\gamma$} -- (t.west);
            \draw[bond] (t.east) -- (1.44,0) node[below,text=black] {$\scriptstyle\alpha$};
            \node[tensor] (delta) at (-0.66,0.77) {$\Delta$};
            \node[tensor] (twist) at (0.66,0.77) {$J$};
            \node[tensor] (mleft) at (-0.66,1.83) {$m$};
            \node[tensor] (mright) at (0.66,1.83) {$m$};
            \draw[hopf] (t.north) -- (delta.south);
            \draw[hopf] ([xshift=-0.17cm]delta.north) -- ([xshift=-0.17cm]mleft.south);
            \draw[hopf] ([xshift=0.17cm]twist.north) -- ([xshift=0.17cm]mright.south);
            \path ([xshift=0.17cm]delta.north) -- ([xshift=-0.17cm]mright.south)
                coordinate[pos=0.43] (understart) coordinate[pos=0.57] (underend);
            \draw[thick] ([xshift=0.17cm]delta.north) -- (understart);
            \draw[hopf] (underend) -- ([xshift=-0.17cm]mright.south);
            \draw[thick,black,->-=0.76] ([xshift=-0.17cm]twist.north) -- ([xshift=0.17cm]mleft.south);
            \draw[hopf] (mleft.north) -- (-0.66,2.43) node[above] {$H$};
            \draw[hopf] (mright.north) -- (0.66,2.43) node[above] {$H$};
        \end{scope}
        \node at (2.50,1.16) {$=$};
        \begin{scope}[xshift=4.90cm]
            \node[tensor] (first) at (-0.66,0) {$T$};
            \node[tensor] (second) at (0.66,0) {$T$};
            \draw[bond] (-1.53,0) node[below,text=black] {$\scriptstyle\gamma$} -- (first.west);
            \draw[bond] (first.east) -- (second.west);
            \draw[bond] (second.east) -- (1.44,0) node[below,text=black] {$\scriptstyle\alpha$};
            \draw[hopf] (first.north) -- (-0.66,2.43) node[above] {$H$};
            \draw[hopf] (second.north) -- (0.66,2.43) node[above] {$H$};
        \end{scope}
    \end{tikzpicture}
\end{gathered}
\label{eq:hopftwistvirtual}
\end{equation}
\endgroup
The row index $\gamma$ of $T_{\gamma\alpha}$ labels the output and the column index $\alpha$ labels the input. 
The tensor $T$ specifies how the symmetry acts on the virtual bond $W$. 

We equip $\operatorname{End}(W)$ with the normalized trace inner product $\langle\widetilde{M},\widetilde{N}\rangle=\operatorname{tr}(\widetilde{M}^\dagger\widetilde{N})/\dim W$ and choose a linear isometry $\iota:\operatorname{End}(W)\to H$ with $\iota(\widetilde{I})=1$.\footnote{For irreducible $W$, each irreducible corepresentation of $H$ occurs in $\operatorname{End}(W)$ with multiplicity at most its dimension \cite[Theorem~2.9]{bichon2006ergodic}. Under $\Delta$, it occurs in $H$ with multiplicity equal to its dimension, so a symmetry-preserving isometric embedding exists.} It maps the virtual pair $\widetilde{E}_{\alpha\beta}=\ket{\alpha}\bra{\beta}$ at one site to the physical state $\ket{\iota(\widetilde{E}_{\alpha\beta})}$. To preserve the symmetry, applying the coproduct after $\iota$ must give the same result as acting on the virtual indices before $\iota$. In a bond basis where $T$ is unitary, this condition reads
\begingroup
\tikzexternaldisable
\begin{equation}
\begin{gathered}
    (\Delta\circ\iota)(\widetilde{E}_{\alpha\beta})
    =(\iota\otimes\mathrm{id})\big[T(\widetilde{E}_{\alpha\beta}\otimes1)T^\dagger\big].\\[5pt]
\begin{tikzpicture}[baseline=(current bounding box.center),
    tensor/.style={draw,thick,fill=white,minimum height=0.40cm,inner sep=2.5pt},
    bond/.style={thick,cyan,->-=0.5},
    hopf/.style={thick,black,->-=0.55}]
    \begin{scope}
        \node[tensor,minimum width=0.66cm] (site) at (0,0.62) {$\iota$};
        \draw[bond] (-1.08,0.62) node[below,text=black] {$\scriptstyle\alpha$} -- (site.west);
        \draw[bond] (site.east) -- (1.08,0.62) node[below,text=black] {$\scriptstyle\beta$};
        \node[tensor,minimum width=0.75cm] (coproduct) at (0,1.45) {$\Delta$};
        \draw[hopf] (site.north) -- (coproduct.south);
        \draw[hopf] ([xshift=-0.18cm]coproduct.north) -- (-0.55,2.47) node[above] {$H$};
        \draw[hopf] ([xshift=0.18cm]coproduct.north) -- (0.55,2.47) node[above] {$H$};
    \end{scope}
    \node at (1.72,1.05) {$=$};
    \begin{scope}[xshift=4.45cm]
        \node[tensor,minimum width=0.66cm] (leftT) at (-1.19,0.62) {$T$};
        \node[tensor,minimum width=0.53cm] (iota) at (0,0.62) {$\iota$};
        \node[tensor,minimum width=0.66cm] (rightT) at (1.19,0.62) {$T^\dagger$};
        \draw[bond] (-2.08,0.62) node[below,text=black] {$\scriptstyle\alpha$} -- (leftT.west);
        \draw[bond] (leftT.east) -- (iota.west) node[midway,below,text=black] {$\scriptstyle\gamma$};
        \draw[bond] (iota.east) -- (rightT.west) node[midway,below,text=black] {$\scriptstyle\eta$};
        \draw[bond] (rightT.east) -- (2.18,0.62) node[below,text=black] {$\scriptstyle\beta$};
        \node[tensor,minimum width=0.70cm] (mult) at (0.99,1.99) {$m$};
        \draw[thick,black,rounded corners=3pt] (leftT.north) |- (-0.65,1.63);
        \draw[thick,black,->-=0.25] (-0.65,1.63) -- (0.40,1.63);
        \draw[thick,black,rounded corners=3pt] (0.40,1.63) -| ([xshift=-0.20cm]mult.south);
        \draw[hopf] (rightT.north) -- ([xshift=0.20cm]mult.south);
        \draw[hopf] (mult.north) -- (0.99,2.47) node[above] {$H$};
        \draw[thick,black] (iota.north) -- (0,1.43);
        \draw[thick,black,preaction={draw=white,line width=3.6pt}]
            (0,1.43) .. controls (0,1.53) and (0.15,1.53) .. (0.15,1.63)
                     .. controls (0.15,1.73) and (0,1.73) .. (0,1.83);
        \draw[hopf] (0,1.83) -- (0,2.47) node[above] {$H$};
    \end{scope}
\end{tikzpicture}
\end{gathered}
\label{eq:hopftwistiota}
\end{equation}
\endgroup
In the right-hand diagram, $\gamma$ and $\eta$ are summed. The first $H$ output comes from $\iota(\widetilde{E}_{\gamma\eta})$, and the second is the product $T_{\gamma\alpha}(T_{\eta\beta})^*$ in $H$. Contracting the cyan virtual legs of neighboring $\iota$ tensors around the ring gives the fixed point MPS $\ket{A_J}$ as follows,
\begingroup
\tikzexternaldisable
\begin{equation}
    \begin{gathered}
        (A_J^h)_{\alpha\beta}
        =\begin{tikzpicture}[baseline=-0.55ex]
            \node[draw,thick,fill=white,minimum width=0.66cm,minimum height=0.40cm,inner sep=2pt] (encode) at (0,0) {$\iota$};
            \draw[thick,cyan,->-=0.5] (-0.93,0) node[below,text=black] {$\scriptstyle\alpha$} -- (encode.west);
            \draw[thick,cyan,->-=0.5] (encode.east) -- (0.93,0) node[below,text=black] {$\scriptstyle\beta$};
            \draw[thick,gray,->-=0.55] (encode.north) -- (0,0.63) node[above,text=black] {$h$};
        \end{tikzpicture}
        =\langle h\mid\iota(\widetilde{E}_{\alpha\beta})\rangle,\\[4pt]
        \ket{A_J}=\;
        \begin{tikzpicture}[baseline=(site1.base),
            encoder/.style={draw,thick,fill=white,minimum width=0.95cm,inner ysep=2pt},
            pair/.style={draw=cyan,thick,rounded corners=4pt}]
            \foreach \x/\site in {0/1,1.65/2,4.95/N} {
                \node[encoder] (site\site) at (\x,0) {$\iota$};
                \draw[thick,black!65,->-=0.55] (site\site.north) -- (\x,0.68)
                    node[above,text=black] {$H_{\site}$};
            }
            \node[minimum width=0.95cm,inner ysep=2pt] (omitted) at (3.30,0) {$\cdots$};
            \foreach \leftsite/\rightsite in {site1/site2,site2/omitted,omitted/siteN} {
                \draw[pair] ([xshift=0.28cm]\leftsite.south) -- ++(0,-0.38)
                    -| ([xshift=-0.28cm]\rightsite.south);
            }
            \draw[pair] ([xshift=0.28cm]siteN.south) -- ++(0,-0.80)
                -| ([xshift=-0.28cm]site1.south);
        \end{tikzpicture}
        ,\qquad
        \langle h_1,\ldots,h_N|A_J\rangle=\operatorname{tr}(A_J^{h_1}\cdots A_J^{h_N}).
    \end{gathered}
    \label{eq:hopftwistmps}
\end{equation}
\endgroup
For physical states orthogonal to the image of $\iota$, the MPS matrices vanish. The state is injective on one site. A commuting-projector parent Hamiltonian penalizes states outside the encoded subspace and enforces a Bell pair on each bond \cite{Inamura2022nigap,lu2026hopfising}. For $J=1\otimes1$, choosing $\widetilde{\pi}(f)=f(1)$ gives $\dim W=1$ and recovers the product state $\ket{1}^{\otimes N}$.

Since $\iota(\widetilde{I})=1$, the physical state $\ket{1}$ corresponds to a normalized virtual Bell pair within each site. A finite-depth NISPT entangler prepares $\ket{A_J}$ from $\ket{1}^{\otimes N}$ by moving these Bell pairs onto the bonds between neighboring sites. A symmetric NISPT entangler must also commute with every symmetry MPO on the ring. This condition applies to the full circuit, even if individual gates break the symmetry. Whether such a symmetric entangler exists depends on the phases it connects \cite{Sahand2024cluster,you2025symmetricentanglers}.

\subsection{Action tensors and string order parameters}
To apply the symmetry MPO $\hCO_\rho$ to $\ket{A_J}$, attach its local tensor in \eqref{eq:hopflocalmpo} to the physical state $\ket{\iota(\widetilde{E}_{\alpha\beta})}$ at one site. By \eqref{eq:hopftwistiota}, $\rho$ acts on the product $T_{\gamma\alpha}(T_{\eta\beta})^*$ on the second leg of the coproduct. Write $\widetilde{T}^\rho_{ai}=\widetilde{\pi}(\rho_{ai})=(\mathrm{id}\otimes\rho_{ai})(T)$ for the resulting bond matrices, with entries $(\widetilde{T}^\rho_{ai})_{\gamma\alpha}=\rho_{ai}(T_{\gamma\alpha})$. Here $a,i$ refer to the MPO bond $V_\rho$, and $\gamma,\alpha$ refer to the MPS bond $W$. Using \eqref{eq:hopftwistmps}, the local MPO action becomes
\begingroup
\tikzexternaldisable
\begin{equation}
\begin{gathered}
    \sum_k (\CO_\rho^{h,k})_{ab} A_J^k
    =\sum_i (\widetilde{T}^\rho_{ai})^{\mathsf T} A_J^h\,\overline{\widetilde{T}^\rho_{bi}} .\\[5pt]
    \begin{tikzpicture}[baseline=(current bounding box.center),hopf mpo diagram,bond/.style={thick,cyan,->-=0.5}]
        \begin{scope}
            \node[tensor] (site) at (0,0) {$\iota$};
            \node[tensor] (op) at (0,1.60) {$\CO_\rho$};
            \draw[bond] (-1.02,0) node[index,below] {$\alpha$} -- (site.west);
            \draw[bond] (site.east) -- (1.02,0) node[index,below] {$\beta$};
            \draw[physical] (site.north) -- (op.south) node[midway,right,index] {$k$};
            \draw[physical] (op.north) -- (0,2.50) node[above,text=black] {$h$};
            \draw[mpo] (-0.99,1.60) node[left] {$a$} -- (op.west);
            \draw[mpo] (op.east) -- (0.99,1.60) node[right] {$b$};
        \end{scope}
        \node at (1.76,1.03) {$=$};
        \begin{scope}[xshift=3.57cm]
            \pic (local) at (0,0) {hopf coproduct mpo};
            \node[tensor] (encode) at (0,0) {$\iota$};
            \draw[bond] (-1.02,0) node[index,below] {$\alpha$} -- (encode.west);
            \draw[bond] (encode.east) -- (1.02,0) node[index,below] {$\beta$};
            \draw[physical] (encode.north) -- (local-cop.south) node[midway,right,index] {$k$};
        \end{scope}
        \node at (6.04,1.03) {$=$};
        \begin{scope}[xshift=9.05cm]
            \node[tensor] (leftT) at (-1.29,0) {$T$};
            \node[tensor] (iota) at (0,0) {$\iota$};
            \node[tensor] (rightT) at (1.29,0) {$T^\dagger$};
            \node[tensor] (leftRho) at (-1.29,1.60) {$\rho$};
            \node[tensor] (rightRho) at (1.29,1.60) {$\rho$};
            \draw[bond] (-2.17,0) node[index,below] {$\alpha$} -- (leftT.west);
            \draw[bond] (leftT.east) -- (iota.west) node[midway,below,index] {$\gamma$};
            \draw[bond] (iota.east) -- (rightT.west) node[midway,below,index] {$\eta$};
            \draw[bond] (rightT.east) -- (2.25,0) node[index,below] {$\beta$};
            \draw[mpo] (leftT.north) -- (leftRho.south);
            \draw[mpo] (rightT.north) -- (rightRho.south);
            \draw[mpo] (-2.17,1.60) node[left] {$a$} -- (leftRho.west);
            \draw[thick,black,->-=0.28] (leftRho.east) -- (rightRho.west) node[pos=0.27,above,index] {$i$};
            \draw[mpo] (rightRho.east) -- (2.25,1.60) node[right] {$b$};
            \draw[thick,gray] (iota.north) -- (0,1.38);
            \draw[thick,gray,preaction={draw=white,line width=3.6pt}]
                (0,1.38) .. controls (0,1.48) and (0.14,1.49) .. (0.14,1.60)
                         .. controls (0.14,1.71) and (0,1.72) .. (0,1.82);
            \draw[physical] (0,1.82) -- (0,2.50) node[above,text=black] {$h$};
        \end{scope}
    \end{tikzpicture}
\end{gathered}
\label{eq:hopftwistmpo}
\end{equation}
\endgroup
The transpose acts on $W$. The left $T$ box contributes $(\widetilde{T}^\rho_{ai})_{\gamma\alpha}$, which multiplies row $\gamma$ of $A_J^h$ to give row $\alpha$ on the left of \eqref{eq:hopftwistmpo}. Comparing with \eqref{eq:oactonA} gives the black and white action tensors
\begin{equation}
    (\phi_\rho)_i=\sum_a\bra{a}\otimes(\widetilde{T}^\rho_{ai})^{\mathsf T},
    \qquad (\bphi_\rho)_i=(\phi_\rho)_i^\dagger .
    \label{eq:hopftwistaction}
\end{equation}
Since $\iota$ is an isometry and $\iota(\widetilde{I})=1$, the Haar trace gives $\lambda(\iota(\widetilde{E}_{\alpha\beta}))=\delta_{\alpha\beta}/\dim W$. Applying $\lambda$ to the first leg of \eqref{eq:hopftwistiota} and using Haar invariance gives $\sum_\gamma T_{\gamma\alpha}(T_{\gamma\beta})^*=\delta_{\alpha\beta}1$. Evaluating $\rho$ on this identity makes the block matrix $[(\widetilde{T}^\rho_{ai})^{\mathsf T}]_{a,i}$ unitary, giving the action-tensor orthogonality in \eqref{eq:actionorthogonality}. On the ring, the contraction leaves a sum over $i$ and gives $\hCO_\rho\ket{A_J}=d_\rho\ket{A_J}$. For two MPOs, the virtual matrices multiply as $(\widetilde{T}^\sigma_{bj})^{\mathsf T}(\widetilde{T}^\rho_{ai})^{\mathsf T}=(\widetilde{T}^\rho_{ai}\widetilde{T}^\sigma_{bj})^{\mathsf T}$. Projecting \eqref{eq:hopftwistvirtual} onto the fusion channel $(\tau,\mu)$ then reproduces the $L$-symbol in \eqref{eq:hopftwistactionfusion}.

The isometry $\iota$ lets us realize each local dressing in \eqref{eq:stringcond} on a single physical qudit. The left dressing sends $\ket{\iota(\widetilde{E}_{\alpha\beta})}$ to $\ket{\iota(\widetilde{E}_{\alpha\beta}(\widetilde{T}^\rho_{ai})^\dagger)}$, and the right dressing sends it to $\ket{\iota(\widetilde{T}^\rho_{bj}\widetilde{E}_{\alpha\beta})}$. We choose both operators to vanish on the orthogonal complement of the image of $\iota$. The triangles denote these physical operators in the following diagrams.
\begingroup
\tikzexternaldisable
\tikzset{
    tensor/.style={draw,thick,fill=white,minimum width=0.66cm,minimum height=0.40cm,inner sep=2.5pt},
    bond/.style={thick,cyan,->-=0.5},
    mpo/.style={thick,black,->-=0.55},
    physical/.style={thick,gray,->-=0.55},
    index/.style={font=\scriptsize,text=black,inner sep=2pt},
    charge/.style={draw,thick,fill=blue!30,regular polygon,regular polygon sides=3,minimum size=3.5mm,inner sep=0pt}}
\begin{equation}
\begin{array}{@{}c@{\qquad}c@{}}
    \displaystyle\sum_k\bra{h}\big[{}^{A_J}(\bar{\CE}_\rho)_i\big]_a\ket{k}\,A_J^k
        = A_J^h\,\overline{\widetilde{T}^\rho_{ai}},
    &
    \displaystyle\sum_k\bra{h}\big[{}^{A_J}(\CE_\rho)_j\big]_b\ket{k}\,A_J^k
        = (\widetilde{T}^\rho_{bj})^{\mathsf T}A_J^h .\\[6pt]
    \begin{tikzpicture}[baseline=(current bounding box.center)]
        \node[tensor] (site) at (0,0) {$\iota$};
        \node[charge,shape border rotate=-30] (dress) at (0,0.99) {};
        \draw[bond] (-0.79,0) node[index,below] {$\alpha$} -- (site.west);
        \draw[bond] (site.east) -- (0.79,0) node[index,below] {$\beta$};
        \draw[physical] (site.north) -- (dress.south) node[midway,right,index] {$k$};
        \draw[physical] (dress.north) -- (0,1.84) node[above,text=black] {$h$};
        \node[left=1pt] at (dress.west) {$i$};
        \draw[mpo] (dress.east) -- (0.78,0.99) node[right] {$a$};
        \node at (1.43,0.62) {$=$};
        \node[tensor] (encode) at (2.60,0) {$\iota$};
        \node[tensor] (tdag) at (4.00,0) {$T^\dagger$};
        \node[tensor] (rep) at (4.00,0.99) {$\rho$};
        \draw[bond] (1.82,0) node[index,below] {$\alpha$} -- (encode.west);
        \draw[bond] (encode.east) -- (tdag.west) node[midway,below,index] {$\eta$};
        \draw[bond] (tdag.east) -- (4.79,0) node[index,below] {$\beta$};
        \draw[physical] (encode.north) -- (2.60,1.84) node[above,text=black] {$h$};
        \draw[mpo] (tdag.north) -- (rep.south);
        \draw[mpo] (3.22,0.99) node[left] {$i$} -- (rep.west);
        \draw[mpo] (rep.east) -- (4.79,0.99) node[right] {$a$};
    \end{tikzpicture}
    &
    \begin{tikzpicture}[baseline=(current bounding box.center)]
        \node[tensor] (site) at (0,0) {$\iota$};
        \node[charge,shape border rotate=30] (dress) at (0,0.99) {};
        \draw[bond] (-0.79,0) node[index,below] {$\alpha$} -- (site.west);
        \draw[bond] (site.east) -- (0.79,0) node[index,below] {$\beta$};
        \draw[physical] (site.north) -- (dress.south) node[midway,right,index] {$k$};
        \draw[physical] (dress.north) -- (0,1.84) node[above,text=black] {$h$};
        \draw[mpo] (-0.78,0.99) node[left] {$b$} -- (dress.west);
        \node[right=1pt] at (dress.east) {$j$};
        \node at (1.43,0.62) {$=$};
        \node[tensor] (t) at (2.64,0) {$T$};
        \node[tensor] (encode) at (4.04,0) {$\iota$};
        \node[tensor] (rep) at (2.64,0.99) {$\rho$};
        \draw[bond] (1.85,0) node[index,below] {$\alpha$} -- (t.west);
        \draw[bond] (t.east) -- (encode.west) node[midway,below,index] {$\gamma$};
        \draw[bond] (encode.east) -- (4.83,0) node[index,below] {$\beta$};
        \draw[mpo] (t.north) -- (rep.south);
        \draw[mpo] (1.85,0.99) node[left] {$b$} -- (rep.west);
        \draw[mpo] (rep.east) -- (3.42,0.99) node[right] {$j$};
        \draw[physical] (encode.north) -- (4.04,1.84) node[above,text=black] {$h$};
    \end{tikzpicture}
\end{array}
\label{eq:hopftwistdress}
\end{equation}
\endgroup
Place the two dressings at sites $I<K$ and contract $a,b$ with the truncated MPO between them. By \eqref{eq:actionorthogonality}, the resulting string has expectation value $\delta_{ij}$ in $\ket{A_J}$. Thus each component with the same index $i$ at both sites, as in \eqref{eq:stringop}, has expectation value one.

\subsubsection{The two-dimensional line of \texorpdfstring{$\Rep(D_8)$}{Rep(D8)}}
We return to the two-dimensional line $\CD$ of \secref{sec:screpd8}, now using $H=\IC[D_8]$ as the local Hilbert space. Write $z=r^2$ and take $t=s$ or $rs$. For either choice of $t$, the subgroup $\langle z,t\rangle\cong\IZ_2\times\IZ_2$ gives the twist $J_t=\tfrac12(1\otimes1+1\otimes t+z\otimes1-z\otimes t)$. Let $\widetilde{X},\widetilde{Z}$ be the Pauli matrices on the virtual qubit $W$. We specify $T$ by evaluating its $H$ leg on the basis functions $\delta_g\in H^*=\IC^{D_8}$, where $\delta_g(h)=\delta_{g,h}$. Writing $\widetilde{T}(\delta_g)=(\mathrm{id}\otimes\delta_g)(T)=\widetilde{\pi}(\delta_g)$, a solution of \eqref{eq:hopftwistvirtual} has the four nonzero components
\begin{equation}
    \begin{aligned}
        \widetilde{T}(\delta_1) &= \tfrac14(\widetilde{I}+\widetilde{Z})(\widetilde{I}+\widetilde{X}), & \widetilde{T}(\delta_z)  &= \tfrac14(\widetilde{I}+\widetilde{Z})(\widetilde{I}-\widetilde{X}),\\
        \widetilde{T}(\delta_t) &= \tfrac14(\widetilde{I}-\widetilde{Z})(\widetilde{I}+\widetilde{X}), & \widetilde{T}(\delta_{zt}) &= \tfrac14(\widetilde{I}-\widetilde{Z})(\widetilde{I}-\widetilde{X}).
    \end{aligned}
    \label{eq:hopftwistd8}
\end{equation}
Thus $T=\sum_{g\in D_8}\widetilde{T}(\delta_g)\otimes g$, with $\widetilde{T}(\delta_g)=0$ for $g\notin\langle z,t\rangle$.
To determine $\iota$, observe that $T(\widetilde{Z}^a\widetilde{X}^b\otimes1)T^\dagger=\widetilde{Z}^a\widetilde{X}^b\otimes z^at^b$ for $a,b=0,1$. Since $\Delta(z^at^b)=z^at^b\otimes z^at^b$, the choice $\iota(\widetilde{Z}^a\widetilde{X}^b)=z^at^b$ satisfies \eqref{eq:hopftwistiota}. The four Pauli matrices $\widetilde{Z}^a\widetilde{X}^b$ are orthonormal for the normalized trace on $W$. In the physical basis $\ket{a,b}=\ket{z^at^b}$, \eqref{eq:hopftwistmps} gives $A_{J_t}^{a,b}=\widetilde{Z}^a\widetilde{X}^b/2$, with zero amplitude on the other four basis states of the qudit.

For the line $\CD$, choose the MPO basis $\ket{1},\ket{2}$ in which $\rho(z)=-I$ and $\rho(t)=\operatorname{diag}(1,-1)$. Then $(\mathrm{id}\otimes\rho)(T)=\sum_g \widetilde{T}(\delta_g)\otimes\rho(g)$ is diagonal in the MPO basis, with blocks $\widetilde{X}$ and $\widetilde{Z}\widetilde{X}$. The black action tensors in \eqref{eq:hopftwistaction} are therefore $(\phi_{\CD})_1=\bra{1}\otimes\widetilde{X}$ and $(\phi_{\CD})_2=\bra{2}\otimes\widetilde{X}\widetilde{Z}$. For $i=j=1$, the bulk component $(\CO_{\CD})_{11}$ acts on $\ket{a,b}$ as $Z_a$, with eigenvalue $(-1)^a$. Multiplying $\widetilde{Z}^a\widetilde{X}^b$ by $\widetilde{X}$ on the right flips $b$, giving the left dressing $X_b$. Multiplication on the left also contributes $(-1)^a$, giving the right dressing $Z_aX_b$. As in \eqref{eq:repd8oddstring}, the action tensors determine the local operators dressing the truncated $\CD$ symmetry MPO. In the present Hopf-algebra realization, the selected MPO component and its dressings give the simple Pauli string
\begin{equation}
    \bra{A_{J_t}}X_{b,I}\left(\prod_{n=I+1}^{K-1} Z_{a,n}\right)Z_{a,K}X_{b,K}\ket{A_{J_t}}=1.
    \label{eq:hopftwistd8string}
\end{equation}
The two subgroups for $t=s$ and $t=rs$ intersect only at $1$ and $z$. On these two basis states the left dressing produces $\ket{t}$ and $\ket{zt}$, both outside the other subgroup. The same string therefore has zero expectation in the other twisted state and in the product state.

\subsubsection{The two SPT phases of \texorpdfstring{$\Rep(A_{12})$}{Rep(A12)}}
Masuoka's $A_{12}$ \cite{masuoka2000galois}, denoted $A_+$ in \cite{chen2026nissb}, is obtained by twisting the coproduct of $\IC[D_{12}]$, where $D_{12}\cong S_3\times\IZ_2$ \cite{david2011kac}. Write $r=(123)$ and $t=(12)$ in $S_3$, and let $z$ generate the central $\IZ_2$. With $p_\pm=(1\pm z)/2$, the unitary twist $F=F^{-1}$ and the coproduct of $A_{12}$ are
\begin{equation}
    F=p_+\otimes1+p_-\otimes t,\qquad
    \Delta_{12}(g)=F(g\otimes g)F
    =gp_+\otimes g+gp_-\otimes tgt,\qquad g\in S_3\times\IZ_2 .
    \label{eq:a12coproduct}
\end{equation}
Thus $z$ and $t$ are group-like, while $\Delta_{12}(r)=rp_+\otimes r+rp_-\otimes r^{-1}$ is noncocommutative. Multiplying the group Clebsch--Gordan embeddings by $(\rho\otimes\sigma)(F)$ gives the embeddings for the tensor product defined by $\Delta_{12}$. The twist equation ensures that both orders of fusing three lines acquire the same unitary, so their $F$-symbols agree in these bases. Hence $\Rep(A_{12})\simeq\Rep(D_{12})$ as fusion categories \cite[Secs.~5.2, 6.1 and 9.2]{david2011kac}. This category is group-theoretical because it is Morita equivalent to $\VEC_{D_{12}}$ \cite{etingof2015tensor}. We use $\Delta_{12}$ to realize its symmetry MPOs on the same twelve-dimensional qudit chain for both SPT phases.

Each site consists of a qutrit and two qubits, with basis $\ket{m,a,b}=\ket{r^m z^a t^b}$, where $m\in\IZ_3$ and $a,b\in\IZ_2$. The qutrit Pauli operators act as $\CZ\ket{m}=e^{2\pi i m/3}\ket{m}$ and $\CX\ket{m}=\ket{m-1\bmod 3}$ in the convention of \appref{app:qdzx}. We use $X,Z$ on the $z^a$ qubit and $\hat X,\hat Z$ on the $t^b$ qubit. The four one-dimensional lines are the characters $\chi_{u,v}(r^m z^a t^b)=(-1)^{ub+va}$, with $u,v=0,1$. Their symmetry operators on a ring of $N$ sites are
\begin{equation}
    \hCO_{\chi_{u,v}}=\prod_{n=1}^N \hat Z_n^{\,u}Z_n^{\,v}.
    \label{eq:a12charmpo}
\end{equation}
For the two-dimensional lines $\rho_v$, choose $\rho_v(r)$ to be the real rotation through $2\pi/3$, $\rho_v(t)=\operatorname{diag}(1,-1)$ and $\rho_v(z)=(-1)^v I$. Inserting $\Delta_{12}$ in the local MPO tensor \eqref{eq:hopflocalmpo} gives
\begin{equation}
    \CO_{\rho_v}=\frac12
    \begin{pmatrix}
        \CZ+\CZ^\dagger & i(\CZ-\CZ^\dagger)X\hat Z\\
        -i(\CZ-\CZ^\dagger)X & (\CZ+\CZ^\dagger)\hat Z
    \end{pmatrix} Z^{\,v},\qquad
    \hCO_{\rho_v}=\operatorname{Tr}_{V_{\rho_v}}\!\left(\CO_{\rho_v;1}\cdots\CO_{\rho_v;N}\right).
    \label{eq:a12doubletmpo}
\end{equation}
The trace contracts the two-dimensional MPO bond around the ring. Conjugation by $\rho_v(t)$ preserves the diagonal entries and reverses the off-diagonal entries of $\rho_v(g)$. Equation~\eqref{eq:a12coproduct} therefore makes the off-diagonal entries flip the $z^a$ qubit, producing the factor $X$. The order $XZ^{\,v}$ evaluates the central character before this flip.

Fusing two symmetry MPOs combines their $S_3$ representations by tensor product and adds their $\IZ_2$ labels modulo two \cite{lu2026hopfising}. The four invertible lines $\chi_{u,v}$ and the two non-invertible lines $\rho_v$ obey
\begin{equation}
    \begin{aligned}
        \hCO_{\chi_{u,v}}\circ\hCO_{\chi_{u',v'}} &= \hCO_{\chi_{u+u',\,v+v'}},\\
        \hCO_{\chi_{u,v}}\circ\hCO_{\rho_w} = \hCO_{\rho_w}\circ\hCO_{\chi_{u,v}} &= \hCO_{\rho_{v+w}},\\
        \hCO_{\rho_v}\circ\hCO_{\rho_w} &= \hCO_{\chi_{0,\,v+w}}+\hCO_{\chi_{1,\,v+w}}+\hCO_{\rho_{v+w}} .
    \end{aligned}
    \label{eq:a12mpofusion}
\end{equation}
All binary labels are added modulo two, and $\hCO_{\chi_{0,0}}$ is the identity. The three terms in the last line are the trivial, sign and two-dimensional representations in the tensor square of the $S_3$ doublet.

To write the MPOs and Hamiltonians in qutrit and ququart Pauli operators, identify $\ket{a,b}$ with $\ket{\ell}$, where $\ell=2a+b$. The ququart shift $X_4$ and clock $Z_4$ act as $X_4\ket{\ell}=\ket{\ell-1\bmod 4}$ and $Z_4\ket{\ell}=i^\ell\ket{\ell}$. The qubit Pauli operators on each site are then
\begin{equation}
    \begin{gathered}
        X=X_4^2,\qquad \hat Z=Z_4^2,\qquad Z=\tfrac12\big[(1-i)Z_4+(1+i)Z_4^\dagger\big],\\
        \hat X=\tfrac12\big[X_4^\dagger(1+Z_4^2)+X_4(1-Z_4^2)\big].
    \end{gathered}
    \label{eq:a12ququart}
\end{equation}

We now construct the two SPT states and their commuting-projector Hamiltonians on a periodic chain with $N\geq2$. Let $P_n=(1+\CZ_n+\CZ_n^\dagger)/3$ project the qutrit at site $n$ onto $m=0$, with site indices taken modulo $N$. The twist $J=1\otimes1$ gives the product state $\ket{0,0,0}^{\otimes N}=\ket{1}^{\otimes N}$, which is the unique ground state of
\begin{equation}
    H_{1\otimes1}=\sum_{n=1}^{N}\left[1-P_n\,\frac{(1+Z_n)(1+\hat Z_n)}{4}\right].
    \label{eq:a12productham}
\end{equation}
The qubit factor is the ququart projector $\frac14\sum_{p=0}^3 Z_{4,n}^p$ onto $\ell=0$.

For the second phase, take $J=F$ in \eqref{eq:hopftwisteq} with coproduct $\Delta_{12}$. Twisting back gives the ordinary group coproduct $g\mapsto g\otimes g$. Its cocommutativity distinguishes the resulting Hopf algebra from $A_{12}$, so $1\otimes1$ and $F$ are gauge-inequivalent twists. These give the two SPT phases of $\Rep(A_{12})$. Indeed, up to conjugation, $\langle z,t\rangle\cong\IZ_2\times\IZ_2$ is the only nontrivial subgroup of $D_{12}$ supporting a nondegenerate cocycle, and it has one such cocycle class \cite{etingof2015tensor}.

On $\langle z,t\rangle$, $\Delta_{12}$ is the ordinary group coproduct and $F$ has the same form as $J_t$ in the preceding example. We use the virtual qubit tensor $T$ of \eqref{eq:hopftwistd8} and the isometry $\iota(\widetilde{Z}^a\widetilde{X}^b)=z^at^b$. Equation~\eqref{eq:hopftwistmps} gives $A_F^{z^at^b}=\widetilde{Z}^a\widetilde{X}^b/2$, with zero amplitude on the other eight basis states.

On two neighboring sites, the product of the MPS matrices is $(-1)^{ba'}\widetilde{Z}^{a+a'}\widetilde{X}^{b+b'}/4$ for labels $(a,b)$ and $(a',b')$. Flipping both $a$ labels gives a sign $(-1)^b$, and flipping both $b$ labels gives $(-1)^{a'}$. Compensating these signs gives the two Pauli stabilizers of each bond and the parent Hamiltonian
\begin{equation}
    \begin{aligned}
        K^a_n &= X_n \hat Z_n X_{n+1},\qquad
        K^b_n = \hat X_n Z_{n+1} \hat X_{n+1},\\
        H_F &= \sum_{n=1}^{N}\left[1-P_nP_{n+1}\,\frac{(1+K^a_n)(1+K^b_n)}{4}\right].
    \end{aligned}
    \label{eq:a12twistham}
\end{equation}
The factor $P_nP_{n+1}$ fixes both qutrits to $m=0$, while the qubit projector enforces $K^a_n=K^b_n=+1$. A bond with either qutrit outside this subspace costs one unit of energy. In the qubit order $\ldots,a_n,b_n,a_{n+1},b_{n+1},\ldots$, the operators $K^a_n$ and $K^b_n$ are the $XZX$ stabilizers centered on $b_n$ and $a_{n+1}$. Applying a Hadamard to every qubit gives the usual $ZXZ$ cluster stabilizers \cite{nat2023gcluster} and maps $\prod_n Z_n$ and $\prod_n\hat Z_n$ to the two sublattice spin flips. Thus the unique periodic ground state $\ket{A_F}$ is the $\IZ_2\times\IZ_2$ cluster state in this basis, with the qutrits in $\ket{0}^{\otimes N}$.

Both Hamiltonians preserve all six symmetry MPOs on the full qudit chain. To check this, each $P_n$ commutes with the MPO tensors, and at $m=0$ the doublet tensor is diagonal and commutes with the product-state projector. On two neighboring sites with $m=0$, its contraction is $(Z_nZ_{n+1})^v\operatorname{diag}(1,\hat Z_n\hat Z_{n+1})$, whose two parities commute with $K^a_n$ and $K^b_n$. When every qutrit has $m=0$, closing the MPO gives $\hCO_{\rho_v}=\hCO_{\chi_{0,v}}+\hCO_{\chi_{1,v}}$. On this subspace, the doublets therefore act as sums of the invertible symmetry operators.

The same tensor $T$ determines the string dressings. Evaluating its Hopf-algebra leg on $\chi_{u,v}$ gives the virtual matrix $\widetilde{Z}^u\widetilde{X}^v$. For $\rho_v$, the matrix $[\widetilde{T}^{\rho_v}_{ai}]_{a,i}$ is diagonal in the MPO basis, with blocks $\widetilde{X}^v$ and $\widetilde{Z}\widetilde{X}^v$. Choose $\rho_0$ and $i=j=2$, so the virtual matrix in both dressings is $\widetilde{Z}$. On the $m=0$ subspace, the bulk component $(\CO_{\rho_0})_{22}$ acts as $\hat Z$. Multiplying $\widetilde{Z}^a\widetilde{X}^b$ by $\widetilde{Z}$ on the right gives the dressing $X\hat Z$ at site $I$, and multiplication on the left gives $X$ at site $K$. The resulting string is the product $K^a_I\cdots K^a_{K-1}$ of the bond stabilizers, so
\begin{equation}
    \bra{A_F}(X_I\hat Z_I)\left(\prod_{n=I+1}^{K-1}\hat Z_n\right)X_K\ket{A_F}=1,\qquad I<K .
    \label{eq:a12string}
\end{equation}
This dressed string has zero expectation in the product state because each dressing flips $\ket{1}$ to $\ket{z}$. The bare string $\prod_{n=I}^{K}\hat Z_n$ on a nonempty proper interval has expectation one in the product state and zero in $\ket{A_F}$. The latter follows because the $b$ labels have uniform weight subject to even total parity, so their parity on a proper interval averages to zero.

\begingroup
\color{black}\hypersetup{citecolor=ceruleanblue,linkcolor=ceruleanblue}
The same $(2,2)$ component of $\rho_0$ gives a local strange charged operator between the two phases. Contracting the product and twisted state action tensors in \eqref{eq:strangecond} gives $\widetilde Z$ on the virtual bond. In the overlap $\bra{1}^{\otimes N}\ket{A_F}$, a site without an insertion contributes $A_F^1=\widetilde I/2$. Inserting the physical Pauli $X$ gives $A_F^z=\widetilde Z/2$ since $\bra{1}X=\bra{z}$. Thus $2X$ satisfies \eqref{eq:strangecond} with the normalization shown there. The overlap is $2^{1-N}$, and the one-point function vanishes because $\operatorname{tr}(\widetilde Z)=0$. In the Pauli normalization, $\widetilde Z^2=\widetilde I$ gives the connected strange correlator
\begin{equation}
    \frac{\bra{1}^{\otimes N}X_I X_K\ket{A_F}}{\bra{1}^{\otimes N}\ket{A_F}}=1,\qquad I\ne K .
    \label{eq:a12strange}
\end{equation}
\endgroup
We checked \eqref{eq:a12string} and \eqref{eq:a12strange} by evaluating the matrix elements in the physical basis for every pair of distinct sites on periodic chains of two to six sites. OpenAI Codex (GPT-6) assisted with the code for these checks.

\paragraph{Self-dual gauging}
The Hopf self-duality of $A_{12}$ \cite[Theorems~6.1 and 9.2]{david2011kac} also gives a self-dual gauging of this fusion category. Gauging with the canonical fiber functor of $\Rep(A_{12})$, given by $J=1\otimes1$, returns $\Rep(A_{12}^*)\simeq\Rep(A_{12})$ \cite{lu2026hopfising}. In the equivalent $\Rep(D_{12})$ description, this is realized by a twisted multiplication on the regular object. The multiplication tensor is part of the symmetric special Frobenius algebra that specifies the gauging, and inequivalent choices on the same object can give different dual symmetries \cite{Bhardwaj2017finitesym,Perez-Lona:2024sds}.

Here the regular object is $\CA=\bigoplus_{u,v=0,1}\chi_{u,v}\oplus\rho_0^{\oplus2}\oplus\rho_1^{\oplus2}$. Realize it as functions on $D_{12}$, with the group action $(h\cdot f)(g)=f(h^{-1}g)$. Applying $F$ from \eqref{eq:a12coproduct} to the two functions by right translation and then multiplying pointwise gives
\begin{equation}
    (f\star k)(g)=\frac{f(g)+f(gz)}{2}\,k(g)
    +\frac{f(g)-f(gz)}{2}\,k(gt),
    \qquad g\in D_{12}.
    \label{eq:a12gaugingalgebra}
\end{equation}
The unit is the constant function $1$. With the delta functions orthonormal and $m(f\otimes k)=f\star k$, the Frobenius coproduct is $m^\dagger$ and the counit is $f\mapsto\sum_g f(g)$, with $m\circ m^\dagger=\mathrm{id}_{\CA}$. Each of the three cosets of $\langle z,t\rangle$ gives a block $M_2(\IC)$, so this multiplication makes $\CA\cong M_2(\IC)^{\oplus3}$ as an ordinary algebra. Gauging it returns $\Rep(D_{12})$. Pointwise multiplication on the same regular object gives $\CA\cong\IC^{12}$ and the dual symmetry $\VEC_{D_{12}}$ \cite{Bhardwaj2017finitesym}. This second gauging corresponds to the fiber functor with $J=F$ in the $A_{12}$ description.

\section{Conclusion and discussion}\label{sec:concl}
In this paper, we connected the interface algebra to strange correlators and string order parameters for NISPTs in 1+1d. The action tensors give an algebraic construction of local strange charged operators and string decorations on closed chains, providing explicit observables for future numerical studies. We also showed how representations of the interface algebra constrain entanglement-spectrum degeneracies.

The strange correlator probes the representation of the interface algebra realized by the chosen MPSs. {\color{black}\hypersetup{citecolor=ceruleanblue,linkcolor=ceruleanblue} To compare the phases, we first minimize the MPS bond dimension using finite-depth circuits of locally symmetric gates and construct the insertions from the resulting action tensors. For states with the same symmetry action on the same Hilbert space, the local insertions satisfying \eqref{eq:strangecond} exist when both tensors are injective after a common finite blocking. Suitable representatives of distinct NISPT phases exhibit long-range strange correlations. The minimum bond dimension need not select a unique representative. Two phases coincide precisely when their full interface algebra admits a one-dimensional representation \cite{kansei2024nispt}.}

For the string order parameter, we truncate the symmetry MPO and dress it with local charge decoration operators satisfying \eqref{eq:stringcond}. These operators are obtained from the action tensors after finite blocking. For the matched fixed point MPS, action-tensor orthogonality fixes the contraction of the dressing indices and gives a nonzero long-distance expectation value. A string constructed for one phase can nevertheless remain nonzero in another phase, so distinguishing phases requires suitable operator choices and selection rules.

\paragraph{Connection between strange correlator and string order parameter}
Both constructions use the same action tensors. Contracting the charge decoration operators of MPSs $A$ and $B$ gives their interface action and hence the strange charged operator:

\begingroup
\color{black}
\begin{equation}\label{eq:stringdecorstrange}
	\begin{tikzpicture}
		\tikzset{baseline=(current bounding box.center)}
		\draw[thick, black, ->-=.8] (0,0) -- (0.5,0);
		\draw[thick, black, ->-=.8] (0,0.5) -- (0.5,0.5);
		\draw[thick, black] (0.5,0) -- (0.5,0.5);
		\draw[fill=black, draw=black, thick] (0.5,0.25) circle (0.08);
		\draw[thick, black, dashed] (0.5,0.25) -- (1,0.25);
		\draw[thick,gray, ->-=0.25, ->-=0.55,->-=0.9] (0,-0.5) -- (0,1);
		\node[draw, regular polygon, regular polygon sides=3, minimum size=3mm, inner sep=0mm, thick,fill=blue!30, shape border rotate=-30] at (0,0) {};
		\draw[thick, cyan, -<-=.3,-<-=0.9] (-0.5,-0.5) -- (0.5,-0.5);
		\filldraw[fill=blue!30, draw=black, thick] (0,-0.5) circle (0.12);
		\node[left] at (-0.1,0) {${}^A(\bar{\CE}_x)_i$};
		\node[below] at (0,-0.6) {$A^{\otimes n}$};
		\node[draw, regular polygon, regular polygon sides=3, minimum size=3mm, inner sep=0mm, thick,fill=yellow!30, shape border rotate=-30] at (0,0.5) {};
		\draw[thick, orange, ->-=.3,->-=0.9] (-0.5,1) -- (0.5,1);
		\filldraw[fill=yellow!30, draw=black, thick] (0,1) circle (0.12);
		\node[left] at (-0.1,0.5) {${}^B(\CE_{\bar{x}})_{i'}$};
		\node[above] at (0,1.0) {$B^{\otimes n}$};
	\end{tikzpicture} = \begin{tikzpicture}
		\tikzset{baseline=(current bounding box.center)}
		\draw[thick, black, ->-=.8] (0.5,0) -- (1,0);
		\draw[thick, black, -<-=.8] (0.5,0.5) -- (1,0.5);
		\draw[thick, black] (1,0) -- (1,0.5);
		\draw[fill=black, draw=black, thick] (1,0.25) circle (0.08);
		\draw[thick, black, dashed] (1,0.25) -- (1.5,0.25);
		\draw[thick, black, ->-=.8] (0.5,0.5) -- (0.5,1);
		\draw[thick, black, -<-=.8] (0.5,0) -- (0.5,-0.5);
		\draw[thick,gray, ->-=0.35, ->-=0.8] (0,-0.5) -- (0,1);
		\draw[thick, cyan, -<-=.2,-<-=0.6,-<-=0.9] (-0.5,-0.5) -- (1,-0.5);
		\filldraw[fill=blue!30, draw=black, thick] (0,-0.5) circle (0.12);
		\node[below] at (-0.1,-0.6) {$A^{\otimes n}$};
		\draw[thick, orange, ->-=.2,->-=0.6,->-=0.9] (-0.5,1) -- (1,1);
		\filldraw[fill=yellow!30, draw=black, thick] (0,1) circle (0.12);
		\node[above] at (-0.1,1.0) {$B^{\otimes n}$};
		\filldraw[fill=white, draw=black, thick] (0.5,-0.5) circle (0.08);
		\filldraw[fill=black, draw=black, thick] (0.5,1) circle (0.08);
		\node[below] at (0.8,-0.5) {$(\bphi^A_{x})_{i}$};
		\node[above] at (0.8,1) {$(\phi^B_{x})_{i'}$};
	\end{tikzpicture} = \begin{tikzpicture}
		\tikzset{baseline=(current bounding box.center)}
		\draw[thick,gray, ->-=0.35, ->-=0.8] (0,-0.5) -- (0,1);
		\draw[thick,gray, ->-=0.35, ->-=0.8] (0.5,-0.5) -- (0.5,1);
		\draw[thick, cyan, -<-=.2,-<-=0.55,-<-=0.9] (-0.5,-0.5) -- (1,-0.5);
		\filldraw[fill=blue!30, draw=black, thick] (0,-0.5) circle (0.12);
		\filldraw[fill=blue!30, draw=black, thick] (0.5,-0.5) circle (0.12);
		\node[below] at (-0.1,-0.6) {$A^{\otimes n}$};
		\draw[thick, orange, ->-=.2,->-=0.55,->-=0.9] (-0.5,1) -- (1,1);
		\filldraw[fill=yellow!30, draw=black, thick] (0,1) circle (0.12);
		\filldraw[fill=yellow!30, draw=black, thick] (0.5,1) circle (0.12);
		\node[above] at (-0.1,1.0) {$B^{\otimes n}$};
		\node[draw, rectangle,minimum width=2mm, minimum height=2mm, inner sep=0mm,thick, fill=dgreen] at (0.5,0.25) {};
		\node[right] at (0.5,0.25) {$\CS_{I,x,(i,i')}^{A|B}$};
	\end{tikzpicture}
\end{equation}
\par\endgroup

Here $\bar{x}$ is the symmetry line $x$ with reversed orientation. The black junction with the dashed leg contracts $x$ and $\bar{x}$ in the identity channel using the same pairing as in \eqref{eq:strangecond}.

\paragraph{Entanglement spectrum}
The interface algebra also constrains entanglement-spectrum degeneracies, extending the symmetry argument for ordinary SPTs \cite{pollmann2010entanglement}. We consider an on-site symmetry realization with a symmetric product reference \cite{jose2023mpoclas,kansei2024nispt,lan2024mpo}. For a normalized MPS in canonical form, a single cut of an infinite chain gives the Schmidt decomposition

\begingroup
\color{black}
\begin{equation}
	\begin{tikzpicture}
		\tikzset{baseline=(current bounding box.center)}
		\draw[thick,gray] (-1.5,0) -- (-1.5,0.5);
		\draw[thick,gray] (-0.5,0) -- (-0.5,0.5);
		\draw[thick,gray] (1.5,0) -- (1.5,0.5);
		\draw[thick,gray] (0.5,0) -- (0.5,0.5);
		\draw[thick,cyan] (-2,0) -- (2,0);
		\node [above] at (-2,0) {$\cdots$};
		\node [above] at (2,0) {$\cdots$};
		\node[below] at (0,0) {$A$};
	\end{tikzpicture} = \sum_{\alpha=1}^{\dim(V)} \lambda_\alpha \begin{tikzpicture}
		\tikzset{baseline=(current bounding box.center)}
		\draw[thick,gray] (-1.5,0) -- (-1.5,0.5);
		\draw[thick,gray] (-0.5,0) -- (-0.5,0.5);
		\draw[thick,cyan] (-2,0) -- (0,0);
		\filldraw[fill=cyan!30, draw=black, thick] (0,0) circle (0.1);
		\node [above] at (-2,0) {$\cdots$};
		\node [above] at (0,0) {$\alpha$};
		\node[below] at (-1,0) {$A_L$};
	\end{tikzpicture} \begin{tikzpicture}
		\tikzset{baseline=(current bounding box.center)}
		\draw[thick,gray] (1.5,0) -- (1.5,0.5);
		\draw[thick,gray] (0.5,0) -- (0.5,0.5);
		\draw[thick,cyan] (2,0) -- (0,0);
		\filldraw[fill=cyan!30, draw=black, thick] (0,0) circle (0.1);
		\node [above] at (2,0) {$\cdots$};
		\node [above] at (0,0) {$\alpha$};
		\node[below] at (1,0) {$A_R$};
	\end{tikzpicture}
\end{equation}
\par\endgroup

where $\lambda_\alpha$ are the Schmidt coefficients and $\dim(V)$ is the virtual bond dimension. The reduced-density-matrix eigenvalues are $\lambda_\alpha^2$, and the corresponding entanglement energies are $-\ln(\lambda_\alpha^2)$ for nonzero $\lambda_\alpha$. To describe the symmetry action on the left Schmidt states, we replace the right half by the symmetric product reference. The diagram \begingroup\color{black}\begin{tikzpicture}
	\draw[thick,cyan] (-0.5,0) -- (0,0);
	\draw[thick,dashed, gray!30] (0,0)--(0.5,0);
	\filldraw[fill=cyan!30, draw=black, thick] (0,0) circle (0.1);
	\node [above] at (0,0) {$\alpha$};
\end{tikzpicture}\endgroup then represents the interface between $A$ and this reference, and the MPO acts as

\begingroup
\color{black}
\begin{equation}
	\begin{tikzpicture}
		\tikzset{baseline=(current bounding box.center)}
		\draw[thick,black] (-2,0.5) -- (2,0.5);
		\draw[thick,gray] (-1.5,0) -- (-1.5,1);
		\draw[thick,gray] (-0.5,0) -- (-0.5,1);
		\draw[thick,gray] (1.5,0) -- (1.5,1);
		\draw[thick,gray] (0.5,0) -- (0.5,1);
		\draw[thick,dashed, gray!30] (0,0)--(0.5,0);
		\draw[thick,cyan] (-2,0) -- (0,0);
		\filldraw[fill=cyan!30, draw=black, thick] (0,0) circle (0.1);
		\filldraw[fill=yellow!30, draw=black, thick] (0.5,0) circle (0.1);
		\filldraw[fill=yellow!30, draw=black, thick] (1.5,0) circle (0.1);
		\filldraw[fill=gray!30, draw=black, thick] (-1.5,0.5) circle (0.12);
		\filldraw[fill=gray!30, draw=black, thick] (-0.5,0.5) circle (0.12);
		\filldraw[fill=gray!30, draw=black, thick] (1.5,0.5) circle (0.12);
		\filldraw[fill=gray!30, draw=black, thick] (0.5,0.5) circle (0.12);
		\node [above] at (-2,0) {$\cdots$};
		\node [above] at (0,0) {$\alpha$};
		\node[below] at (-1,0) {$A_L$};
		\node[above] at (-2,0.5) {$x$};
	\end{tikzpicture} = \sum_{i,i'=1,\cdots d_x}\begin{tikzpicture}
		\tikzset{baseline=(current bounding box.center)}
		\draw[thick,gray] (-1.5,0) -- (-1.5,0.5);
		\draw[thick,gray] (-0.5,0) -- (-0.5,0.5);
		\draw[thick,gray] (2.5,0) -- (2.5,0.5);
		\draw[thick,gray] (1.5,0) -- (1.5,0.5);
		\draw[thick,dashed, gray!30] (0.5,0)--(1.5,0);
		\draw[thick,cyan] (-2,0) -- (0.5,0);
		\draw[thick,black] (0,0) -- (0,0.5);
		\draw[thick,black] (0,0.5) -- (1,0.5);
		\draw[thick,black] (1,0) -- (1,0.5);
		\filldraw[fill=cyan!30, draw=black, thick] (0.5,0) circle (0.1);
		\filldraw[fill=yellow!30, draw=black, thick] (1.5,0) circle (0.1);
		\filldraw[fill=yellow!30, draw=black, thick] (2.5,0) circle (0.1);
		\filldraw[fill=white, draw=black, thick] (0.0,0) circle (0.08);
		\filldraw[fill=black, draw=black, thick] (1,0) circle (0.08);
		\node [above] at (-2,0) {$\cdots$};
		\node [above] at (0.5,0) {$\alpha$};
		\node[below] at (-1,0) {$A_L$};
		\node[below] at (0,0) {$i$};
		\node[below] at (1,0) {$i'$};
		\node [above] at (0.5,0.5) {$x$};
	\end{tikzpicture}
\end{equation}
\par\endgroup

The yellow product-state tensors carry no virtual bond, so the action on the left Schmidt states is represented by the interface algebra elements $\CI^{A|\text{Tri}}_{x,(i,i')}$. The reduced density matrix commutes with this action, and each eigenspace decomposes into irreducible representations of the interface algebra. Thus, if all irreducible representations have dimension greater than one, every nonzero eigenvalue is degenerate. When one-dimensional representations are allowed, a chosen representative may still realize higher-dimensional irreducible representations and exhibit degeneracy; this degeneracy alone does not distinguish SPT phases. This argument requires a consistent half-chain symmetry action and a symmetric product reference and does not directly apply to a general non-on-site realization \cite{david2024nononsite,lu2024nononsite,sahand2025nononsite}.

We illustrated these constructions with ordinary $\IZ_2\times\IZ_2$ SPTs, the three $\Rep(D_8)$ NISPTs, and intrinsically mixed $G\times\Rep(G)$ SPTs. For the $\Rep(D_8)$ Even and Odd states, the local operators dressing the $\CD$ strings carry different $\IZ_2\times\IZ_2$ symmetry charges. Their interfaces with the Tri product state have only two-dimensional irreducible representations, giving at least twofold entanglement-spectrum degeneracy in the single-cut setting above. The analogous interface algebras for the nontrivial $\Rep(D_{16})$ NISPTs also have only two-dimensional irreducible representations \cite{lu2025symset}.

For $G=S_3$, we gave explicit spin-chain Hamiltonians and interface algebras for the product phase and two intrinsically mixed phases. Two distinct phases share the same sign-string expectation value, while a suitable collection of string operators distinguishes all three fixed point states. The interface algebras also require the nonzero reduced-density-matrix eigenvalues to have multiplicities that are multiples of two and six in phases $\IZ_2$ and $S_3$, respectively, relative to the product reference. These examples show how action tensors give the strange charged operators and string order parameters, and how representations of the interface algebra constrain entanglement-spectrum degeneracies.

{\color{black}We also constructed fixed point $\Rep(H)$ SPT states, their action tensors and string order parameters from unitary Drinfeld twists of a finite-dimensional Hopf $C^*$-algebra $H$. For $\Rep(A_{12})$, the two twists give Hamiltonians for the product and twisted states. The same component of a two-dimensional symmetry line gives the dressed string in \eqref{eq:a12string} and the local Pauli insertion in \eqref{eq:a12strange}. The string distinguishes the two states, and the connected strange correlator between them equals one at every nonzero separation.}

We also discussed a geometric extension to higher dimensions, where a symmetry surface can give rise to a loop-like strange insertion. A general construction requires the interface symmetry data for operators of different codimensions and their junctions, and is left for future work.

\section*{Acknowledgements}
We are grateful to Yimu Bao, Arkya Chatterjee, Yichul Choi, Chao-Ming Jian, 	
Michael Hermele, Yabo Li, Conghuan Luo, Nathanan Tantivasadakarn, Apoorv Tiwari, Tomohiro Soejima, Zhengdi Sun, 	
Ashvin Vishwanath, Yifan Wang, Xinping Yang, Zhehao Zhang for interesting discussions. Research of D.C.L. is supported by the Simons Collaboration on Ultra-Quantum Matter, which is a grant from the Simons Foundation 651440. Y.Z.Y. is supported by the NSF Grant No. DMR-2238360. This research was supported in part by grant NSF PHY-2309135 to the Kavli Institute for Theoretical Physics (KITP).

\section*{AI Use Disclosure}
D.C.L. used OpenAI Codex (GPT-6) to assist with literature comparisons, scientific writing, checking and completing derivations and drawing TikZ diagrams. The authors take responsibility for the paper.

\newpage
\appendix

\section{Fusion and associator of \texorpdfstring{$\IZ_2$}{Z2} Kramers-Wannier duality on lattice}\label{app:z2KW}
The 1+1d transverse-field Ising model has global $\IZ_2$ symmetry, and it has self-duality symmetry at the critical point \cite{sahand2024kw}, under which,
\begin{equation}
	\hCO_\CN X_i = Z_{i-1}Z_{i} \hCO_\CN, \quad \hCO_\CN Z_{i-1}Z_{i} = X_{i-1} \hCO_\CN.
\end{equation}
In this case, the MPO representation of the $\mathbb{Z}_2$ symmetry operator $a$ is
\begin{equation}
    \CO_a=\begin{tikzpicture}[tikzfig]
    \tikzset{baseline=(current bounding box.center)}
	\begin{pgfonlayer}{nodelayer}
		\node [style=red dot] (0) at (0, 0.5) {$a$};
		\node [style=none] (1) at (0, 1.25) {};
		\node [style=none] (2) at (0, -0.25) {};
	\end{pgfonlayer}
	\begin{pgfonlayer}{edgelayer}
		\draw (1.center) to (0);
		\draw (0) to (2.center);
	\end{pgfonlayer}
\end{tikzpicture}
\end{equation}
and the MPO representation of the Kramers-Wannier operator $\CN$ is
\begin{equation}
    \CO_\CN = {\color{black}\sqrt{2}}\,\begin{tikzpicture}[tikzfig]
        \tikzset{baseline=(current bounding box.center)}
	\begin{pgfonlayer}{nodelayer}
		\node [style=green dot] (0) at (0, 0.75) {};
		\node [style=had] (1) at (0.75, 0.75) {};
		\node [style=green dot] (2) at (1.5, 0.75) {};
		\node [style=had] (3) at (2.25, 0.75) {};
		\node [style=none] (4) at (3, 0.75) {};
		\node [style=none] (5) at (-0.75, 0.75) {};
		\node [style=none] (6) at (0, 0) {};
		\node [style=none] (7) at (0.75, 1.25) {};
		\node [style=none] (8) at (0, 2) {};
	\end{pgfonlayer}
	\begin{pgfonlayer}{edgelayer}
		\draw (5.center) to (0);
		\draw (0) to (1);
		\draw (1) to (2);
		\draw (2) to (3);
		\draw (3) to (4.center);
		\draw (0) to (6.center);
		\draw [bend right=45, looseness=0.50] (8.center) to (7.center);
		\draw [bend left=15, looseness=0.75] (7.center) to (2);
	\end{pgfonlayer}
\end{tikzpicture}
\end{equation}

{\color{black}Each one-site tensor for $\CN$ or $\CN^\dagger$ contributes $\sqrt{2}$, giving the factors $4$ and $2$ in the zipping identities below.} They satisfy the fusion rule,
\begin{align}
    &\hCO_a\times \hCO_b=\hCO_{a+b},\; \hCO_a\times \hCO_\CN=\hCO_\CN \times \hCO_a = \hCO_\CN,\\
    &\hCO_\CN  \times \hCO_\CN=\hCO_{T^{-1}}\sum_a \hCO_a,\; \hCO_\CN\times \hCO_{\CN^{\dag}}=\sum_a \hCO_a
\end{align}
where $\hCO_{T^{-1}}$ is the MPO for $T^{-1}$ that implements the lattice translation from site $i$ to site $i-1$, and $\hCO_{\CN^{\dag}}$ is represented as up-side-down of $\hCO_{\CN}$.  
Because the fusion category is anomalous, the MPOs can never satisfy the on-site condition \cite{jose2023mpoclas,lan2024mpo,kansei2024nispt}. However, it can satisfy a weaker ``zipping'' condition, i.e. exist 3-leg a defect fusion tensor, such that
\begin{equation}

\end{equation}
We can derive that,
\begin{equation}\label{eq:kwz2j1}
    %
\end{equation}

Then we can derive,

\begin{equation}
    (F^{\CN \CN^\dag \CN}_\CN)_{ab}=%
\propto(-1)^{ab}
\end{equation}

The fusion of $\CO_\CN$ with itself becomes more complex due to the necessity of incorporating a lattice translation,

\begin{equation}
    \textcolor{black}{4}\,%
\end{equation}

So the fusion junction is

\begin{equation}
%
\end{equation}

Now we derive $(F^{\CN \CN \CN}_{\CN T^{-1}})_{aT^{-1},bT^{-1}}$. If we fuse the first two $\CN$ lines and then fuse the $a$ line with $\CN$, we obtain,

\begin{equation}\label{equ:FNNN1}
%
\end{equation}

If we fuse the last two $\CN$ together first and then fuse $\CN$ with the $a$ line, we obtain,

\begin{equation}\label{equ:FNNN2}
%
\end{equation}

Then we can derive the $(F^{\CN \CN \CN}_{\CN T^{-1}})_{aT^{-1},bT^{-1}}$ by comparing \eqref{equ:FNNN1} and \eqref{equ:FNNN2},

\begin{equation}
%
\propto (-1)^{ab}
\end{equation}
The other $F$-symbols can be derived from the fusion junctions \eqref{eq:kwz2j1}, \eqref{eq:kwz2j2} and \eqref{z2ddfuse}.

\section{Categorical data of \texorpdfstring{$\Rep(D_8)$}{Rep(D8)} and its NISPTs}\label{app:repd8all}
\subsection{Derivation of fusion and splitting junctions for \texorpdfstring{$\Rep(D_8)$}{Rep(D8)} symmetry operators}\label{app:repd81}

The fusion and splitting junctions in the category $\Rep(D_8)$ can be systematically derived following the standard procedure outlined in \cite{kansei2024nispt}. The five objects in $\Rep(D_8)$ are listed as $\{\hCO_{(a,b)}, \hCO_\CD\}$ where $a, b \in \{0,1\}$. Their corresponding matrix product operator (MPO) representations are constructed as follows,

\begin{equation}
    \CO_{(a,b)}=

\end{equation}

We summarize the splitting and fusion junctions into three cases. First, the fusion/splitting junctions between two objects $\CO_{(a_1,b_1)}$ and $\CO_{(a_2,b_2)}$ are
\begin{equation}
%
=(-1)^{a_1b_2}
\end{equation}
Second, the junctions between $\CO_{(a,b)}$ and $\CO_\CD$ are
\begin{equation} 
%
\end{equation}
Therefore, the splitting/fusion junctions are
\begin{equation}
    %
\end{equation}
Finally, we derive the junctions between two $O_D$ operators. Notice that
\begin{equation}
    %
.
\end{equation}
We can also notice that
\begin{equation}
\resizebox{0.94\linewidth}{!}{$\displaystyle
%
\end{equation}
So we obtain the fusion/splitting junctions,
\begin{equation}
    %
\end{equation}

We can easily check the orthogonal relation between them. For example
\begin{equation}\frac{1}{2}
    %
=\delta_{a_1,a_2}\delta_{b_1,b_2}
\end{equation}

\subsection{Derivation of action tensors for \texorpdfstring{$\Rep(D_8)$}{Rep(D8)} symmetry invariant states}\label{app:repd82}

There are 3 NISPT states under $\Rep(D_8)$ symmetry, the MPS form of them are,
\begin{equation}
 %
\end{equation}
We derive the action tensors following the notation in \cite{kansei2024nispt}.

For the product state, if we act $\CO_{(a,b)}$, we obtain,
\begin{equation}
    %
.
\end{equation}
If we act $\CO_\CD$, we obtain,
\begin{equation}
    %
\end{equation}
Therefore we derive,
\begin{equation}
    %
\end{equation}
For the Even state, if we act $\CO_{(a,b)}$, we obtain,
\begin{equation}
%
.
\end{equation}
If we act $\CO_{\CD}$, we obtain,
\begin{equation*}2
    %
,\bar{i}=i+\pi.
\end{equation}
The minus sign is the scalar phase of the four-site $\CD$ action; it cancels after blocking eight sites, as explained in \secref{sec:repd8mps}.
Where we used the rule,
\begin{equation}
   \text{CNOT}_{i,j}=\ket{0}_i\bra{0}_i\otimes I_j+\ket{1}_i\bra{1}_i\otimes X_j,
\end{equation}
which means,
\begin{equation}\sqrt{2}\;
    %
\end{equation}
Therefore the action tensors of the Even state are
\begin{equation}
    %
\end{equation}
Similarly, we can also derive the action tensors of the Odd state,
\begin{equation}
        %
\end{equation}
We can easily check the orthogonal relations between them.

\subsection{Derivation of \texorpdfstring{$F$}{F}-symbols and \texorpdfstring{$L$}{L}-symbols for \texorpdfstring{$\Rep(D_8)$}{Rep(D8)} SPTs}\label{app:FLRepD8}

We first derive the $F$-symbols. The $F$-symbols can be expressed in terms of the following closed diagrams,

\begin{equation}
{F_{w}^{xyz}}_{(u,\mu,\nu),(v,\rho,\sigma)} =\frac{1}{D}\;%
\end{equation}
where $D$ is the MPO bond dimension of the line $w$. We derive the $F$-symbols for $\Rep(D_8)$ symmetry operators,
\begin{equation}
{F^{{(a_1,b_1)},{(a_2,b_2)},{(a_3,b_3)}}_{{(a_1+a_2+a_3,b_1+b_2+b_3)}}}_{{(a_1+a_2,b_1+b_2)},{(a_2+a_3,b_2+b_3)}}=1
\end{equation}
\begin{equation}
{F^{{(a_1,b_1)},\CD,{(a_2,b_2)}}_{\CD}}_{\CD,\CD}
=\frac{1}{2}
%
=\frac{1}{2}(-1)^{a_2b_1+a_1b_2}
\end{equation}

We then derive the $L$-symbols. The $L$-symbols can be expressed in terms of the following closed diagrams,
\begin{equation}
 (L_{xy}^{z})_{(i,j),(k,\mu)}=\frac{1}{D}\;
    %
 
\end{equation}
In the $\Rep(D_8)$ $L$-symbols below, the Pauli matrix entries $X_{i,i'},Y_{i,i'},Z_{i,i'}$ give coefficients in the action-tensor labels $i,i'$. The factor $1/\sqrt{2}$ in the $\CD\otimes\CD$ coefficients comes from the normalized fusion junction.\par
\begin{equation}
    L^{{(a_1,b_1)},{(a_2,b_2)}}_{\text{Tri}}=(-1)^{a_1b_2}
\end{equation}

\begin{equation}
{L^{{(a,b)},{\CD}}_{\text{Tri}}}_{i,i'}
=\frac{1}{2}
%
\right.
\end{equation}

The $L$-symbol has gauge freedom and the gauge transformation of the $L$-symbols is given by \cite{kansei2024nispt},
\begin{equation}
L^{x;i,y;j}_{(z;k)}\rightarrow {}^U L^{x;i,y;j}_{(z;k)}=\sum_{i',j',k'}(U_x)_{i,i'}(U_y)_{j,j'}(U_z)^{-1}_{k'k}L^{x;i',y;j'}_{(z;k')}
\end{equation}

For the Tri$|$Even table in \secref{sec:repd8mps}, the endpoint indices are ordered as (Tri, Even), opposite to their order in the simplified $\CD$ interface diagram. Starting from the action tensors used above, change the basis of their labels with $U_{\CD}^{\text{Tri}}=-H$, $U_{\CD}^{\text{Even}}=ZH$, and $U_{(a,b)}^{\text{Tri}}=U_{(a,b)}^{\text{Even}}=1$. Conjugating the resulting interface matrices by $\widetilde{H}\widetilde{Z}$ on the Even MPS bond then gives the first representation in the table.

\section{Qudit ZX-Calculus rules}\label{app:qdzx}
We summarize the ZX-calculus rules used in the tensor contractions. Following \cite{wang2021zxqudit}, we work in a $d$-dimensional Hilbert space with orthonormal basis $\{\ket{i}:i=0,1,\dots,d-1\}$ and use the following diagrammatic conventions.

\begin{enumerate}
    \item $Z$-spider:$\quad

$
\end{enumerate}
We introduce the diagrams for the generalized Pauli operators, also known as shift and clock operators. $Z$ and $X$ are $d\times d$ matrices, acting on the states as $X\ket{n}=\ket{n-1},Z\ket{n}=\omega^n\ket{n},\omega=e^{i\frac{2\pi}{d}}$. They satisfy the algebra $XZ = \omega ZX$. $Z$ and $X$, their eigenstates, and their algebra have the following ZX-calculus representations. 

\begin{itemize}
    \item $
        Z=\sum_{j=0}^{d-1}{\omega^j}\ket{j}\bra{j}=%
$
\end{itemize}
The generalized controlled-$Z$ and controlled-$X$ gates are represented by
\begin{itemize}
\item Controlled-$Z$ gate:
$\sqrt{d}\;%
$
\end{itemize}
We now present key rewrite rules for ZX diagrams. Proper attention to arrow directions is essential, since correct orientation determines the validity of all subsequent diagram transformations.
\begin{enumerate}
    \item Spider fusion:
    $%
$ and similarly for the X-spiders.

    \item Hadamard cancellation:
    $%
$ and similarly with Z/X-spiders exchanged.

\item $K_j$ commutation:
$%
$ and similarly with Z/X-spiders exchanged.

\item Color change:
\[\begin{gathered}
%
$

\item Rewrite rules for antipode $S$ operator
\begin{itemize}
\item The basic rules of $S$ operator,
\begin{equation}
%
\end{equation}

\item It can change the orientation of the arrow between different colors,
\begin{equation}
%
\end{equation}

\end{itemize}

\end{enumerate}

\section{On-site MPO representation of \texorpdfstring{$\text{TY}(\mathbb{Z}_N\times \mathbb{Z}_N,\chi_{\text{off-diag}},+1)$}{TY(ZN x ZN, chi-off-diag, +1)}}\label{app:znzn}
We derive an on-site MPO realization of $\TY(\IZ_N\times\IZ_N,\chi_{\text{off-diag}},+1)$. There are $N^2$ invertible symmetry lines and one non-invertible line. {\color{black}We draw one two-site cell for each state or operator. The periodic contraction includes the indicated scalar once per cell.} The invertible symmetry operators are
 \begin{equation}\prod_{k}X^{i}_{2k} X^{j}_{2k+1}=
\equiv \CO_{(i,j)}
 \end{equation}
 where $i,j=0,1,\dots,N-1$ and $\omega=e^{2\pi i/N}$. We use the qudit ZX conventions of \appref{app:qdzx}. The non-invertible operator $\CO_{\text{KW}}$ is \cite{nat2023prestate}
 \begin{equation}\CO_{\text{KW}}=\textcolor{black}{N}\,
 %
 \end{equation}
 The cluster state
 \begin{equation}\ket{\text{cluster}}=
 %
 \end{equation}
 {\color{black}has eigenvalue $1$ under each invertible symmetry operator and eigenvalue $N$ under $\CO_{\text{KW}}$.}

 We can conjugate the symmetry operators by the entangler

 \begin{equation}
U_T= \prod_{i} \cz_{2i,2i+1}\cz^{-1}_{2i+1,2i+2}=\textcolor{black}{N}\,
 %
 \end{equation}
 {\color{black}In the product-state basis the non-invertible symmetry operator is $\CO_\CD=U_T^\dagger\CO_{\text{KW}}U_T$.}
 \begin{equation}
 \textcolor{black}{N}\,
 %
 \equiv \CO_\CD
 \end{equation}
 {\color{black}Acting with $U_T^\dagger$ on the cluster state gives the product state with each qudit in $N^{-1/2}\sum_{q=0}^{N-1}\ket{q}$.}
 \begin{equation}
 %
 \end{equation}

The fusion/splitting junctions between $\CO_{(i_1,j_1)}$ and $\CO_{(i_2,j_2)}$ are

\begin{equation}
    %
=
{\omega^{-i_1j_2}}
\end{equation}

The fusion/splitting junctions between $\CO_{(i,j)}$ and $\CO_\CD$ {\color{black}follow from the diagrams below. The common factor $N^2$ from the two $\CD$ cells cancels on both sides.}

\begin{equation}
\resizebox{0.94\linewidth}{!}{$\displaystyle
\begin{aligned}
%
\end{aligned}
$}
\end{equation}

Therefore, the fusion/splitting junctions between $\CO_{(i,j)}$ and $\CO_\CD$ are
\begin{equation}
    %
\end{equation}

Similarly, we can derive the fusion/splitting junctions between $\CO_\CD$ and $\CO_\CD$,

\begin{equation}
%
\end{equation}
From the fusion junctions, we calculate the $F$-symbols,
\begin{equation}
	F^{\CD g \CD}_{h} =F^{g \CD h}_{\CD} = {\omega^{g_1 h_2 + g_2 h_1}},\quad F^{\CD \CD \CD}_{\CD,(g,h)} = \frac{1}{N}\omega^{-g_1 h_2 - g_2 h_1}
\end{equation}
with all other allowed coefficients equal to $1$. These MPOs realize $\TY(\IZ_N\times\IZ_N,\chi_{\text{off-diag}},+1)$, where
\begin{equation}
	{\chi_{\text{off-diag}}((g_1,g_2),(h_1,h_2))=\omega^{g_1 h_2 + g_2 h_1}}
\end{equation}
The three-$\CD$ $F$-matrix contains $\chi_{\text{off-diag}}^{-1}/N$, consistent with the junction diagrams and the convention $XZ=\omega ZX$ \cite{tambara1998tensor,Gelaki2009grpty}.

We can obtain the action tensor of the $\TY(\IZ_N\times \IZ_N,\chi_{\text{off-diag}},+1)$ trivial state by definition,

\begin{equation}
    \begin{tikzpicture}
    \tikzset{baseline=(current bounding box.center)}
    \draw[thick, orange, -<-=.3,-<-=0.9] (-0.5,0) -- (0.5,0);
    \draw[thick, black] (0.0,0.0) -- (0.0,0.5);
    \draw[thick, black,->-=0.6] (-0.5,0.5) -- (0.0,0.5);
    \filldraw[fill=black, draw=black, thick] (0,0) circle (0.08);
    \node[above] at (-0.5,0) {Trivial};
    \node[above] at (-0.5,0.5) {${(i, j)}$};
\end{tikzpicture}\ = \
1,\quad
\begin{tikzpicture}
    \tikzset{baseline=(current bounding box.center)}
    \draw[thick, orange, -<-=.3, -<-=.9] (-0.5,0) -- (0.5,0);
    \draw[thick, black] (0,0) -- (0,0.5);
    \draw[thick, black, ->-=.6] (0.,0.5) -- (0.5,0.5);
    \node[above] at (-0.5,0) {Trivial};
    \node[above] at (0.5,0.5) {${(i, j)}$};
    \filldraw[fill=white, draw=black, thick] (0,0) circle (0.08);
    \end{tikzpicture}
    =
   1
\end{equation}

\begin{equation}
    \begin{tikzpicture}
    \tikzset{baseline=(current bounding box.center)}
    \draw[thick, orange, -<-=.3,-<-=0.9] (-0.5,0) -- (0.5,0);
    \draw[thick, black] (0.0,0.0) -- (0.0,0.5);
    \draw[thick, black,->-=0.6] (-0.5,0.5) -- (0.0,0.5);
    \filldraw[fill=black, draw=black, thick] (0,0) circle (0.08);
    \node[above] at (-0.5,0) {Trivial};
    \node[above] at (-0.5,0.5) {${\CD}$};
    \node[below] at (0.,0.) {$m$};
\end{tikzpicture}=\frac{1}{\sqrt{N}}
\begin{tikzpicture}
\tikzset{baseline=(current bounding box.center)}
	\begin{pgfonlayer}{nodelayer}
		\node [style=red dot] (0) at (1, 0) {$K_m$};
		\node [style=none] (1) at (0, 0) {};
	\end{pgfonlayer}
	\begin{pgfonlayer}{edgelayer}
		\draw [style=narrow] (0) to (1.center);
	\end{pgfonlayer}
\end{tikzpicture}
,\quad
\begin{tikzpicture}
    \tikzset{baseline=(current bounding box.center)}
    \draw[thick, orange, -<-=.3, -<-=.9] (-0.5,0) -- (0.5,0);
    \draw[thick, black] (0,0) -- (0,0.5);
    \draw[thick, black, ->-=.6] (0.,0.5) -- (0.5,0.5);
    \node[above] at (-0.5,0) {Trivial};
    \node[above] at (0.5,0.5) {${\CD}$};
    \node[below] at (0.,0.) {$m$};
    \filldraw[fill=white, draw=black, thick] (0,0) circle (0.08);
    \end{tikzpicture}
=\frac{1}{\sqrt{N}}
\begin{tikzpicture}
\tikzset{baseline=(current bounding box.center)}
	\begin{pgfonlayer}{nodelayer}
		\node [style=red dot] (0) at (0, 0) {$K_m^\dag$};
		\node [style=none] (1) at (1, 0) {};
	\end{pgfonlayer}
	\begin{pgfonlayer}{edgelayer}
		\draw [style=narrow] (1.center) to (0);
	\end{pgfonlayer}
\end{tikzpicture}
\end{equation}

\section{Module functors and interface algebra}\label{app:functor}
Let $M$ and $N$ be the indecomposable module categories describing two NISPTs of $\CC$. Each has one simple object up to isomorphism. The $L$-symbols relate two successive symmetry actions to the action of the fused line, as shown for $M$:
\begin{equation}
	\begin{tikzpicture}
		\tikzset{baseline=(current bounding box.center)}
		\draw[thick, black] (0.5,1) -- (0,0.25);
		\draw[thick, black] (1,1) -- (0,-0.5);
		\draw[thick,cyan] (0,-1) -- (0,1);
		\filldraw[fill=black, draw=black, thick] (0,0.25) circle (0.05);
		\filldraw[fill=black, draw=black, thick] (0,-0.5) circle (0.05);
		\node[above] at (0.5,1) {$y$};
		\node[above] at (1,1) {$x$};
		\node[above] at (0,1) {$M$};
		\node[left] at (0,0.25) {$j$};
		\node[left] at (0,-0.5) {$i$};
	\end{tikzpicture} =\sum_{z,k,\mu} ({}^M L_{xy}^z)_{(i,j),(k,\mu)} \begin{tikzpicture}
		\tikzset{baseline=(current bounding box.center)}
		\draw[thick, black] (0.5,1) -- (0.5,0.25);
		\draw[thick, black] (1,1) -- (0,-0.5);
		\draw[thick,cyan] (0,-1) -- (0,1);
		\filldraw[fill=black, draw=black, thick] (0.5,0.25) circle (0.05);
		\filldraw[fill=black, draw=black, thick] (0,-0.5) circle (0.05);
		\node[above] at (0.5,1) {$y$};
		\node[above] at (1,1) {$x$};
		\node[above] at (0,1) {$M$};
		\node[right] at (0.5,0.25) {$\mu$};
		\node[left] at (0,-0.5) {$k$};
		\node[right] at (0.25,-0.25) {$z$};
	\end{tikzpicture}
\end{equation}

A $\CC$-module functor specifies how a symmetry line passes through the interface consistently with these fusion relations \cite{kitaev2012gapbdy,etingof2015tensor}. The red circle denotes the interface, and the coefficients $(\omega_N^{M,x})_{(i,i')}$ are operators acting on its states:
\begin{equation}
	\begin{tikzpicture}
		\tikzset{baseline=(current bounding box.center)}
		\draw[thick, black] (1,1) -- (0,-0.5);
		\draw[thick,cyan] (0,0.25) -- (0,1);
		\draw[thick,orange] (0,0.25) -- (0,-1);
		\filldraw[fill=red!50, draw=black, thick] (0,0.25) circle (0.1);
		\filldraw[fill=black, draw=black, thick] (0,-0.5) circle (0.05);
		\node[above] at (1,1) {$x$};
		\node[above] at (0,1) {$M$};
		\node[left] at (0,-0.5) {$i'$};
		\node[below] at (0,-1) {$N$};
	\end{tikzpicture} =\sum_i (\omega_N^{M,x})_{(i,i')} \begin{tikzpicture}
		\tikzset{baseline=(current bounding box.center)}
		\draw[thick, black] (0.5,1) -- (0,0.25);
		\draw[thick,cyan] (0,-0.5) -- (0,1);
		\draw[thick,orange] (0,-0.5) -- (0,-1);
		\filldraw[fill=black, draw=black, thick] (0,0.25) circle (0.05);
		\filldraw[fill=red!50, draw=black, thick] (0,-0.5) circle (0.1);
		\node[above] at (0.5,1) {$x$};
		\node[above] at (0,1) {$M$};
		\node[left] at (0,0.25) {$i$};
		\node[below] at (0,-1) {$N$};
	\end{tikzpicture}
\end{equation}
The indices $i$ and $i'$ label the action tensors on the $M$ and $N$ sides, respectively. Moving $x$ and $y$ through the interface in succession must agree with first fusing them:
\begin{equation}
    \sum_{i,j} ({}^M L_{xy}^{z})_{(i,j),(k,\mu)} (\omega_N^{M,x})_{(i,i')} (\omega_N^{M,y})_{(j,j')}
    = \sum_{k'} (\omega_N^{M,z})_{(k,k')} ({}^N L_{xy}^{z})_{(i',j'),(k',\mu)}.
    \label{eq:modulefunctorcoherence}
\end{equation}
The identity line acts as $(\omega_N^{M,\mathbf{1}})_{(1,1)}=I$, where $I$ is the identity on the interface states. Using the unitarity of the $L$-symbols, \eqref{eq:modulefunctorcoherence} gives the multiplication rule \eqref{eq:interalg}. Thus, $\CI_{x,(i,i')}^{M|N}\mapsto(\omega_N^{M,x})_{(i,i')}$ defines a representation of the interface algebra on these states \cite{kitaev2012gapbdy,kansei2024nispt}.

The phases described by $M$ and $N$ coincide precisely when their fiber functors are equivalent. This is equivalent to the existence of a one-dimensional representation of the interface algebra \cite{kansei2024nispt}.

\section{Action tensors and interface algebra for \texorpdfstring{$\Rep(H_8)$}{Rep(H8)}}\label{app:h8actions}
We give the action tensors for the product state $\ket{P}$ and the nontrivially $\IZ_2\times\IZ_2$-twisted state $\ket{C}$ discussed in \secref{sec:interalg}, using the same symmetry MPO on an eight-dimensional local Hilbert space. We use the orthonormal basis $\{\ket{1},\ket{x},\ket{y},\ket{xy},\ket{z},\ket{xz},\ket{yz},\ket{xyz}\}$ and the Hopf-algebra presentation \cite{lu2026hopfising}
\begin{equation}
    x^2=y^2=1,\qquad xy=yx,\qquad zx=yz,\qquad zy=xz,\qquad z^2=\frac{1+x+y-xy}{2}.
    \label{eq:h8relations}
\end{equation}
The coproduct is determined by
\begin{equation}
\begin{aligned}
    \Delta(x)&=x\otimes x,\qquad \Delta(y)=y\otimes y,\\
    \Delta(z)&=\frac{1}{2}(1\otimes1+y\otimes1+1\otimes x-y\otimes x)(z\otimes z),
\end{aligned}
    \label{eq:h8coproduct}
\end{equation}
with $\Delta(1)=1\otimes1$. Writing $\Delta(k)=\sum_{h,\ell}\Delta_k^{h,\ell}h\otimes\ell$, the right coaction gives the local symmetry MPO tensor
\begin{equation}
    \CO_\rho^{h,k}=\sum_\ell\Delta_k^{h,\ell}\rho(\ell),
    \label{eq:h8mpotensor}
\end{equation}
where $h,k,\ell$ run over the eight Hopf basis elements and $\rho$ is an irreducible representation of $H_8$. The indices $h,k$ are the physical indices, while the matrix indices of $\rho$ are the MPO virtual indices, giving bond dimension $d_\rho$.

The product state is $\ket{P}=\ket{1}^{\otimes N}$, with MPS tensor $P^1=1$ and $P^h=0$ for $h\ne1$. The twisted state has a two-dimensional MPS virtual space and tensors
\begin{equation}
\begin{aligned}
    C^1&=\frac{\widetilde{I}}{2},\qquad C^x=\frac{\widetilde{X}}{2},\qquad C^y=\frac{\widetilde{Z}}{2},\qquad C^{xy}=\frac{\widetilde{X}\widetilde{Z}}{2},\\
    C^z&=C^{xz}=C^{yz}=C^{xyz}=0,
\end{aligned}
    \label{eq:h8twistedmps}
\end{equation}
where $\widetilde{X},\widetilde{Y},\widetilde{Z}$ are Pauli matrices and $\widetilde{I}$ is the identity on the MPS virtual qubit. The matrices $\widetilde{X}$ and $\widetilde{Z}$ represent the two generators of the twisted group algebra of $\IZ_2\times\IZ_2$; their anticommutation realizes the nontrivial cocycle class. Both MPSs are injective on one site: $P$ has bond dimension one, while the nonzero matrices in \eqref{eq:h8twistedmps} span $M_2(\IC)$.

Following the orientation in \eqref{eq:oactonA}, $(\phi_\rho^A)_i$ maps the MPO virtual space tensored with the MPS virtual space to the MPS virtual space, for $A=P,C$. We choose $(\bphi_\rho^A)_i=(\phi_\rho^A)_i^\dagger$. In components, the action-tensor equation is
\begin{equation}
    \sum_k\CO_\rho^{h,k}\otimes A^k
    =\sum_i(\bphi_\rho^A)_i A^h(\phi_\rho^A)_i.
    \label{eq:h8actioncondition}
\end{equation}
\begin{samepage}
For the four one-dimensional representations, the index $i$ is omitted. Their representation matrices and action tensors are
\begingroup
\renewcommand{\arraystretch}{1.25}
\setlength{\arraycolsep}{9pt}
\begin{equation}
\begin{array}{|c|c|c|c|c|c|}
    \hline
    \rho & \rho(x) & \rho(y) & \rho(z) & \phi_\rho^P & \phi_\rho^C\\
    \hline
    1 & 1 & 1 & 1 & 1 & \widetilde{I}\\
    \hline
    s_1 & -1 & -1 & -i & 1 & \widetilde{Y}\\
    \hline
    s_2 & -1 & -1 & i & 1 & \widetilde{Y}\\
    \hline
    s_3 & 1 & 1 & -1 & 1 & \widetilde{I}\\
    \hline
\end{array}
    \label{eq:h8invertibleactions}
\end{equation}
\endgroup
\end{samepage}
For the non-invertible line $\CD$, we choose a basis of the MPO virtual space in which $\rho_\CD(x)=-\rho_\CD(y)=\operatorname{diag}(1,-1)$ and $\rho_\CD(z)$ exchanges $\ket{0}$ and $\ket{1}$. The two action tensors, labeled here by $i=0,1$, are
\begin{equation}
\begin{aligned}
    (\phi_\CD^P)_0&=\bra{0},&\qquad (\phi_\CD^P)_1&=\bra{1},\\
    (\phi_\CD^C)_0&=\bra{0}\otimes\widetilde{X},& (\phi_\CD^C)_1&=\bra{1}\otimes\widetilde{Z}.
\end{aligned}
    \label{eq:h8noninvertibleactions}
\end{equation}
The bras contract the MPO virtual index, while the Pauli matrices in the second line act on the MPS virtual qubit. Taking adjoints gives the corresponding $\bphi$ tensors. They obey $(\phi_\rho^A)_i(\bphi_\rho^A)_j=\delta_{ij}\widetilde{I}_A$, where $\widetilde{I}_A$ is the identity on the MPS bond of $A$, as in \eqref{eq:actionorthogonality}.

To check \eqref{eq:h8actioncondition}, take $g=x^a y^b$ with $a,b\in\{0,1\}$. The coproduct is $\Delta(g)=g\otimes g$, and $\rho_\CD(g)=\operatorname{diag}((-1)^b,(-1)^a)$. The non-invertible action follows from $\widetilde{X}C^g\widetilde{X}=(-1)^b C^g$ and $\widetilde{Z}C^g\widetilde{Z}=(-1)^a C^g$. The invertible actions follow from $\widetilde{Y}C^g\widetilde{Y}=(-1)^{a+b}C^g$. The coproduct preserves the spans of $\{1,x,y,xy\}$ and $\{z,xz,yz,xyz\}$ separately; since both MPS tensors vanish on the latter, the remaining four physical components of \eqref{eq:h8actioncondition} vanish on both sides.

The twisted comodule algebra defining $\ket{C}$ is equivariantly Morita equivalent to the trivial algebra defining $\ket{P}$ \cite{tambara2000representations,vangrinsven2025comodule,lu2026hopfising}. Thus, the two states represent the same $\Rep(H_8)$ SPT phase, as discussed in \secref{sec:interalg}.

\begingroup
\color{black}
\hypersetup{citecolor=ceruleanblue,linkcolor=ceruleanblue}
\paragraph{Interface algebra and its representations}
We now calculate the representations of $\mathfrak I^{P|C}$ in the action-tensor bases above. We suppress the superscript $P|C$ on $\CI_{\rho,(i,j)}$ and omit the indices for the invertible lines. The algebra has eight basis elements: $\CI_1,\CI_{s_1},\CI_{s_2},\CI_{s_3}$ and the four $\CI_{\CD,(i,j)}$, with $i,j=0,1$. Contracting the action tensors according to \eqref{eq:Lsymbol} and substituting the resulting $L$-symbols into \eqref{eq:interalg}, we find that $\CI_{s_3}$ is central and $\CI_{s_3}^2=\CI_1$. The two central projectors
\begin{equation}
    p_\pm=\frac{\CI_1\pm\CI_{s_3}}{2}
    \label{eq:h8interfaceprojectors}
\end{equation}
satisfy $p_+p_-=0$ and $p_++p_-=\CI_1$. The diagonal generators $\CI_{\CD,(0,0)},\CI_{\CD,(1,1)}$ lie in the $p_+$ part of the algebra, while $\CI_{\CD,(0,1)},\CI_{\CD,(1,0)}$ lie in the $p_-$ part. Their products are
\begin{equation}
\begin{aligned}
    \CI_{\CD,(0,0)}^2&=\CI_{\CD,(1,1)}^2=p_+,\\
    \CI_{\CD,(0,0)}\CI_{\CD,(1,1)}&=-\CI_{\CD,(1,1)}\CI_{\CD,(0,0)}=\frac{i}{2}(\CI_{s_1}+\CI_{s_2}),\\
    \CI_{\CD,(0,1)}^2&=\CI_{\CD,(1,0)}^2=p_-,\\
    \CI_{\CD,(0,1)}\CI_{\CD,(1,0)}&=\CI_{\CD,(1,0)}\CI_{\CD,(0,1)}=\frac{1}{2}(\CI_{s_1}-\CI_{s_2}).
\end{aligned}
    \label{eq:h8interfaceproducts}
\end{equation}
The two generators in the $p_+$ part anticommute and square to its identity, giving $M_2(\IC)$. The two generators in the $p_-$ part commute and square to its identity. Assigning each of them an independent sign therefore gives four one-dimensional representations.

\begin{samepage}
The four one-dimensional representations $\lambda_{\epsilon,\eta}$, labeled by independent signs $\epsilon,\eta=\pm1$, and the two-dimensional representation $R$ are
\begingroup
\renewcommand{\arraystretch}{1.25}
\setlength{\arraycolsep}{15pt}
\begin{equation}
\begin{array}{|c|c|c|}
    \hline
    & \lambda_{\epsilon,\eta} & R\\
    \hline
    \CI_1 & 1 & \widetilde{I}\\
    \hline
    \CI_{s_1} & \epsilon\eta & -\widetilde{Y}\\
    \hline
    \CI_{s_2} & -\epsilon\eta & -\widetilde{Y}\\
    \hline
    \CI_{s_3} & -1 & \widetilde{I}\\
    \hline
    \CI_{\CD,(0,0)} & 0 & \widetilde{X}\\
    \hline
    \CI_{\CD,(0,1)} & \epsilon & 0\\
    \hline
    \CI_{\CD,(1,0)} & \eta & 0\\
    \hline
    \CI_{\CD,(1,1)} & 0 & \widetilde{Z}\\
    \hline
\end{array}
    \label{eq:h8interfacerepresentations}
\end{equation}
\endgroup
\end{samepage}
Each assignment satisfies the multiplication rule \eqref{eq:interalg}. The four choices of $(\epsilon,\eta)$ are inequivalent, and $R$ is irreducible because $\widetilde{X}$ and $\widetilde{Z}$ generate $M_2(\IC)$. Since $4\times1^2+2^2=8=\dim\mathfrak I^{P|C}$, these are all the irreducible representations. Thus
\begin{equation}
    \mathfrak I^{P|C}\cong\IC^{\oplus4}\oplus M_2(\IC),\qquad d_{\min}(P|C)=1,
    \label{eq:h8mixedinterfacedecomposition}
\end{equation}
in agreement with the self-interface algebra in \cite{kansei2024nispt}. The action-tensor contraction between the particular MPSs $P$ and $C$ realizes the two-dimensional representation $R$. The full interface algebra also has four one-dimensional representations, allowing a nondegenerate symmetry-preserving interface.

The string and entanglement diagnostics also depend on the chosen representatives. The bare truncated $\CD$ MPO has expectation values $2$ and $0$ in $\ket{P}$ and $\ket{C}$, whereas the string dressed using the $C$ action tensors, summed over its two components, has expectation values $0$ and $2$, respectively. These fixed microscopic strings distinguish the representatives, not their SPT phases. For a single cut, the Schmidt probabilities are $(1)$ and $(1/2,1/2)$. The twofold degeneracy is allowed by the two-dimensional representation, but is not enforced throughout the phase because one-dimensional representations also exist.

{\color{black}\hypersetup{citecolor=ceruleanblue,linkcolor=ceruleanblue}On an even periodic chain, a depth-two circuit of locally $\Rep(H_8)$-symmetric gates maps the cluster state $\ket{C}$ to the product state $\ket{P}$ within the same SPT phase. The ordinary $\IZ_2\times\IZ_2$ SPT entangler $U_T=\prod_i\cz_{i,i+1}$ connects distinct phases and preserves the symmetry only as a full circuit on the ring. In our locally symmetric circuit, the first layer acts on disjoint pairs $(1,2),(3,4),\ldots$. For each choice of the outer MPS indices $i,k=0,1$, the gate exchanges the normalized two-site cluster vector $\ket{A_{ik}}$ with the product vector $\ket{\ell_i,r_k}$, where
\[
\begin{gathered}
    \ket{A_{ik}}=\sqrt{2}\sum_{g,h}(C^g C^h)_{ik}\ket{g,h},\\
\begin{aligned}
    \ket{\ell_0}&=(\ket{xz}+\ket{xyz})/\sqrt{2}, & \ket{r_0}&=-\ket{yz},\\
    \ket{\ell_1}&=(\ket{xyz}-\ket{xz})/\sqrt{2}, & \ket{r_1}&=\ket{z}.
\end{aligned}
\end{gathered}
\]
The gate acts as the identity on the orthogonal complement of these eight vectors. By \eqref{eq:h8coproduct}, the two-site coaction acts by the same matrices on both sets of vectors, so the gate is locally symmetric. Contracting the remaining MPS indices gives a product of dimers $(\ket{r_0,\ell_0}+\ket{r_1,\ell_1})/\sqrt{2}$ on bonds $(2,3),\ldots,(N,1)$. The dimer and $\ket{1,1}$ both have trivial total coaction, so a locally symmetric gate can exchange them. Applying this gate on each dimer in the second layer gives the product state $\ket{P}=\ket{1}^{\otimes N}$.}
\par\endgroup

\bibliographystyle{utphys}
\bibliography{main}
\end{document}